\documentclass[twocolumn, showpacs,amsmath,amssymb,pra,superscriptaddress,floatfix,9pt]{revtex4}

\usepackage{graphicx}
\usepackage{dcolumn}
\usepackage{bm}

\usepackage{amssymb}
\usepackage{amsmath,dsfont}
\usepackage{amsfonts}
\usepackage{graphicx}
\usepackage{hyperref}
\hypersetup{
    bookmarks=true,         
    unicode=false,          
    pdftoolbar=true,        
    pdfmenubar=true,        
    pdffitwindow=false,     
    pdfstartview={FitH},    
    pdftitle={My title},    
    pdfauthor={Author},     
    pdfsubject={Subject},   
    pdfcreator={Creator},   
    pdfproducer={Producer}, 
    pdfkeywords={keyword1} {key2} {key3}, 
    pdfnewwindow=true,      
    colorlinks=false,       
    linkcolor=red,          
    citecolor=green,        
    filecolor=magenta,      
    urlcolor=cyan           
}

\usepackage{color,verbatim, braket}

\usepackage{xcolor}
\definecolor{red}{rgb}{0.7,0,0}
\definecolor{green}{rgb}{0.,0.35,0.}
\definecolor{blue}{rgb}{0.2,0.2,0.7} 
\definecolor{black}{rgb}{0.15,0.15,.15}

\begin{document}

\title{Driven-dissipative dynamics of a strongly interacting  Rydberg gas}

\date{\today}
\author{A. W. Glaetzle}
\affiliation{Institute for Theoretical Physics,
  University of Innsbruck, and Institute
  for Quantum Optics and Quantum Information of the Austrian Academy
  of Sciences, A-6020 Innsbruck, Austria}
  \affiliation{ITAMP, Harvard-Smithsonian Center for Astrophysics, Cambridge, Massachusetts 02138, USA}
  \affiliation{Joint Quantum Institute, National Institute of Standards and Technology, and University of Maryland, Gaithersburg, Maryland 20899, USA}
  
\author{R. Nath}
\affiliation{Institute for Theoretical Physics,
  University of Innsbruck, and Institute
  for Quantum Optics and Quantum Information of the Austrian Academy
  of Sciences, A-6020 Innsbruck, Austria}

\author{B. Zhao}
\affiliation{Institute for Theoretical Physics,
  University of Innsbruck, and Institute
  for Quantum Optics and Quantum Information of the Austrian Academy
  of Sciences, A-6020 Innsbruck, Austria}

\author{G. Pupillo}
\affiliation{Institute for Theoretical Physics,
  University of Innsbruck, and Institute
  for Quantum Optics and Quantum Information of the Austrian Academy
  of Sciences, A-6020 Innsbruck, Austria}
\affiliation{ISIS (UMR 7006) and IPCMS (UMR 7504), Universit\'e de Strasbourg and CNRS,
Strasbourg, France}

\author{P. Zoller}
\affiliation{Institute for Theoretical Physics,
  University of Innsbruck, and Institute
  for Quantum Optics and Quantum Information of the Austrian Academy
  of Sciences, A-6020 Innsbruck, Austria}
  \affiliation{ITAMP, Harvard-Smithsonian Center for Astrophysics, Cambridge, Massachusetts 02138, USA}
\affiliation{Joint Quantum Institute, National Institute of Standards and Technology, and University of Maryland, Gaithersburg, Maryland 20899, USA}

\begin{abstract}
We study the non-equilibrium many-body dynamics of a cold gas of ground state alkali atoms weakly admixed by Rydberg states with laser light. On a timescale shorter than the lifetime of the dressed states, effective dipole-dipole or van der Waals interactions between atoms can lead to the formation of strongly correlated phases, such as atomic crystals. Using a semiclassical approach, we study the long-time dynamics where decoherence and dissipative processes due to spontaneous emission and blackbody radiation dominate, leading to heating and melting of atomic crystals as well as particle losses. These effects can be substantially mitigated by performing active laser cooling in the presence of atomic dressing.
\end{abstract}

\maketitle

\section{Introduction}\label{sec:intro}
Rydberg states are highly excited electronic states of atoms and  molecules with large principle quantum numbers $n$ \cite{Gallagher2005, Saffman2010, Comparat2010, Seaton1983, Low2012}.  The remarkable properties of Rydberg states include their size $r\sim a_0 n^2$ with $a_{0}$ the Bohr radius, implying huge polarizabilities $\alpha\sim n^7$ and large electric dipole moments between Rydberg states, and thus strong coupling to external electric DC and microwave AC fields. Rydberg states can be excited from atomic or molecular ground states by laser light via absorption of one or more photons, with Rabi frequencies scaling as $\sim n^{-3/2}$. On the other hand, Rydberg states are long-lived, with lifetimes $\tau$ scaling as $\tau\sim n^3$ ($\tau\sim n^5$) for low (high) angular momentum states. For typical present experiments with $n\sim 50$ these can be as large as tens of $\mu$s. Furthermore, atoms prepared in Rydberg states interact strongly, and these interactions can be controlled and enhanced by external fields. In particular, the van der Waals (vdW) interaction between $ns$-states scales as $V_{\rm VdW}\sim (e a_0)^4 n^{11}/r^6$, with $e$ the electron charge, and can be made both attractive and repulsive. In the presence of external DC or AC electric fields the Rydberg-Rydberg interaction is a dipolar interaction, $V_{\rm dd}\sim (e a_0 n^2)^2 / r^3$. The latter is both {\it long-range} and {\it anisotropic}. Additional tunability can be achieved using, e.g., F\"orster resonances. These phenomena have been explored in recent experiments \cite{Anderson1998, Singer2004, Tong2004,  Liebisch2005, Afrousheh2006, Carroll2006, Heidemann2007, Vogt2007, Amthor2007, VanDitzhuijzen2008, Johnson2008, Pritchard2010, Sevincli2011a,  Dudin2012, Nipper2012}.

The remarkable properties of Rydberg states, and the tunability and strength of interactions between Rydberg atoms is reflected in novel many-particle physics of Rydberg gases \cite{Weimer2008, Low2009, Pohl2010, Weimer2010, Honer2010, Pupillo2010, Henkel2010, Maucher2011, Sela2011, Schachenmayer2011, Weimer2012}, and provide the basis for applications in quantum information processing, e.g.~ in implementing (fast) quantum gates \cite{Jaksch2000, Lukin2001,Urban2009a, Gaetan2009, Isenhower2010,Wilk2010}. An underlying principle is the  {\it Rydberg-blockade mechanism}, as first proposed in Ref. \cite{Jaksch2000} and \cite{Lukin2001}.  In the blockade regime the presence of a single atom excited to a Rydberg state shifts the energy levels of the surrounding atoms within a characteristic radius $r_b$ ($\sim \mu$m), such that the laser excitation probability to the Rydberg state for any other atom within a volume $\sim r_b^3$ is strongly suppressed. This excitation will be delocalized among all particles within $r_b$ forming a ``superatom'', with a collectively enhanced Rabi frequency $\sqrt{N_b} \Omega$. Here,  $N_b$ is the number of atoms within $r_b^3$ constituting a superatom and $\Omega$ is the single particle Rabi frequency. Various phenomena in the fields of many-body physics, quantum information applications and quantum optics related to the dipole-blockade mechanism have been discussed recently in Refs. \cite{Brion2007,  Saffman2008, Moller2008, Muller2009, Olmos2009a, Weimer2010a, Han2010, Zhao2010, Kuznetsova2011, Gorshkov2011}. 

Present experiments on Rydberg gases as a many-body system created by laser excitation from BECs \cite{Heidemann2008, Viteau2011}, MOTs \cite{Anderson1998, Singer2004, Tong2004, Liebisch2005, Afrousheh2006, Carroll2006, Vogt2007, Amthor2007, Heidemann2007, VanDitzhuijzen2008,  Johnson2008, Pritchard2010, Amthor2010, Sevincli2011a, Nipper2012}, atomic vapor cells \cite{Kubler2010}  or optical lattices \cite{Dudin2012} have mainly explored the {\em frozen gas regime}. This corresponds to a {\em short time dynamics}, where the atomic motion can be ignored and (resonant) laser excitation of the Rydberg levels leads to large Rydberg-Rydberg interactions. In this limit atomic dissipation, e.g.~spontaneous emission from Rydberg states, is negligible  \cite{Mourachko1998} and the dynamics maps to effective spin models, described by {\em Hamiltonian} dynamics \cite{Weimer2008, Pohl2010, Lesanovsky2011, Schachenmayer2011}.

An alternative regime is the {\em Rydberg-admixed gas regime} \cite{Honer2010, Pupillo2010, Henkel2010, Maucher2011, Li2012}: there the idea is to admix the strong Rydberg-Rydberg interactions by off-resonant laser light weakly to the atomic ground states, thus providing an effectively much smaller, but still tunable vdW or dipolar ground state dynamics. That is, instead of kDebye dipole moments of Rydberg states one obtains effective ground state dipoles in the range of tens of Debyes  \cite{Santos2000}. Thus kinetic energies and effective interactions can become comparable, while at the same time spontaneous emission from the Rydberg state due to off-resonant laser tuning is strongly reduced. This effectively extends the lifetime of the gas, making it reminiscent of the dynamics of dipolar gases of polar molecules, as reviewed in \cite{Baranov2008, Carr2009, Lahaye2009, Boninsegni2012, Baranov2012}.

An interesting question emerging from the above discussion is, whether it is possible to form interesting condensed matter phases with these engineered {\em atomic} dipolar gases; an example is provided by the formation of  (stable) {\em dipolar crystals of cold atomic gases} in analogy to dipolar crystals discussed for polar molecules \cite{Buchler2007}. In contrast to polar molecules 
\footnote{On relevant experimental timescales decoherence due to spontaneous emission between rotational states of polar molecules is negligible. However, in a high temperature environment, $T\sim$ 1 K, blackbody radiation accounts for the dominant heating mechanism.}
for Rydberg-admixed gases dissipation and decoherence due to spontaneous emission is non-negligible for long times, and this raises the question of describing the long-time non-equilibrium dynamics of heating in such gases. 

Understanding the long-term dynamics of Rydberg dressed atoms can also have important applications beyond many-body dynamics {\it per se}. For example, recent proposals have investigated the possibility to efficiently couple cold atomic Rydberg gases to comparatively hot molecular ensembles via long-range dipolar or vdW interactions in order to achieve Doppler \cite{Huber2012} and collisional Sisyphus-like \cite{Zhao2012} cooling schemes for molecules. 

\begin{figure}[tb]
\centering
\includegraphics[width=0.95\columnwidth]{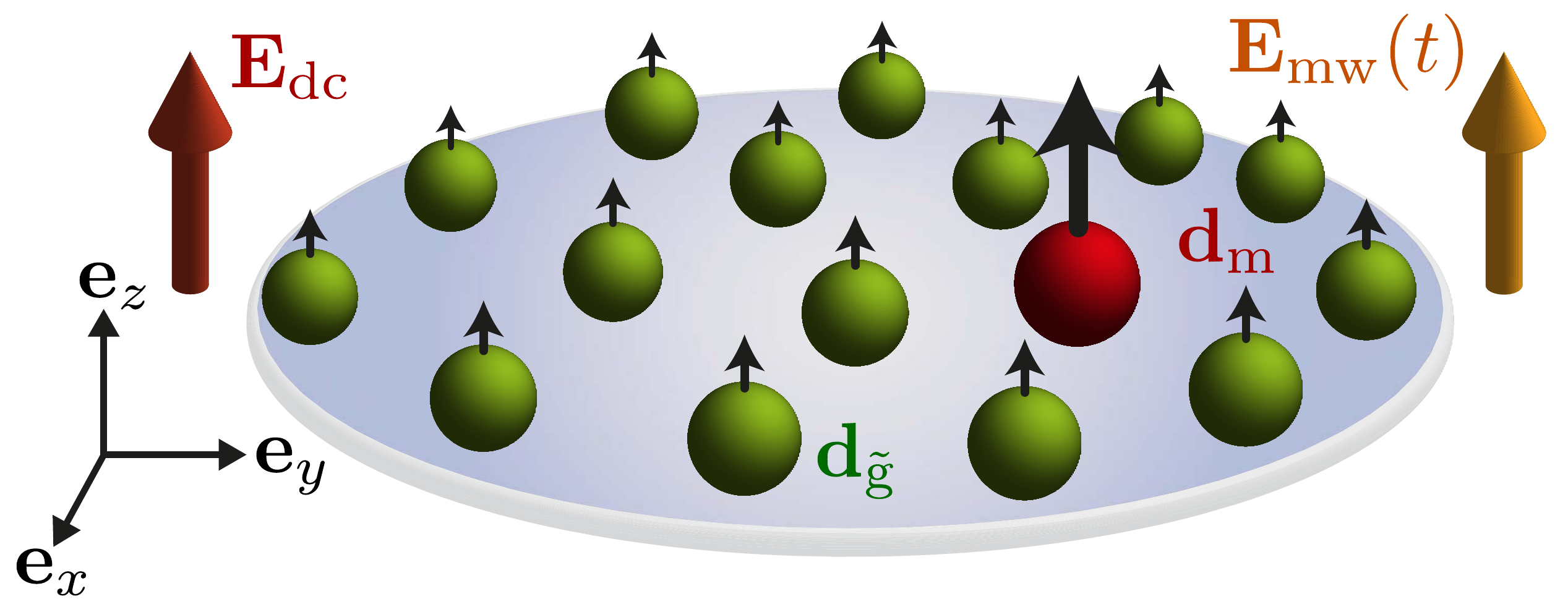}
\caption{(color online) Sketch of the proposed setup: Rydberg-dressed ground state atoms confined in the $x$-$y$-plane, e.g. by an optical lattice (not shown), are polarized perpendicular to the plane by a DC-electric field, $\mathbf{E}_{\rm dc}$, or an AC-microwave field, $\mathbf{E}_{\rm mw}(t)$. Atoms can either be in the dressed ground state $\ket{\tilde g}$ (green particles) with a dipole moment $\mathbf{d}_{\tilde g}$ or in one of the intermediate Rydberg states $\ket{m}$ (red particles) with a dipole moment $\mathbf{d}_m$ (see text), resulting in strong dipole-dipole interactions.}
\label{fig:plate}
\end{figure}

The paper is organized as follows: In Sec.~\ref{sec:summary} we give an overview of various atomic configurations studied, identify the main questions, and summarize the main results of the paper. Technical details of our calculations can be found in the remainder of the paper: In Sec.~\ref{sec:model} we introduce the model and the notation which we will use throughout the paper.  In Sec.~\ref{sec:singleHam}  we discuss the Hamiltonian for a single Rydberg-dressed atom in the presence of external static and electromagnetic fields. The interaction Hamiltonian of two Rydberg-dressed atoms in the presence of external fields is considered in Sec.~\ref{sec:twoparticleham}, where Born-Oppenheimer potentials are derived for both DC-electric and AC-microwave fields. In Sec.~\ref{sec:2dvalidity} we study the validity of the 2D treatment. In Sec.~\ref{sec:lasercooling} we derive Fokker-Planck equations for laser cooled ground state atoms in the presence of Rydberg dressing. We find an additional two-body diffusion term due to the interaction between two Rydberg-dressed atoms. The dynamics of decay from the Rydberg state can be modeled by coupled Fokker-Planck equations for atoms effectively in the dressed and laser cooled ground state and atoms in one of the intermediate excited states. In Sec.~\ref{sec:decoherencesingle} we study numerically spontaneous emission and blackbody radiation of a single Rydberg dressed atom. Population of intermediate excited states after decay from the Rydberg state will lead to a fluctuating dipole moment and heating of the motion due to photon recoil. The case of an interacting {\it ensemble} of Rydberg dressed atoms is numerically studied in Sec.~\ref{sec:decoherencetwo}. We find that the fluctuations of the dipole moment lead to strong mechanical effects and heating of the external motion, which strongly depends on the atomic density and the dressing scheme used. We study numerically the effect of laser cooling and melting in the presence of an optical lattice with the aim to extending the lifetime of the gas.

\begin{figure}[tb]
\centering
\includegraphics[width=0.7\columnwidth]{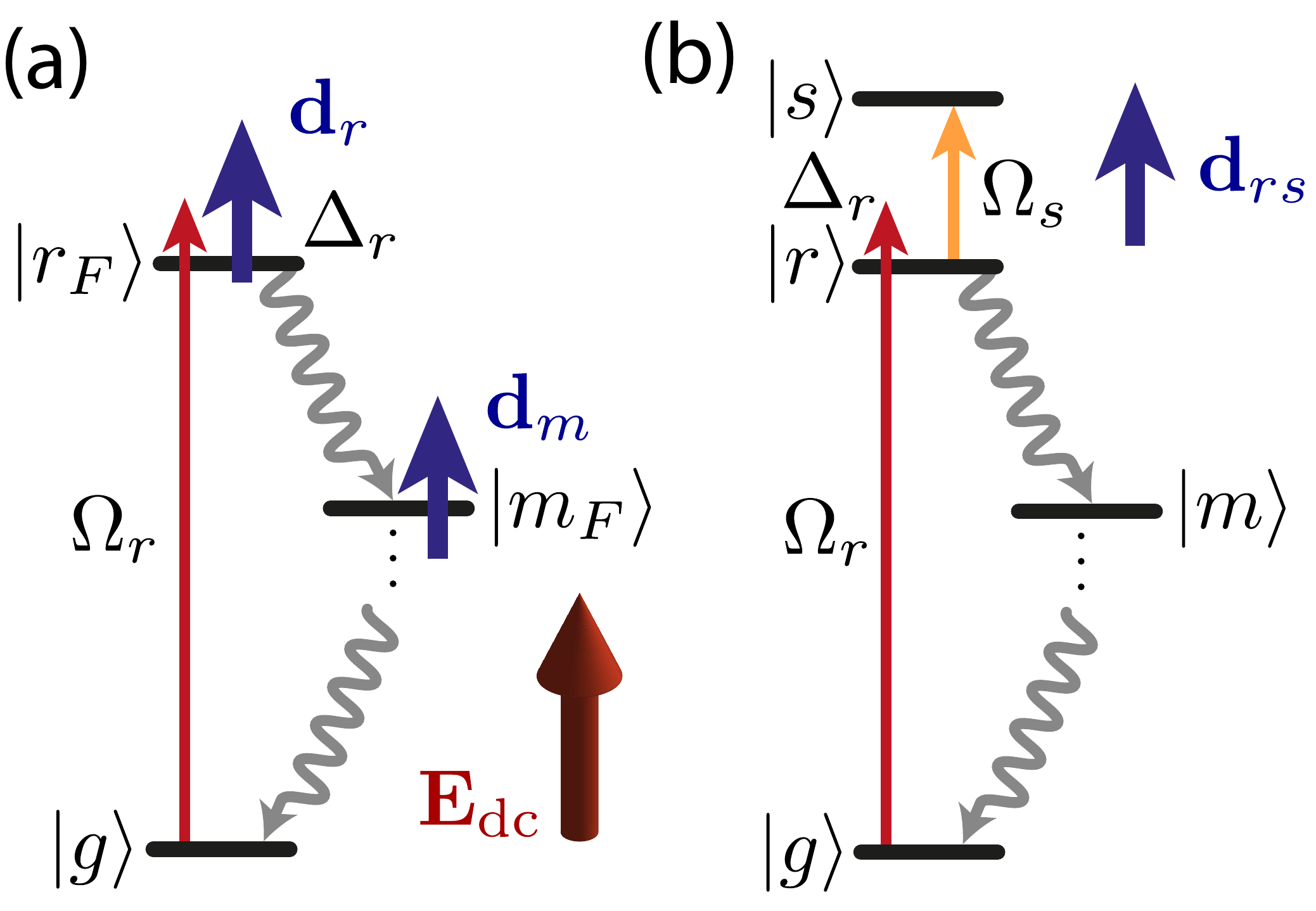}
\caption{(color online) Qualitative sketch of the energy levels (black lines), lasers (thin solid arrows) and decay paths (wiggly arrows) for (a) the \textit{DC-electric field} dressing scheme and (b) the \textit{AC-microwave} dressing scheme. In both schemes the ground state $\ket g$  is weakly dressed with a Rydberg-state $\ket{r_F}$ ($\ket r$) using a far detuned laser (red arrow) with Rabi frequency $\Omega_r$ and detuning $\Delta_r\gg\Omega_r$. In the long-time limit cascaded decay from the Rydberg states $\ket r$, $\ket s$ or $\ket{r_F}$  (gray wiggly arrows) will populate intermediate Rydberg states $\ket{m_{(F)}}$ due to spontaneous emission and blackbody radiation. In panel~(a) a static electric field, $\mathbf{E}_{\rm dc}=F \mathbf{e}_z$, polarizes the atoms leading to new Stark-split eigenstates, indicated with an index $F$. Both, the Rydberg state $|r_F\rangle$ and intermediate states $|m_F\rangle$ will obtain an intrinsic dipole moment $\mathbf{d}_r$ and $\mathbf{d}_m$, respectively (blue thick arrows).  In panel ~(b) a near resonant AC-microwave field, $\mathbf{E}_{\rm mw}(t)$, (yellow arrow) with Rabi frequency $\Omega_s$ couples two Rydberg states $\ket r$ and $\ket s$ and induces an oscillating dipole moment proportional to $\mathbf{d}_{rs}$. The intermediate Rydberg states $|m\rangle$ have no dipole moment.  }
\label{fig:levelscheme0}
\end{figure}

\section{Overview of results}\label{sec:summary}
Before presenting a detailed discussion of the dynamics of strongly interacting Rydberg dressed atoms, we find it worthwhile to summarize the main features of Rydberg dressing in a gas of (realistic) atoms, and the resulting many-particle dynamics. The processes we summarize here are derived in detail in Sec.~\ref{sec:model} - \ref{sec:dissipation}.

Motivated by proposals for many-body physics with dipolar gases \cite{Micheli2007, Buchler2007}, here we focus on the dynamics of a Rydberg-dressed dipolar gas. We assume that the system is initially prepared in a crystalline state, e.g.~by preparing a Mott insulator state of atoms in an optical lattice, and turning on adiabatically the Rydberg admixture while switching off the optical lattice, so that a dipolar crystal is formed. Similar to the setups of self-assembled crystals for polar molecules studied in Refs.~\cite{Micheli2007, Buchler2007}, we assume a two-dimensional (2D) configuration, i.e.~ that atoms are confined to the ($x-y$)-plane by an external optical field, e.g., an optical lattice, in the $z$-direction (see Fig.~\ref{fig:plate}), such that in-plane dipole-dipole interactions are purely repulsive. As explained below, this will minimize collisional losses linked to the attractive part of dipole interactions. This analysis is readily extended to, and actually simplified in the case of isotropic  repulsive vdW interactions in three dimensions \cite{Gorshkov2008}.

The main question we want to address is how heating and dissipation affect the many-body dynamics due to decay from the Rydberg state in the long-time limit. In fact, for times long enough, spontaneous emission and blackbody radiation will inevitably  redistribute the population from the Rydberg state to various different excited states, which invalidates the simple two-level approximation for the internal dynamics of a single atom, mostly discussed so far \cite{Weimer2008, Weimer2010, Sela2011, Low2009, Pohl2010, Schachenmayer2011, Weimer2012, Honer2010, Pupillo2010, Henkel2010, Maucher2011}. Given the fact that atoms in these other excited states can interact very differently from those prepared in the dressed ground state, e.g.~they can have different dipole moments, decay from the Rydberg state can lead to strong mechanical effects and collisional losses in an ensemble of interacting Rydberg dressed atoms. In the following we will investigate this complex many-body dynamics. In particular, we will explore how to mitigate and control these effects using laser-cooling or in-plane optical lattices for a realistic scenario where each atom comprises a large number of internal states.

\subsection{Atomic configuration}
The atomic configurations we have in mind are summarized in Fig.~\ref{fig:levelscheme0}: (i) in panel (a) a DC-electric field with strength $F$ polarizes each atom by splitting its energy levels into the Stark structure; the new Stark-split Rydberg state $|r_F\rangle$ obtains an intrinsic dipole moment $d_r$ which can be either {\it parallel} or {\it antiparallel} with respect to the external field. (ii) in panel (b) an AC-(microwave)-field of Rabi frequency $\Omega_s$ is used to strongly couple two Rydberg states $|r\rangle$ and $|s\rangle$, which induces an oscillating dipole proportional to the transition dipole moment $d_{rs}$. 

In both configurations of Fig.~\ref{fig:levelscheme0} we weakly admix the ground state $|g\rangle$ of each atom  with the Rydberg state $|r_F\rangle$ or $|r\rangle$, respectively, using an off-resonant continuous wave laser with Rabi frequency $\Omega_r$ and detuning $\Delta_r\gg\Omega_r$.  This immediately results in an effective dipole moment $d_{\tilde g}\sim(\Omega_r/2\Delta_r)^2 d_0$ into the dressed ground state $|\tilde g\rangle \sim |g\rangle + (\Omega_r/2\Delta_r) |r\rangle$, which can be tuned using the laser parameters, see Sec.~\ref{sec:ham}. Here, $d_0=d_r$ or $d_0=d_{rs}$ for DC-electric fields or AC-microwave fields, respectively, see Sec.~\ref{sec:twoparticleham}. This dressed ground state $|\tilde g\rangle$ has now a finite, albeit comparatively long, lifetime $\sim 1/\Gamma_{\tilde g}$, where $\Gamma_{\tilde g}\sim(\Omega_r/2\Delta_r)^2 \Gamma_r$ with $\Gamma_r$ the decay rate of the Rydberg state. 

For timescales which are comparable to, or even larger than $1/\Gamma_{\tilde g}$ population in each of the Rydberg states $\ket{r_F}$ or $\ket{r}$ and $\ket{s}$  will be redistributed due to spontaneous emission and blackbody radiation among several excited states $|m\rangle$. In the long-time limit, after, e.g., a spontaneous emission event from the Rydberg states, the atomic state will in general not return to the ground state directly, but via a cascade process where several $\ket{ m }$-states are populated. Since each $\ket{m}$-state has a finite lifetime, the cascade will not happen instantaneously. Population of these intermediate states can induce strong mechanical  effects on the gas dynamics, leading to heating and losses, as explained below in Sec.~\ref{sec:dissipation}. It turns out that these effects depend crucially on how the dipole moment $d_0$ in the Rydberg states is created, see Fig.~\ref{fig:levelscheme0}(a) and (b).

\begin{figure*}[tb]
\centering
\includegraphics[width=8 cm]{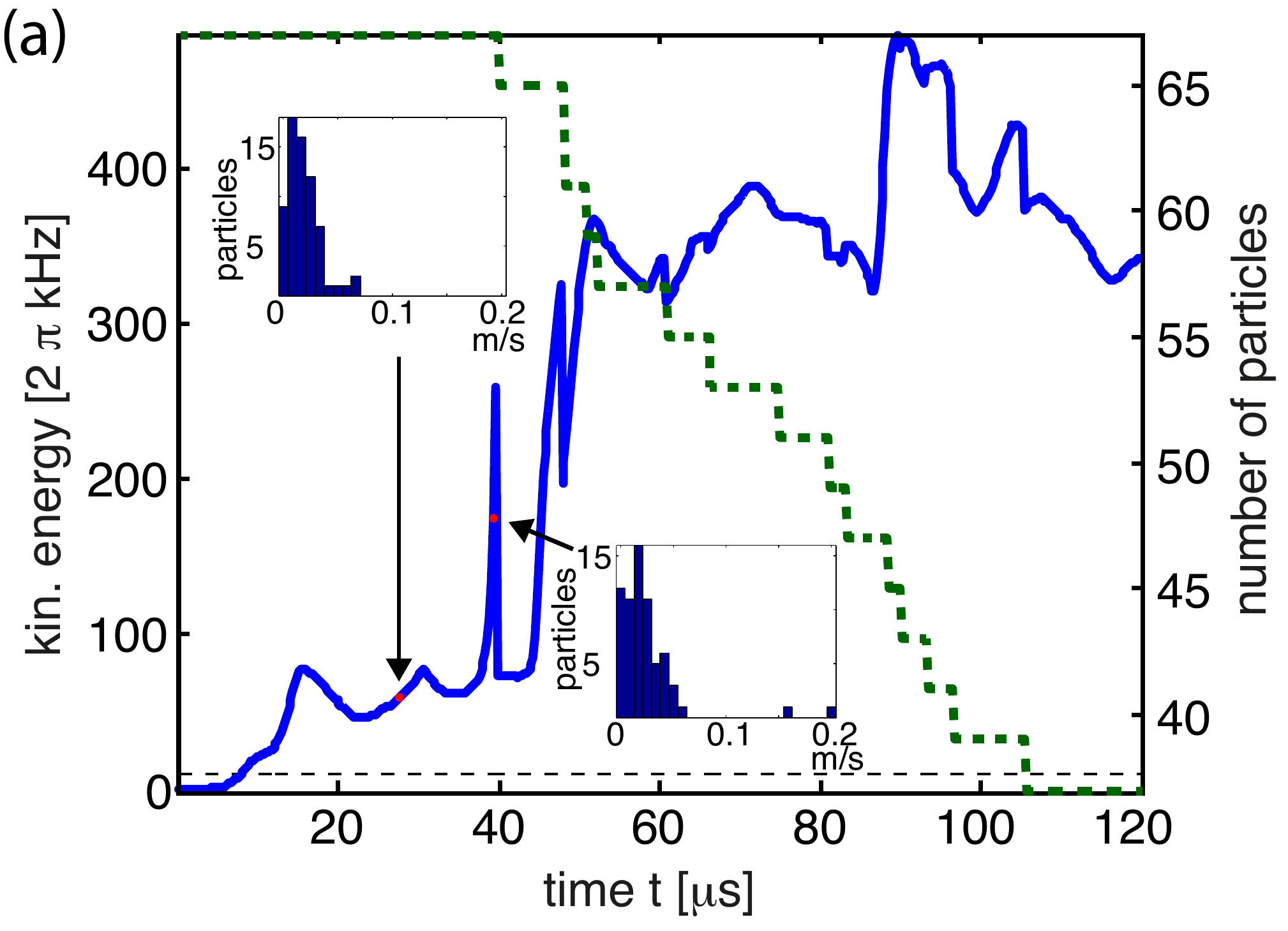}\qquad
\includegraphics[width=8 cm]{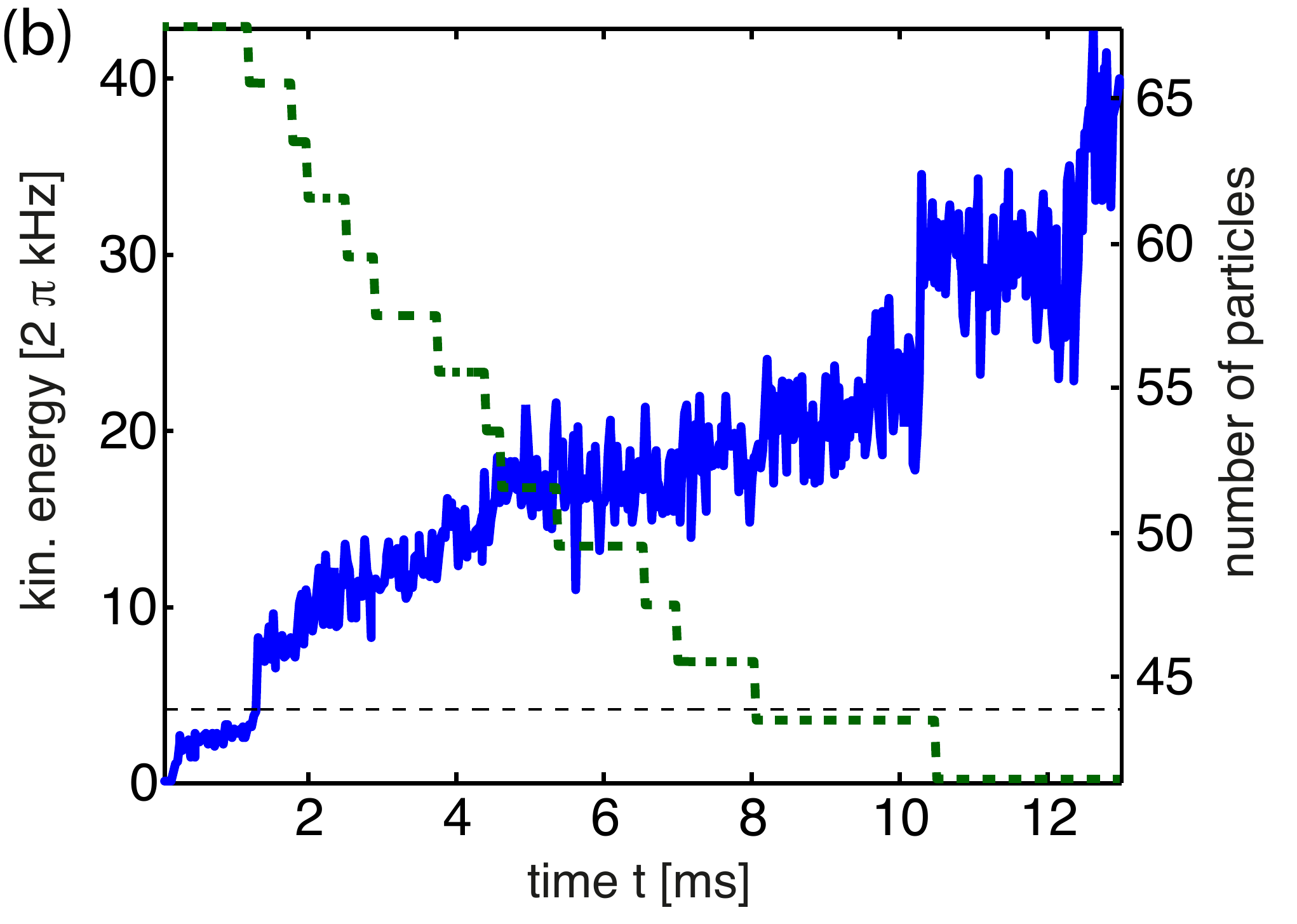}
\caption{(color online) Single trajectory of a semi-classical molecular dynamics simulation studying the non-equilibrium and melting dynamics of a 2D Rydberg-dressed crystal as a function of time $t$. Heating due to decay from the Rydberg states and population of intermediate states $|m_{(F)}\rangle$ leads to a rapid increase of the mean kinetic energy, $E_{\rm kin}(t)$ (solid blue line, left axis) and a reduction of the particle number due to losses (dotted green line, right axis).  In panel~(a) the dipole moment in the Rydberg state is created using the DC-electric field dressing scheme of Sec.~\ref{sec:sumdc} and the same atomic parameters as given there. Initially we choose a density $n_{2D}=1\,\mu\text{m}^{-2}$. Two insets show histograms of the momentum distribution at $t=27\;\mu$s and $t=40\;\mu$s. In panel~(b) the dipole moment in the Rydberg state is created using the AC-microwave dressing scheme of Sec.~\ref{sec:sumac} with the same atomic parameters as given in the text. The initial density is $n_{2D}=0.2\,\mu\text{m}^{-2}$.  In both panels the simulation time corresponds to the effective lifetime of the dressed ground state atoms, $\tau_{\tilde g}=125\,\mu$s  or $\tau_{\tilde g}=12.8$ ms, respectively. The thin dashed line is the initial melting temperature, $T_M$, of the crystal.}
\label{fig:singletrajectory}
\end{figure*}

We analyze these effects in detail by performing semi-classical molecular dynamics simulations for an ensemble of interacting Rydberg-dressed atoms confined to a 2D geometry including a large number of internal states. In Figs.~\ref{fig:singletrajectory} (a) and (b), we provide representative examples for the cases (i) and (ii) above, respectively, for a system comprising $N=67$ atoms in a box with hard walls. The external dynamics is treated classically, while quantum jumps from the Rydberg state account for spontaneous emission and blackbody radiation, leading to a time-dependent dipole moment of the atoms. In the simulation we use model-atoms with several thousands of internal electronic states with the level structure and mass $M$ of $^{85}$Rb. At time $t=0$ the atoms are  prepared in a triangular crystal structure with a given density $n_{2D}$ (see figure caption) at zero temperature $T=0$. In Fig.~\ref{fig:singletrajectory}, the thin dashed lines are the initial melting temperatures $T_M = 0.089$ $d_{\tilde g}^2 n_{2D}^{3/2} / k_B$ as determined in Ref.~\cite{Kalia1981}, with $k_B$ Boltzmann's constant. The solid blue and dotted green lines correspond to the mean atomic kinetic energy $E_{\rm kin}(t)$ and the total atom number $N(t)$, respectively, plotted as a function of time $t$. In each plot, the simulation time corresponds to $1/\Gamma_{\tilde g}$, for the given choice of parameters.

\subsection{Rydberg dressing with a DC-electric field}\label{sec:sumdc}
Figure~\ref{fig:singletrajectory}(a) shows an example where we choose a very strong DC-field with strength $F=3$ kV/cm in order to polarize $^{85}$Rb atoms. Such a high electric field is motivated by studying mixtures of polar molecules and Rydberg atoms where both species are polarized by the same field \cite{Glaetzle}. In order to avoid, e.g., field-ionization with threshold scaling as $F_{\rm ion} \sim n^{-4}$, the choice of  Rydberg states is thus limited to those with a low principal quantum number, e.g. $n\sim 16$. Here, we use  $|r\rangle=|16d,m=0\rangle$, with lifetime $\tau_r\sim 5.5\, \mu$s and field-induced dipole moment $d_r\sim 680$~Debye. We choose a dressing laser with $\Omega_r/\Delta_r=0.42$, resulting in $d_{\tilde g}\sim 30$~Debye and a lifetime $\tau_{\tilde g}\sim 125\;\mu$s.

Figure~\ref{fig:singletrajectory}(a) shows that $E_{\rm kin}(t)$ (blue thick solid line) rapidly increases and exceeds $T_M$ after about 10~$\mu$s. This is due to spontaneous emission events, and the ensuing population of various $|m\rangle$-states, which have different dipole moments (see also scheme in Fig.~\ref{fig:levelscheme0}(a)). Population of these intermediate states lead to a picture of fluctuating dipoles, where the energy of local crystal distortions due to dipole-dipole interactions is rapidly redistributed among all particles as heat. The corresponding momentum distribution of the atoms is shown in the first inset. The heating effect from the fluctuating dipole moments is dominant over single-particle recoil effects (with a characteristic rate of a few hundreds of kHz/ms, as discussed in Sec.\ref{sec:decoherencesingle} below). 

In addition, the blue line in panel (a) of Fig.~\ref{fig:singletrajectory} shows rapid large variations, or ``spikes", of $E_{\rm kin}$, e.g.~at $t \sim 40$~$\mu$s, which correspond to rare events, where one of the atoms populates a $|m\rangle$-state with either a large dipole moment parallel to the DC-electric field and long lifetime resulting in a strong repulsive interaction, or an anti-parallel dipole moment resulting in strong in-plane dipolar attraction. This out-of-equilibrium situation is reflected in a temporary change in the distribution of momenta of the atoms, as shown in the second inset by the appearance of small peaks at high-momenta. Such rare events result in collisional losses of the most energetic particles, similar to a filtering process, as well as  small local ``explosions", where a large number of particles can be lost. 

The heating and loss processes described above are interaction-dependent (since the interaction strength scales as $\sim n_{2D}^{3/2}$) and thus decrease considerably for a smaller atomic density. In Sec.~\ref{sec:decoherencetwo} below we show how active laser cooling and an additional in-plane optical lattice can affect these processes.

\subsection{Rydberg dressing with a  AC-microwave field}\label{sec:sumac}
In Fig.~\ref{fig:singletrajectory}(b) we choose a strong and resonant AC-microwave field to couple the Rydberg states $|r\rangle=|50s\rangle$ and  $|s\rangle=|49p\rangle$ (see level scheme of Fig.~\ref{fig:levelscheme0}(b)), which have a transition dipole moment of $d_{rs}\sim 5.9$~kDebye. A dressing laser with $\Omega_r/\Delta_r=0.14$ yields dressed ground states atoms with $d_{\tilde g}\sim 30$~Debye and $\tau_{\tilde g}\sim 12.8$ ms. The comparatively long lifetime is due to the choice of a state with a larger principal quantum number $n \sim 50$. Since  $F=0$ intermediate Rydberg states $|m\rangle$ have essentially no dipole moment and we consider them as non-interacting. 

Figure~\ref{fig:singletrajectory}(b) shows that $E_{\rm kin}(t)$ increases with $t$. This is again due to fluctuations of the atomic dipole moment between the value $d_{\tilde g}$ and zero, corresponding to the atom being in the dressed ground state $|\tilde g\rangle$ or in one of the intermediate states $|m\rangle$, respectively, together with photon recoil after a decay event. As discussed before, heating comes from the redistribution of the energy associated to local crystal distortions due to dipole fluctuations among all particles. For $t\lesssim 1.5$ ms the Rydberg dressed atoms are in a crystal-like phase, with $E_{\rm kin}(t)<k_BT_M$.

In comparison with the DC-electric field case of Fig.~\ref{fig:singletrajectory}(a) the heating rate $E_{\rm kin}(\tau_{\tilde g})\gamma_{\tilde g}$ of the AC-microwave-dressing scheme is approximately one order of magnitude smaller while the particle loss rate $N(\tau_{\tilde g})\gamma_{\tilde g}$ is approximately equal. This is because the initial density is smaller and the intermediate states have negligible dipole moments for AC-microwave scheme. \

In the remainder of this work we derive a microscopic model for the long-time heating dynamics of Rydberg dressed atoms and discuss quantitatively  the dependence of the heating and particle loss rates on the system parameters, e.g., the atomic density. We state under what conditions standard laser cooling can be performed in the presence of Rydberg dressing, and numerically investigate how it counters heating effects.

\section{The model}\label{sec:model}
The purpose of this section is to introduce the physical model and the notation that we will use throughout the paper. The system is illustrated in Fig. \ref{fig:plate}: $N$ identical (alkali) atoms with  momenta $\mathbf{\hat p}_i$ at positions $ \mathbf{\hat r}_i$ ($1\leq i \leq N$) are confined to a quasi two-dimensional (2D) geometry in the ($x$-$y$)-plane by a strong confinement along the $z$-axis.

The Hamiltonian dynamics of the system is governed by 
\begin{equation}
H=\sum_{i=1}^N H^{(i)}+\sum_{ i<j }^N H^{(ij)}_\mathrm{int},  \label{eq:Ham}
\end{equation}
which consists of a sum over single-particle terms $H^{(i)}$ (see Sec.~\ref{sec:ham}) and two-particle interaction terms
\begin{equation}\label{eq:Hamdipdip}
H_{\rm int}^{(ij)}=\frac{\mathbf{\hat d}_i\cdot \mathbf{\hat d}_j-3(\mathbf{\hat d}_i\cdot \mathbf{\hat r})(\mathbf{\hat d}_j\cdot \mathbf{\hat r})}{4\pi\epsilon_0 r^3},
\end{equation}
which account for the dipole-dipole interaction between two atoms. Here, $\mathbf{r}=\mathbf{r}_i-\mathbf{r}_j=\mathbf{\hat r}\;r$ is the relative distance between the atoms, $\mathbf{\hat d}_i$ the dipole operator of the $i$-th atom and $\epsilon_0$ is the vacuum permittivity.

\subsection{Internal level structure and setup}
We denote $|\alpha\rangle=|n_\alpha,\ell_\alpha,j_\alpha,m_\alpha\rangle$ and $\hbar\omega_\alpha=E_{n_\alpha \ell_\alpha j_\alpha}$ as the unperturbed eigenfunctions and eigenenergies, respectively, of the atomic Hamiltonian $H_{at}$.  Here, $n_\alpha$ is the principal quantum number, $\ell_\alpha$ the orbital angular momentum, $j_\alpha$ the total angular momentum and $m_\alpha$  projection of the total angular momentum along a specified axis. 

In this manifold of states we focus on three specific states, see Fig.~\ref{fig:levelscheme}: (i) the energetic ground state of the atom denoted by $\ket g$. All internal energies will be measured relative to the energy of this state, e.g.~$\hbar\omega_g=0$. (ii) a highly excited Rydberg states $\ket r$ with energy $\hbar\omega_r$ and (iii) a lower-lying excited state $\ket e$ with energy $\hbar\omega_e \ll \hbar \omega_{r}$, which will be utilized for laser cooling. All other states $\ket{\alpha}$ can be occupied via spontaneous emission or blackbody radiation from the high-lying Rydberg states. As we have already mentioned in Sec.~\ref{sec:summary}, the ground state of each atom is off-resonantly coupled to a high-lying Rydberg state $\ket r$ using a continuous wave laser  with a Rabi frequency $\Omega_r$ and detuning $\Delta_r$($\gg \Omega_r$), see Fig.~\ref{fig:levelscheme}. Typically, in alkali atom experiments, the atoms are excited to the Rydberg state via two-photon transitions in which  $\Delta_r$ and $\Omega_r$ are the effective detuning and Rabi frequency. In addition, we use a laser field  with Rabi frequency $\Omega_e$ and detuning $\Delta_e$ to couple $|g\rangle$ to a nearby excited state $|e\rangle$ ($\hbar\omega_e\ll\hbar\omega_r$) forming a closed cycle for laser-cooling. 

Besides the general setup described above, we consider two different scenarios to create a strong dipole moment in the Rydberg state: (i) an homogeneous electric DC field of strength  $F$ polarizes the atoms perpendicular to the 2D plane by splitting the energy levels into the Stark structure, see Fig.~\ref{fig:levelscheme}(a). The new Stark-split eigenstates $\ket{\alpha_F}$ obtain a large intrinsic dipole moment, e.g.~$\mathbf{d}_0=\langle r_F|\hat{\mathbf{d}}|r_F \rangle$. (ii) a resonant microwave field with Rabi frequency $\Omega_s$ couples  $|r\rangle$ to a nearby Rydberg state $|s\rangle$.  Thereby the atom acquires a large oscillating dipole moment proportional to the transition dipole moment   $\mathbf{d}_0=\langle r|\hat{\mathbf{d}}|s \rangle$. In both scenarios, the dipole moments can be as large as kilo-Debyes. The corresponding dipole-dipole interaction between two atoms in the Rydberg states is governed by $H_{\rm int}$ of Eq.~\eqref{eq:Hamdipdip}. 

\subsection{Master equation dynamics}\label{sec:mastereq}
The dynamics of the driven-dissipative many-body system
made of interacting Rydberg-dressed atoms is described by the following
master equation
\begin{equation}
\begin{split}  \label{eq:mastereq}
\dot \rho=-i\left[H,\rho\right]+\mathcal{L}\rho,
\end{split}%
\end{equation}
where the system density operator $\rho$ acts both on the Hilbert space of internal $\ket{\alpha_1,\alpha_2,\ldots \alpha_N}$ and external $\{(\mathbf{r}_1,\;\mathbf{p}_1),\ldots (\mathbf{r}_N,\;\mathbf{p}_N)\}$
degrees of freedom. This equation is readily written down as an extension of the familiar master equations of laser cooling for multi-level atoms including interaction between atoms \cite{Cirac1992, Dalibard1985b}. 
We denote $\cal{L}$ as the Lindblad operator, accounting for the decoherence due to spontaneous emission and blackbody radiation 
\begin{equation}
\begin{split}
\mathcal{L}\rho=\sum_{i=1}^N\sum_{\alpha,\beta}\Gamma_{\alpha\beta}\mathcal{D}^{(i)}_{\alpha\beta}\rho,
\end{split}%
\end{equation}
which results from the coupling between the atomic system and the vacuum modes of the electromagnetic field, which have
been adiabatically eliminated. The decay rate from state $|\alpha\rangle$ to $|\beta\rangle$,
\begin{equation}
\begin{split}
\Gamma_{\alpha\beta}=\Gamma^{\text{(SE)}}_{\alpha\beta}+\Gamma^{\text{(BBR)}}_{\alpha\beta},
\end{split}
\end{equation}
is obtained as the sum of respective contributions from spontaneous emission (SE) and blackbody radiation (BBR) \cite{Beterov2009a}, with
\begin{subequations}
\begin{align}
&\Gamma^{\text{(BBR)}}_{\alpha\beta}=\bar n_{\alpha\beta}(T) A_{\alpha\beta},\label{decayc1}\\
&\Gamma^{\text{(SE)}}_{\alpha\beta}=\left\{
\begin{array}{ll}
A_{\alpha\beta}, & \hbox{$\omega_\alpha>\omega_\beta$;} \\
0, & \hbox{$\omega_\alpha\leqslant \omega_\beta$.}%
\end{array}
\right.
\label{decayc2}
\end{align}
\end{subequations}
Here, $A_{\alpha\beta}$ is the Einstein A-coefficient and $\bar n_{\alpha\beta}$ is the photon distribution, with
\begin{subequations}
\begin{align}
&A_{\alpha\beta}=\frac{|\omega_{\alpha}-\omega_{\beta}|^3}{3 c^3\pi\hbar \epsilon_0}|\braket{\alpha|\mathbf{\hat d}|\beta}
|^2,\\
&\bar n_{\alpha\beta}(T)=\frac{1}{e^{\hbar|\omega_{\alpha}-\omega_{\beta}|/k_BT}-1%
},
\end{align}
\end{subequations}
respectively. We use the notation
\begin{equation}\begin{split}\label{eq:Lind}
\mathcal{D}_{\alpha\beta}^{(i)}\rho&=
\int\mathrm{d}^2\hat{\mathbf{k}} N_{\alpha\beta}(\hat{\mathbf{k}}) e^{-i k_{\alpha\beta} \hat{\mathbf{k}}\cdot\mathbf{\hat r}_i}\sigma_{\beta\alpha}^{(i)}\rho\sigma_{\alpha\beta}^{(i)}e^{ik_{\alpha\beta} \hat{\mathbf{k}}\cdot\mathbf{\hat r}_i}\\
&-\frac{1}{2}\sigma_{\alpha\alpha}^{(i)}
\rho-\frac{1}{2}\rho\sigma_{\alpha\alpha}^{(i)}\\
\end{split}\end{equation}
to denote a general Lindblad term accounting for population redistribution
from level $\ket{\alpha_i}$ to $\ket{\beta_i}$ of the $i$-th atom ($1\leqslant  i\leqslant N$) 
with $\sigma_{\alpha\beta}^{(i)}=|\alpha_i\rangle\langle\beta_i|$.
With $N_{\alpha\beta}(\hat{\mathbf{k}})$ we denote the angular distribution
of spontaneous emission from level $\alpha$ to $\beta$, which we assume to
be a normalized and even function.
The terms $e^{\pm ik_{\alpha\beta} \hat{\mathbf{k}}\cdot\mathbf{\hat r}_i}$ describe the recoil of the atom due to a spontaneously emitted photon, where $k_{\alpha\beta}=(\omega_\alpha-\omega_\beta)/c$ is the wave number of the corresponding transition, with $c$ the speed of light and $\hat{\mathbf{k}}$ a unit-vector in the direction of spontaneous emission.

\section{Hamiltonian dynamics}\label{sec:ham} 
We now turn to analyze the single
particle and two-particle-interaction Hamiltonians of Eq.~\eqref{eq:Ham}. In
particular, we derive the effective Born-Oppeheimer (BO) potential surfaces for
two interacting Rydberg-dressed atoms in the presence of either a DC-electric-field (Sec.~\ref{sec:dcdressing}) or an AC-microwave-field (Sec.~
\ref{sec:mwdressing}).

\begin{figure}[tb]
\centering
\includegraphics[width=0.95\columnwidth]{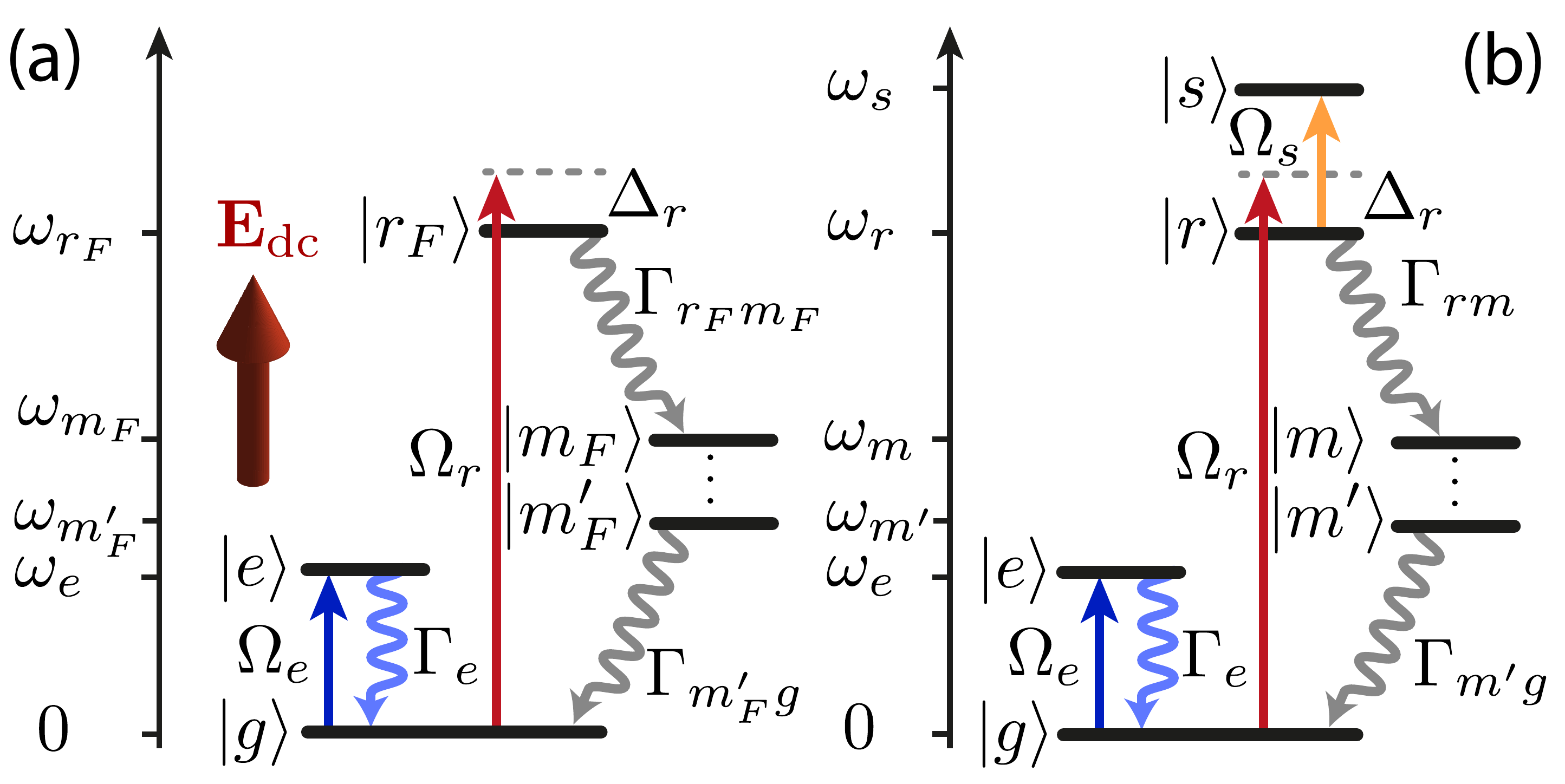}
\caption{(color online) Energies $\omega_\alpha$, states $|\alpha\rangle$ (solid black lines), lasers (solid arrows) and decay paths (grew wiggly arrows) for (a) the \textit{DC-electric field} and (b) the {\it AC-microwave} dressing scheme. Additionally to Fig.~\ref{fig:levelscheme0} a lower-lying excited state $\ket{e}$ is coupled in both schemes to the ground state $\ket{g}$ using a near-resonant laser (blue arrow) with Rabi frequency $\Omega_e$. State $|e\rangle$ decays directly to the ground state (blue wiggly arrow) with a rate $\Gamma_e$ yielding a closed cycle for laser-cooling.  Cascaded decay from the Rydberg states $\ket r$, $\ket{r_F}$ (gray wiggly arrows) and  $\ket s$ (not shown) will populate intermediate Rydberg states $\ket{m_{(F)}}$ and $\ket{m'_{(F)}}$ according to the decay rates $\Gamma_{\alpha\beta}$ (see text).}
\label{fig:levelscheme}
\end{figure}

\subsection{Single particle Hamiltonian}\label{sec:singleHam}
The single particle Hamiltonian $H^{(i)}$ of Eq.~\eqref{eq:Ham} consists of
five terms
\begin{equation}
H^{(i)}(t)=\frac{\mathbf{\hat p}_i^2}{2M}+H^{(i)}_\mathrm{at}+H^{(i)}_\mathrm{dc}
+H^{(i)}_\mathrm{laser}(t)+H^{(i)}_\mathrm{trap}.  \label{eq:Single}
\end{equation}
The first term in Eq.~\eqref{eq:Single} accounts for the kinetic energy of the $
i$-th atom with mass $M$. The second term
\begin{equation}
\begin{split}
H_\mathrm{at}^{(i)}=\sum_{\alpha}\hbar\omega_{\alpha}\sigma_{\alpha
\alpha}^{(i)},  \label{eq:At}
\end{split}
\end{equation}
accounts for the internal atomic energy levels, in the absence of external fields. 
The third term in Eq.~\eqref{eq:Single} reads
\begin{equation}
\begin{split}  \label{eq:Hdc}
&H_\mathrm{dc}^{(i)}=- \mathbf{\hat d}^{(i)}\cdot \mathbf{E}_{dc}=- \hat d_z^{(i)}F,
\end{split}%
\end{equation}
and describes the interaction of an atom with a static electric field, where $\mathbf{\hat d}^{(i)}$ is the atomic dipole operator of the $i$-th atom. The effect of a static electric field $\mathbf{E}_{dc} = F\mathbf{e}_z$ is to polarize the atoms along the field direction $\mathbf{e}_z$, by splitting the energies into the Stark structure \cite{Zimmerman1979}, with $F$ the strength of the field. The new Stark-split eigenstates $\ket{\alpha_F}$ have an intrinsic dipole moment, which in the linear Stark regime is approximately given by $d_\alpha = (3/2) e a_0 n_\alpha (n_1 - n_2)$ with $a_0$ Bohr's radius and $n_1$ and $n_2$  the parabolic quantum numbers of state $\ket{\alpha_F}$. In this regime the energy levels are shifted proportional to the field strength, e.g.~$\Delta E_\alpha= d_\alpha F$. Coupling between adjacent $n$-manifolds can be neglected for field strength $F<F_{\rm IT}$, where $F_{\rm IT}\sim n^{-5}$ is the Inglis-Teller-limit~\cite{Gallagher2005}. 

The term $H^{(i)}_\mathrm{laser}$ in Eq.~\eqref{eq:Single} describes the
interaction of an atom with (e.g., microwave or optical) laser fields and consists of three terms
\begin{equation}
\begin{split}
H_\mathrm{laser}^{(i)}(t)=H_\mathrm{eg-laser}^{(i)}(t)+H_\mathrm{rg-laser}%
^{(i)}(t)+H_\mathrm{sr-laser}^{(i)}(t),  \label{eq:Laser}
\end{split}%
\end{equation}
where each term has the form
\begin{equation}
\begin{split}\label{eq:Laseralphabeta}
H_{\alpha\beta-\text{laser}}^{(i)}(t)=\frac{\hbar\Omega_\alpha}{2}\sigma_{\beta%
\alpha}^{(i)} e^{-i(\mathbf{k}_{L\alpha} \mathbf{r}_i-\omega_{L\alpha} t)}+%
\mathrm{h.c.}.
\end{split}%
\end{equation}
Here $k_{L\alpha}=\omega_{L\alpha}/c$ is the wave-number of the laser with $\omega_{L\alpha}$ the frequency of the laser,
h.c. denotes the hermitian conjugate, and $\Omega_\alpha$ is the Rabi frequency. The first term $H_\mathrm{eg-laser}%
^{(i)}$ in Eq.~\eqref{eq:Laser} describes the coupling of the atom to the
cooling laser on the ($\ket{g_{(F)}}$-$\ket{e_{(F)}}$)-transition, with Rabi frequency $\Omega_e$ and frequency $\omega_{Le}$ detuned by $\Delta_e$. The term $H_\mathrm{rg-laser}^{(i)}$
describes the coupling of the atom to the Rydberg-dressing laser on the ($\ket{g_{(F)}}$-$\ket{r_{(F)}}$)-transition, with Rabi frequency $\Omega_r$ and a frequency
$\omega_{Lr}$ detuned by $\Delta_r$. In the following we are interested in the regime of large
detuning  $\Delta_r\gg\Omega_r$ in order to weakly admix the Rydberg state to the ground state. Finally, $H_\mathrm{sr-laser}^{(i)}$ describes the coupling to a microwave field
strongly mixing the Rydberg-states $\ket r$ and $\ket s$, with Rabi
frequency $\Omega_s$ and laser frequency $\omega_{Ls}$ detuned by $\Delta_s$. 

In addition $H^{(i)}_\mathrm{trap}$ of Eq.~\eqref{eq:Single} accounts for external trapping potentials.

\subsection{Two-particle Hamiltonian: Born-Oppenheimer potentials}\label{sec:twoparticleham}
In this section we study the interaction between {\it two} Rydberg dressed atoms in the presence of an external static electric field (Sec.~\ref{sec:dcdressing}) or a microwave field (Sec.~\ref{sec:mwdressing}).  For relative distances between the atoms larger than the size of the Rydberg atom $r\sim a_0n_r^2$  and in the presence of external fields the atoms interact via dipole-dipole interaction governed by the Hamiltonian of Eq.~\eqref{eq:Hamdipdip}.
In particular, we derive the BO potential surfaces which, in the adiabatic approximation, play the role of effective interaction potentials \cite{Buchler2007} (Eqs.~\eqref{eq:Pert} \eqref{eq:EggMWRcMW}). They form the basis for the analysis of the time-dependent
dynamics of Rydberg-dressed atoms which we discuss below in Sec.~\ref{sec:dissipation} and~\ref{sec:lasercooling}.

\subsubsection{DC-electric field}\label{sec:dcdressing}
We consider two atoms in the presence of a static DC-field of strength $F$, driven by
the Rydberg-dressing laser, in the configuration of Fig.~\ref{fig:levelscheme}(b). The microwave and the cooling laser are absent, i.e. $\Omega_e=\Omega_s=0$. The new Stark-split eigenstates $\ket{g_F}\sim \ket{g}$ and $\ket{r_F}$ are obtained by diagonalizing $H_\text{at}+H_\text{dc}$ and the
detuning of the dressing laser $\Delta_{r}$ is defined relative to the
shifted energy levels. 
To obtain the BO potentials we first neglect dissipation and treat the position operators  $\mathbf{\hat r}_i$ as parameters $\mathbf{r}_i$. Within this limit each atom can be described by a two-state model consisting of $\ket{g}$ and $\ket{r_F}$, coupled by a dressing laser. This is valid for distances larger than $r_n\sim(D/\Delta E_n)^{1/3}$, where diabatic crossings between BO-surfaces of neighboring $n$-manifolds can be neglected. Here, $D=d_0^2/(4\pi\epsilon_0)$ is the dipolar coupling strength and $\Delta E_n\gg \hbar\Delta_r$ is the energy separation between neighboring states. 
In a rotating frame the single particle Hamiltonian describing this model system reduces to
\begin{equation}
H^{(i)}=-\hbar\Delta_{r} \sigma_{rr}^{(i)} +\frac{\hbar\Omega_{r}}{2}
(\sigma_{rg}^{(i)}+\sigma_{gr}^{(i)}),  \label{eq:SingleRed}
\end{equation}
where the operator $\sigma_{\alpha\beta}^{(i)}=\ket{\alpha_F}\bra{\beta_F}$
acts on the new Stark-split eigenstates, $i\in\{1,2\}$, and $\Delta_r =
\omega_{Lr}-\omega_{r}$ is the detuning from the ($\ket{r_F} - \ket{g}$
)-resonance. Position dependent phases of Eq.~\eqref{eq:Laseralphabeta} will be included in Sec.~\ref{sec:lasercooling}, where they lead to recoil kicks from laser absorption and spontaneous emission, and Doppler shifts.

Since the DC-electric field aligns the dipoles of the atoms
along the direction of the field and the dominant dipole moment is $\mathbf{d}_0=\langle r_F|\hat{\mathbf{d}}|r_F\rangle$, the interaction term in Eq,~\eqref{eq:Hamdipdip} is
\begin{equation}\label{eq:inthamdc}
H_\text{int}^{(ij)}=\frac{D(1-3\cos^2\vartheta)}{|
\mathbf{r}_i-\mathbf{r}_j|^3}\left[\sigma_{rr}^{(i)}\otimes\sigma_{rr}^{(j)}\right],
\end{equation}
where $\vartheta$ is the angle between the dipole axis and the radial vector between two atoms. When the atoms are confined in a 2D-plane, e.g.~by a strong optical field, the angle is fixed to $\vartheta=\pi/2$ resulting in a purely repulsive interaction.
Within this model, the total Hamiltonian [Eq.~\eqref{eq:Ham}] for two atoms in the basis $\{\ket {g, g}, \ket {r_F, g}, \ket {g, r_F}, \ket {r_F, r_F}\}$ reads as
\begin{equation}
H=\hbar
\begin{pmatrix}
0 & \frac{1}{2}\Omega_r & \frac{1}{2}\Omega_r & 0 \\
\frac{1}{2}\Omega_r & -\Delta_r & 0 & \frac{1}{2}\Omega_r \\
\frac{1}{2}\Omega_r & 0 & -\Delta_r & \frac{1}{2}\Omega_r \\
0 & \frac{1}{2}\Omega_r & \frac{1}{2}\Omega_r & V(r)-2 \Delta_r \\
&  &  &
\end{pmatrix}%
,  \label{eq:Matrix}
\end{equation}
with
\begin{equation}
\begin{split}\label{eq:Vr}
 \hbar V(r)=\braket{r_F,r_F|H_\text{int}^{(12)}|r_F,r_F}=
\frac{D}{r^3}.
\end{split}
\end{equation} 

\begin{figure}[tb]
\includegraphics[width=8 cm]{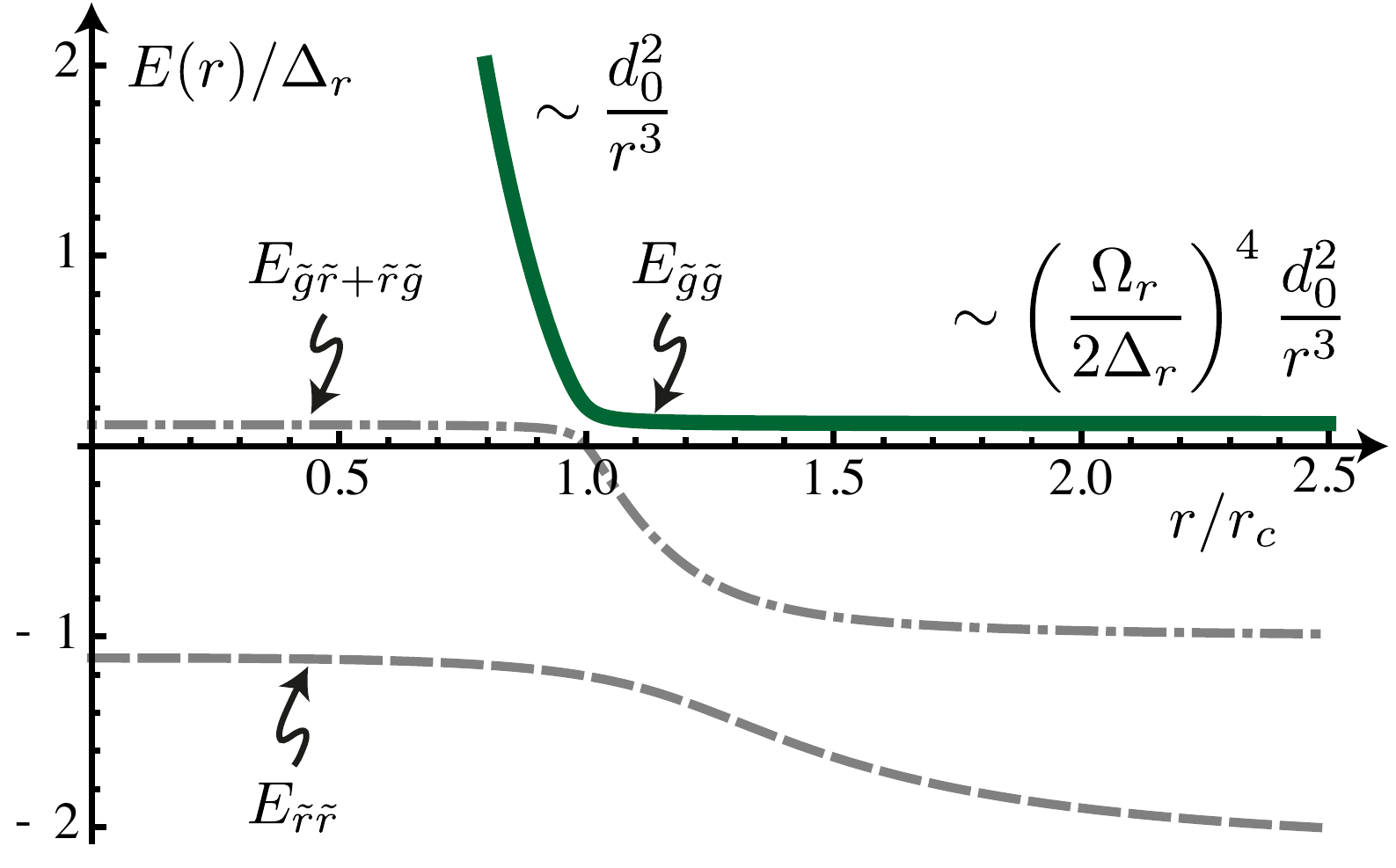}
\caption{(color online) Energy eigenvalues $E(\mathbf{r})$ (dressed BO potential surfaces) of Rydberg-dressed atoms confined in a 2D geometry obtained by diagonalizing the Hamiltonian of Eq.~\eqref{eq:Matrix}. Here, $\mathbf{r}=r(\cos\varphi,\sin\varphi,0)$ is the 2D coordinate in the plane with $z=0$.  Atoms are polarized by the DC-electric field dressing scheme of Sec.~\ref{sec:dcdressing} and $\Delta_r=2 \Omega_r$.  Energy surfaces are labeled using arrows: $E_{\tilde g\tilde g}$ (solid green), $E_{\tilde g\tilde r+\tilde r\tilde g}$ (dashed-dotted) and $E_{\tilde r\tilde r}$ (dashed). At the Condon point $r=r_c$ an avoided crossing leads to a rapid change of the ground state interaction potential $E_{\tilde g\tilde g}$: for $r>r_c$ atoms prepared in the dressed ground state $|\tilde g\rangle$ are weakly interacting, $E_{\tilde g\tilde g}\sim (\Omega_r/2\Delta_r)^4 d_0^2/r^3$, while for $r<r_c$ the potential inherits the character of the Rydberg-Rydberg interaction, $E_{\tilde g\tilde g}\sim  d_0^2/r^3$.
}
\label{fig:E44}
\end{figure}

The BO potentials for the scattering
of two Rydberg-dressed atoms are obtained by diagonalizing the Hamiltonian of Eq.~\eqref{eq:Matrix} for fixed relative position and zero kinetic energy, see Fig.~\ref{fig:E44}. We assume that the linewidths of the corresponding states are smaller than the energy separation between them, and that the relevant kinetic energies are small enough, such that Landau-Zener transitions between different BO surfaces can be neglected (secular approximation). In this case the resulting position dependent eigenvalues act as potentials in each state manifold \cite{Buchler2007,Weiner1999}. Asymptotically (e.g., at large distances where the dipole interaction is negligible), the new dressed eigenstates are $\ket{\tilde g}=\ket{g}+(\Omega_r/2\Delta_r)\ket{r_F}$ and $\ket{\tilde r_F}=\ket{r_F}-(\Omega_r/2\Delta_r)\ket{g}$. We note that the dynamics governed by the Hamiltonian of Eq.~\eqref{eq:Matrix}  for atoms initially in the ground state is restricted to the symmetric subspace made of the three states which asymptotically connect to the states $\ket {\tilde g,  \tilde g}$, $(\ket {\tilde r_F, \tilde g} + \ket {\tilde g, \tilde r_F})/\sqrt{2}$, and $\ket {\tilde r_F, \tilde r_F}$,  with energies slightly perturbed by the (Rydberg-dressing) laser field; the antisymmetric state $(\ket {\tilde r_F, \tilde g} - \ket {\tilde g, \tilde r_F})/\sqrt{2}$ is decoupled, and does not contribute to the dynamics.

In the following, we focus on blue detuning, e.g.~$\Delta_r > 0$. Fig.~\ref{fig:E44} shows the BO-potentials as a function of the interparticle distance $r$ for a specific choice of parameters. Because of the choice of blue-detuning, the figure shows that the ground state BO-potential $E_{\tilde g\tilde g}$ (solid green line) corresponding to the energy of the two-particle asymptotically in the state $\ket {\tilde g, \tilde g}$ has the highest energy. The energy of this BO-potential can be calculated perturbatively up to fourth order in the
small parameter $\Omega_r\ll|V(r)-2\Delta_r|$ as
\begin{subequations}\label{eq:Pert}
\begin{align}
E_{\tilde g \tilde g}(r)&=\frac{\Omega_{r}^2}{2\Delta_{r}} -\frac{
\Omega_{r}^4}{4\Delta_{r}^3}- \frac{\Omega_{r}^4 }{4\Delta_{r}^2\left(V-2%
\Delta_{r}\right)}\\
&\approx\frac{\Omega_{r} ^2}{2\Delta_{r} }-\frac{\Omega_{r} ^4%
}{8\Delta_{r} ^3}+\left(\frac{ \Omega_{r}}{2\Delta_{r} }\right)^4V(r). 
\end{align}
\end{subequations}
where the second line is valid in the limit $V(r)\ll\Delta_r$, i.e. $r\gg r_c$ (see below). While the first two terms on the r.h.s. of Eq.~\eqref{eq:Pert} are simple
light shifts, the equation shows that at large distances the effective
ground state BO-potential has an effective spatial dependence $V_{\tilde g\tilde g}(r)= (\Omega_r/\Delta_r)^4
d_r^2/r^3$. As explained in Refs.~\cite{Honer2010, Pupillo2010, Henkel2010, Maucher2011, Li2012}, this means that by
dressing the particles with the laser field we have achieved a
dipole-dipole interaction for atoms prepared in their (dressed) ground state, and
the strength of the interaction is tunable by varying the ratio $%
\Omega_{r}/\Delta_{r}$.\newline

Figure~\ref{fig:E44} shows that the energy of the dressed two-particle ground state state $E_{\tilde g\tilde g}$ (solid green line) is strongly affected by the dipole-dipole
interactions, and for $V(r) \sim 2\Delta$ avoided crossings occur among the
BO-potentials of all symmetric states.  Note that the energy of the antisymmetric
state $(\ket {\tilde r_F \tilde g_F} - \ket {\tilde g_F \tilde r_F})/\sqrt{2}$
is uncoupled and not shown. In particular, there is a resonant Condon point
at
\begin{equation}
\begin{split}  \label{eq:rc}
r_c=\left(\frac{d_0^2}{8\pi\epsilon_0\hbar\Delta_r}\right)^{1/3}
\end{split}%
\end{equation}
between the ground state BO-potential and other energy surfaces. As a
consequence, there is a sudden change in the slope of the energy surface for
$r \sim r_c$, where the ground state BO-potential inherits the character of the
one that asymptotically connects to the energy of the state $\ket {\tilde r_F,
\tilde r_F}$, and becomes strongly repulsive. This effect has been discussed
in Refs.~\cite{Gorshkov2008, Buchler2007, Weiner1999} in the context of  so-called blue-shielding
techniques, where the strong repulsion for $r < r_c$ does not allow for
particles to come close to each other in a scattering event, thus preventing
collisional losses due to, e.g., collision-induced ionization at short
distance. In the reminder of this work, we will be mostly interested in
confining the dynamics to distances $r > r_c$.\newline

\subsubsection{AC-microwave-field}\label{sec:mwdressing}
In this section we consider the Hamiltonian dynamics
of two atoms in the presence of a linearly polarized, {\it near-resonant} microwave field with Rabi
frequency $\Omega_s$ coupling the Rydberg states $\ket r$ and $\ket s$, see
Fig. \ref{fig:levelscheme}(b). The ground state is again weakly dressed with
the state $\ket r$ using an off-resonant laser with Rabi frequency $\Omega_r$
and a large detuning $\Delta_r\gg\Omega_r$. The DC-electric field and
the cooling laser are absent, i.e. $\Omega_e=0$ and $F=0$. The explicit choice of a {\it near-resonant} microwave field is to obtain large dipoles for atoms in the dressed ground state, which scale as $\sim(\Omega_r/\Delta_r)^2$, similar to the DC-electric field case of Sec.~\ref{sec:dcdressing}. Again neglecting
dissipation and the external degrees of freedom for a moment we are left
with a three level system consisting of the ground state $\ket g$ and the
two Rydberg states $\ket r$ and $\ket s$. Since the electric DC-field is
absent, these states are the bare eigenstates of $H_\text{at}$. Below we show that,
by a judicious choice of system´s parameters, it is possible to obtain an interaction
strength and Condon radius of similar magnitude as in the previous scheme of Sec.~\ref{sec:dcdressing} above.

In a frame rotating with the laser frequencies the single particle
Hamiltonian of Eq.~\eqref{eq:Ham} for this model system reduces to
\begin{equation}
\begin{split}
H^{(i)}=&-\hbar\Delta_r \sigma_{rr}^{(i)}-\hbar(\Delta_r+\Delta_s)\sigma_{ss}^{(i)} \\
&+\frac{\hbar\Omega_r}{2}(\sigma_{rg}^{(i)}+\sigma_{gr}^{(i)})+\frac{\hbar\Omega_s}{2}
(\sigma_{sr}^{(i)}+\sigma_{rs}^{(i)}),  \label{eq:HamSingleMW}
\end{split}
\end{equation}
where the operator $\sigma_{\alpha\beta}^{(i)}=\ket \alpha\bra \beta$ acts
on the bare eigenstates, $i\in\{1,2\}$, and $\Delta_s =
\omega_{Ls}-\omega_{s}$ is the detuning of the microwave laser from the ($%
\ket r - \ket s$)-resonance. The corresponding dipole-dipole interaction Hamiltonian of Eq.~\eqref{eq:Hamdipdip} reduces to
\begin{equation}
H_\text{int}^{(ij)}=\frac{d_{0}^2}{4\pi\epsilon_0}\frac{1-3\cos^2\vartheta}{|%
\mathbf{r}_i-\mathbf{r}_j|^3}\left[\sigma_{sr}^{(i)}\sigma_{rs}^{(j)}+%
\sigma_{rs}^{(i)}\sigma_{sr}^{(j)}\right],  \label{eq:HamIntMW}
\end{equation}
with $\vartheta$ defined as before.

In the case of a \textit{near-resonant}
microwave field, where $\Delta_s\ll \Omega_s$ it is convenient to perform a
unitary transformation  \cite{Klimov2002}
\begin{equation}
U_{rs}^{(i)}=\exp\left\{\frac{1}{2}\tan^{-1}\left(
\frac{\Omega_s}{\Delta_s}\right)\left(\sigma_{rs}^{(i)}-\sigma_{sr}^{(i)}
\right)\right\},  \label{eq:unitary}
\end{equation}
which diagonalizes the Hamiltonian in the subspace  $\{\ket r_i, \ket s_i\}$. The corresponding  new eigenstates  are
$\ket{\pm}_i=a_\mp\ket{s}_i\pm a_\pm\ket{r}_i$ with
\begin{equation}
a_\pm=\frac{1}{\sqrt{2}}\left(1\pm\frac{\Delta_s}{\sqrt{\Delta_s^2+\Omega_s^2}}\right)^{1/2},
\end{equation}
with the corresponding eigenenergies 
\begin{equation}
\begin{split}
E_\pm=-\Delta_r-\frac{1}{2}\left(\Delta_s\mp\sqrt{\Delta_s^2+\Omega_s^2}
\right).
\end{split}
\end{equation}
\begin{figure}[tb]
\centering
\includegraphics[width=6 cm]{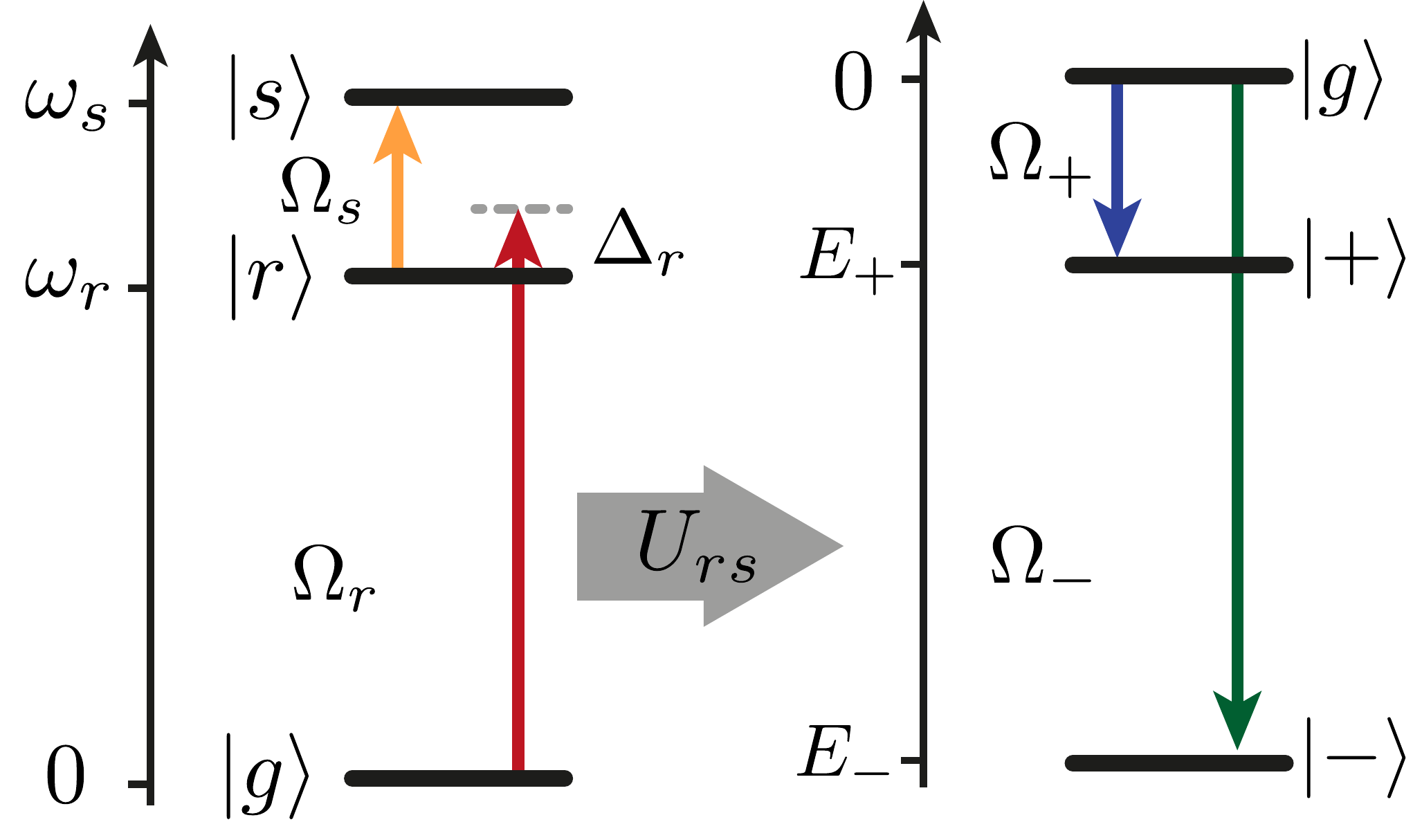}
\caption{(color online) Illustration of the AC-microwave dressing scheme: (left) energies $\omega_\alpha$, states $|\alpha\rangle$ (solid black lines) and lasers (solid arrows) of the system described by the Hamiltonian of Eq.~\eqref{eq:HamSingleMW}. The unitary transformation $U_{rs}$ of Eq.~\eqref{eq:unitary} diagonalizes the Rydberg subspace $\{|r\rangle,|s\rangle\}$. (right) In a rotating frame the new states $|+\rangle$ and $|-\rangle$ with energies $E_\pm$ are separated by an energy difference $\sqrt{\Omega_s^2+\Delta_s^2}$ and individually coupled to the ground state with Rabi frequencies $\Omega_\pm$  according to the Hamiltonian of Eq.~\eqref{eq:HamSinlgeACRot} .}
\label{fig:unitary}
\end{figure}

After the transformation, the single particle Hamiltonian of Eq.~\eqref{eq:HamSingleMW} in the basis  $\{\ket g,\ket +, \ket -\}$ is
\begin{equation}
U_{rs}^{(i)} H^{(i)} U_{rs}^{(i)\dag}=\hbar
\begin{pmatrix}
0 & \frac{1}{2}\Omega_+ & -\frac{1}{2}\Omega_- \\
\frac{1}{2}\Omega_+ & E_+ & 0 \\
-\frac{1}{2}\Omega_- & 0 & E_- \\
\end{pmatrix},\label{eq:HamSinlgeACRot}
\end{equation}
with the effective Rabi frequencies $\Omega_\pm=a_\pm\Omega_r$. In this transformed picture the new eigenstates $\ket \pm$ are 
coupled to the ground state by  lasers with Rabi-frequencies $%
\Omega_\pm$ and have a dipole moment $\mathbf{d}_\pm=\langle \pm|\mathbf{\hat d}|\pm\rangle=\pm\langle r|\mathbf{\hat d}|s\rangle=\pm\mathbf{d}_0$, see Fig.~\ref{fig:unitary}. 
\begin{figure}[tb]
\centering
\includegraphics[width=8 cm]{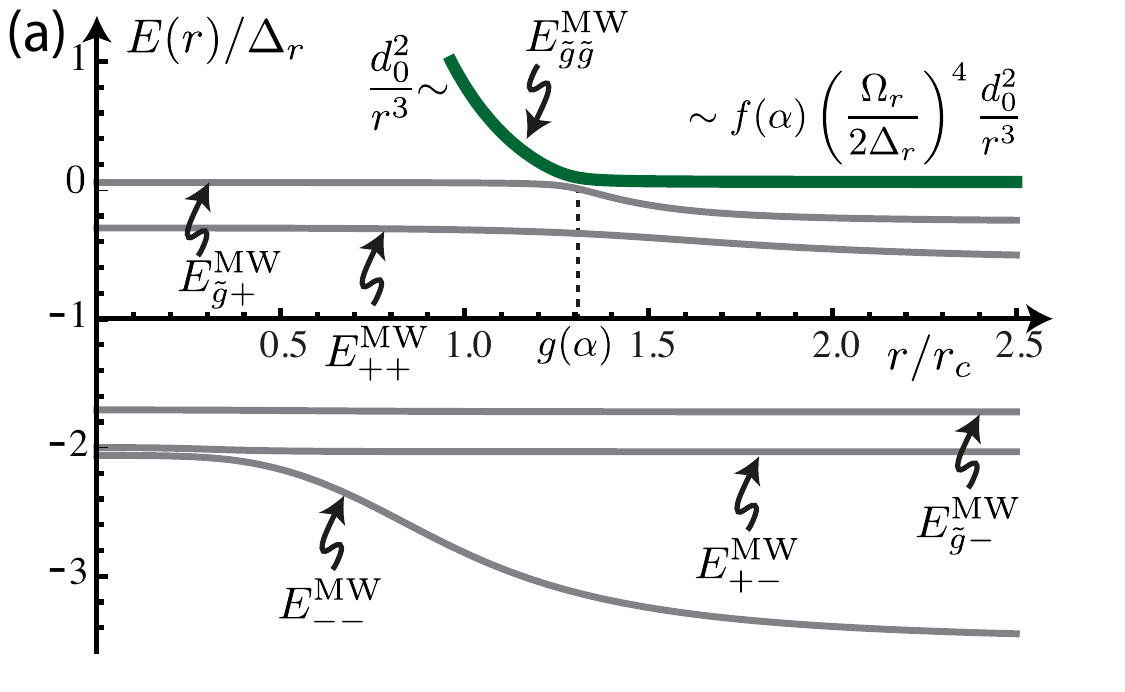}
\includegraphics[width=8 cm]{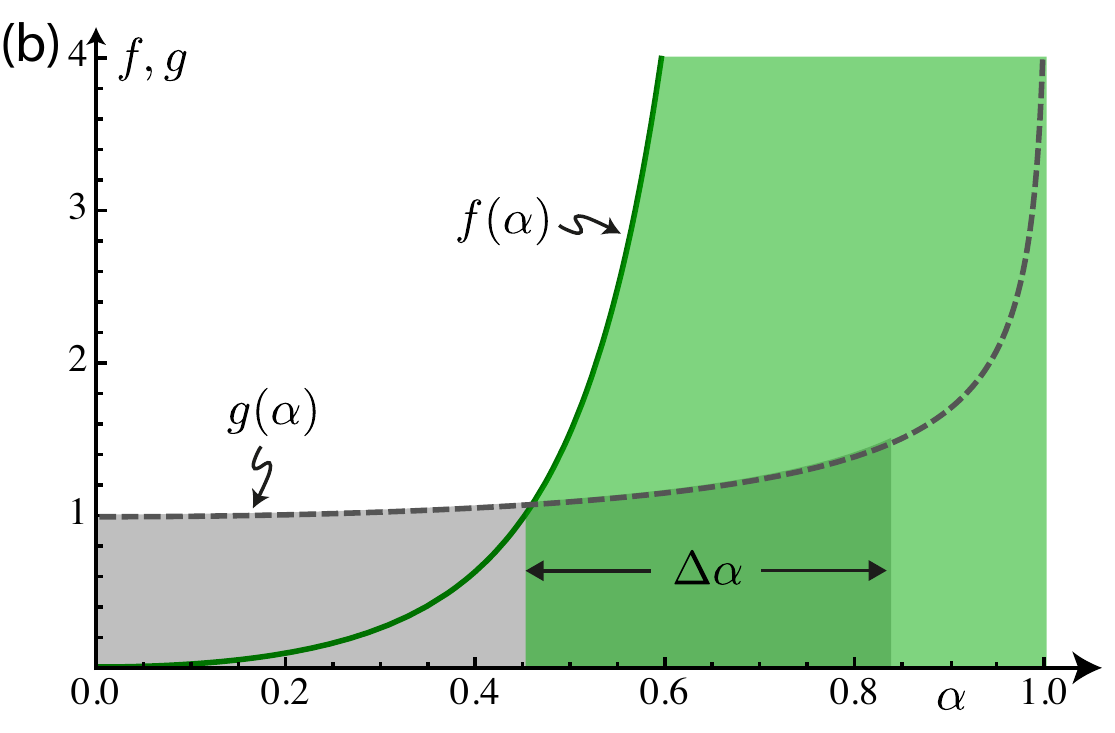}
\caption{(color online) (a) Energy eigenvalues $E(\mathbf{r})$ (dressed BO potential surfaces) of Rydberg-dressed atoms confined in a 2D geometry obtained by diagonalizing the Hamiltonian of Eq.~\eqref{eq:MatrixAC}. Here, $\mathbf{r}=r(\cos\varphi,\sin\varphi,0)$ is the 2D coordinate in the plane with $z=0$. Atoms are polarized by the AC-microwave dressing scheme of Sec.~\ref{sec:dcdressing} and $\Delta_r=4\Omega_r$, $\Omega_s=1.5 \Delta_r$ and $\Delta_s=0$.  Energy surfaces are labeled using arrows. The energies $E_{++}^{\rm MW}(r)$ and $E_{--}^{\rm MW}(r)$ are strongly affected by the dipole interaction and at the Condon point $r_c^{\rm MW}=g(\alpha)r_c$ an avoided crossing leads to a rapid change of the  ground state interaction potential $E_{\tilde g\tilde g}^{\rm MW}(r)$ (thick green line): for $r>r_c$ atoms prepared in the dressed ground state $|\tilde g\rangle$ are weakly interacting, $E_{\tilde g\tilde g}^{\rm MW}\sim f(\alpha)(\Omega_r/2\Delta_r)^4 d_0^2/r^3$, while for $r<r_c$ the potential inherits the character of the Rydberg-Rydberg interaction, $E_{\tilde g\tilde g}^{\rm MW}\sim  d_0^2/r^3$. The figure shows that the Condon radius is slightly shifted to larger values compared to the DC-dressing scheme of Fig.~\ref{fig:E44}. 
Panel (b) shows the dependence of the interaction strength, $f(\alpha)$, (green and solid line) and the Condon radius, $g(\alpha)$ (grey and dashed line) defined in Eq.~\eqref{eq:EggMWRcMW} on $\alpha$. There exists a region $\Delta\alpha$ where $f(\alpha)>1$ and $g(\alpha)<1.5$.}
\label{fig:EResAlpha}
\end{figure}

Again, the dynamics of the symmetric and antisymmetric states are decoupled. For two particles the dynamics of the symmetric subspace is governed by the Hamiltonian $H=H^{(1)}+H^{(2)}+H^{(12)}_{\rm int}$ which, represented in the basis $\{|g,g\rangle,\frac{1}{\sqrt{2}}(|g,+\rangle+|+,g\rangle),\frac{1}{\sqrt{2}}(|g,-\rangle+|-,g\rangle),\frac{1}{\sqrt{2}}(|-,-\rangle+|+,+\rangle),\frac{1}{\sqrt{2}}(|+,-\rangle+|-,+\rangle),\frac{1}{\sqrt{2}}(|-,-\rangle-|+,+\rangle)\}$, reads
\begin{widetext}
\begin{equation} 
H_{\rm sym}=\hbar
\begin{pmatrix}
0	  			&\frac{\Omega_r}{2}				&-\frac{\Omega_r}{2}		&0						&0 						&0\\
\frac{\Omega_r}{2}	&-\Delta_r+\frac{\Omega_s}{2}		&0						&\frac{\Omega_r }{2\sqrt{2}}	&-\frac{\Omega_r }{2\sqrt{2}}	&-\frac{\Omega_r }{2\sqrt{2}}\\
-\frac{\Omega_r}{2}	&0							&-\Delta_r-\frac{\Omega_s}{2}	&-\frac{\Omega_r }{2\sqrt{2}}	&\frac{\Omega_r }{2\sqrt{2}}	&-\frac{\Omega_r }{2\sqrt{2}}\\
0				&\frac{\Omega_r }{2\sqrt{2}}		&-\frac{\Omega_r }{2\sqrt{2}}	&-2\Delta_r				&0						&-\Omega_s\\
0				&-\frac{\Omega_r }{2\sqrt{2}}		&\frac{\Omega_r }{2\sqrt{2}}	&0						&-2\Delta_r				& 0 \\
0				&-\frac{\Omega_r }{2\sqrt{2}}		&-\frac{\Omega_r }{2\sqrt{2}}	&-\Omega_s				&0						& V-2\Delta_r\\
\end{pmatrix},\label{eq:MatrixAC}
\end{equation}
\end{widetext}
where we assumed exact resonance of the microwave field, i.e. $\Delta_ s=0$.

The BO-potentials are obtained by diagonalizing the Hamiltonian $H_{\rm sym}$, leading to the new dressed eigenstates which parametrically depend on $r$. An analytic expression for the BO-potential of two atoms in the dressed ground state can be obtained perturbatively in the limit $\Omega_r\ll\{\Delta_r,\Omega_s\}$
\begin{multline}
E_{gg}^{\rm MW}=\frac{2\Delta_r\Omega_r^2}{4\Delta_r^2-\Omega_s^2}\left[ 1- \frac{
\Omega_r^2 (4\Delta_r^2+3\Omega_s^2)}{\left(4 \Delta_r^2-\Omega_s^2\right)^2}\right.\\
\left.
-\frac{\Omega_r^2 [
\Omega_s^4-16\Delta_r^4-(4\Delta_r^2+3\Omega_s^2)(\Omega_s^2-4\Delta_r^2)]}{\left(4 \Delta_r^2-\Omega_s^2\right)^2( 2
V \Delta_r-4 \Delta_r^2+\Omega_s^2)}\right].  \label{eq:EggMW}
\end{multline}
Note, that there are two resonances at $\Delta_r=\Omega_s/2$ and $2 V(r_c^{\rm MW}) \Delta_r-4 \Delta_r^2+\Omega_s^2=0$. The first one corresponds to the level crossing between the states $\ket{ g-}$ and $\ket{--}$ (red detuning) or $\ket{ g+}$ and $\ket{++}$ (blue detuning)  with the two-particle ground state $\ket{ g g}$. The second one corresponds to a Condon point $r_c^{\rm MW}$, similar to the one we have discussed before for the DC-electric field case. 

We now focus on the case $\Omega_s/2\lesssim\Delta_r$ in which the single-particle  $\ket{+}$-state with energy $ E_+\simeq-\Delta_r+\frac{1}{2}\Omega_s\approx -\epsilon$ ($0<\epsilon\ll \Delta_r$) gets almost degenerates with the ground state $\ket{g}$, while the state $\ket -$ with energy $E_-\simeq-\Delta_r-\frac{1}{2}\Omega_s\approx -2\Delta_r$ is separated by a large energy gap of $2\Delta_r$. In order to investigate the behavior of Eq.~\eqref{eq:EggMW} near this resonance we set $%
\Omega_s=\alpha 2\Delta_r$ with $\alpha< 1-\Omega_r/\Delta_r$.
The latter inequality comes from the fact that we assume $\Omega_r$ is the
smallest frequency scale for the non-degenerate perturbation theory to be
valid. Therefore, these two states must not be exactly degenerate. With this assumption we find for the ground state energy and the Condon radius
\begin{equation}
\begin{split}\label{eq:EggMWRcMW}
&E_{gg}^{\rm MW}=E_{\rm const.}+f(\alpha)\left(\frac{\Omega_r}{2\Delta_r}\right)^4 V(r),\\
&r_c^{\rm MW}=g(\alpha)\sqrt[3]{\frac{d_0^2}{8\pi\epsilon_0\hbar\Delta_r}},
\end{split}%
\end{equation}
with $E_{\rm const.}=\Omega_r^2/[2\Delta_r(1-\alpha^2)]-\Omega_r^4(1+\alpha^2)/[8\Delta_r^3(1-\alpha^2)^3]$, $f(\alpha)=2\alpha^2/(1-\alpha^2)^4$ and $g(\alpha)=1/(1-\alpha^2)^{1/3}$. In the limit $\alpha\rightarrow 0$ the state $|s\rangle$ is not coupled to $|r\rangle$ and we obtain the same light shifts as in Eq.~\eqref{eq:Pert}. Note that in this limit the interaction strength vanishes, $f(\alpha)\rightarrow 0$. This is due to the fact that in the absence of a DC-electric field the bare state $|r\rangle$ has no intrinsic dipole moment.

Figure \ref{fig:EResAlpha}(a) shows the BO-potentials for the symmetric states
as a function of the interparticle distance $r$ for a specific set of
parameters near resonance, as discussed above. 
The energy of the dressed two-particle states $\ket{--}$ and $\ket{ ++ }$ is strongly shifted by the
dipole-dipole interaction, and avoided crossings occur among the
BO-potentials of all symmetric states. This leads to a sudden change
in the slope of the energy surface of, e.g.~$E_{ g g}$, at the
Condon radius $r_c^{\rm MW}$.

The functions $f(\alpha)$ and $g(\alpha)$ versus $\alpha$ are shown in Fig.~\ref{fig:EResAlpha}(b). Increasing $\alpha$ towards 1 will on one hand increase the effective ground state interaction potential according to $f(\alpha)$. On the other hand it will increase the Condon radius $r_c^{\rm MW}$ according to $g(\alpha)$. Fig.~\ref{fig:EResAlpha}(b) shows that there is a region, $\Delta\alpha$, for which $f(\alpha)>1$ but $1<g(\alpha)<1.5$ between $0.45<\alpha< 0.84$. Operating in this region leads to formal similar interaction strength and Condon radii as for the DC-electric field dressing scheme, e.g.~Eq.~\eqref{eq:Pert} and Eq.~\eqref{eq:rc}.

\subsection{Validity of the 2D treatment}\label{sec:2dvalidity}
In this section we examine in detail under what criteria we can treat the system of interacting Rydberg dressed atoms as purely 2D in nature. As mentioned before, the atoms are trapped  in the ($x-y$)-plane by a strong harmonic confinement along the $z$-direction.
The resulting three dimensional (3D) potential in relative coordinates reads
\begin{equation}
\begin{split}\label{eq:3DPotFull}
V_{3D}(\mathbf{r})=\frac{1}{4}M\omega_\perp^2z^2+\hbar E_{gg}^{}(\mathbf{r}).
\end{split}%
\end{equation}
where the first term is the harmonic confinement with trapping-frequency $\omega_\perp$. The second term, $E_{g g}(\mathbf{r})$, is the BO potential of two interacting dressed ground state atoms, obtained by numerically diagonalizing the Hamiltonian of \eqref{eq:Matrix}, taking into account the full 3D characteristic of the dipole-dipole interactions. 

\begin{figure}[tb]
\centering
\includegraphics[width=7 cm]{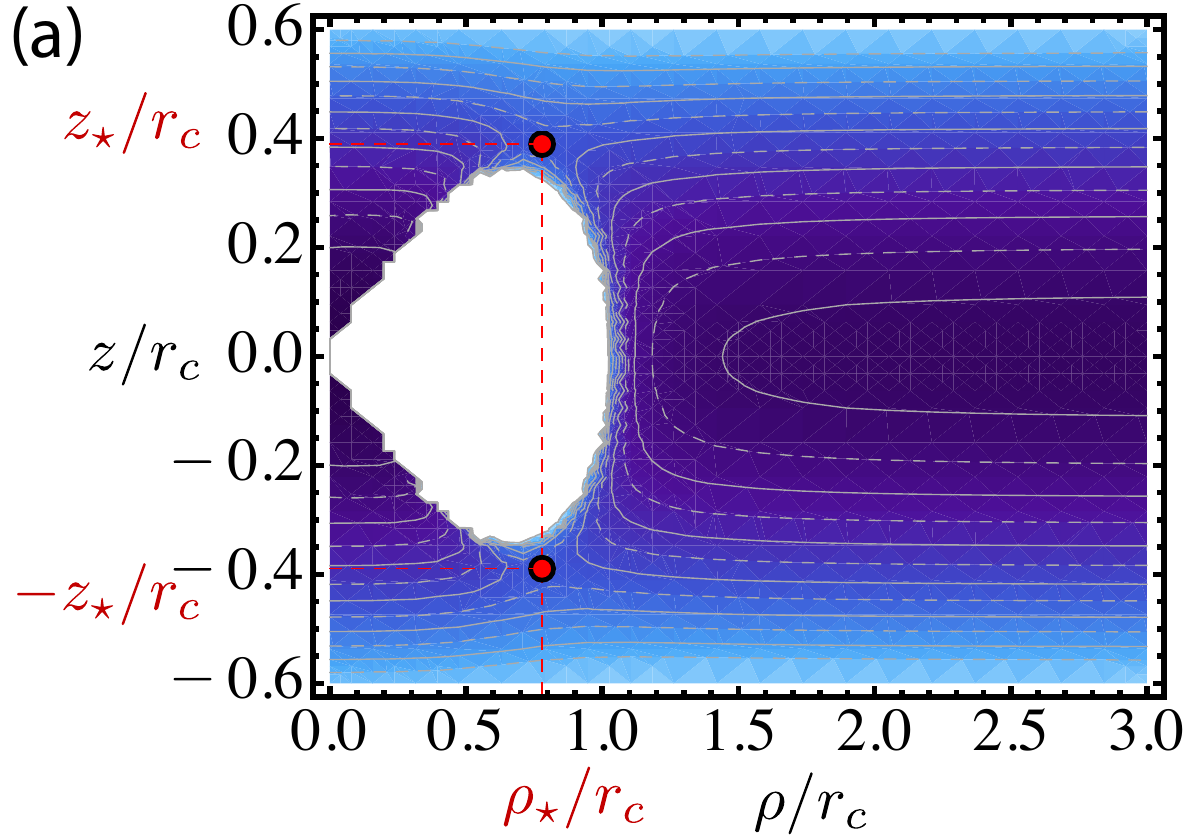}
\includegraphics[width=7 cm]{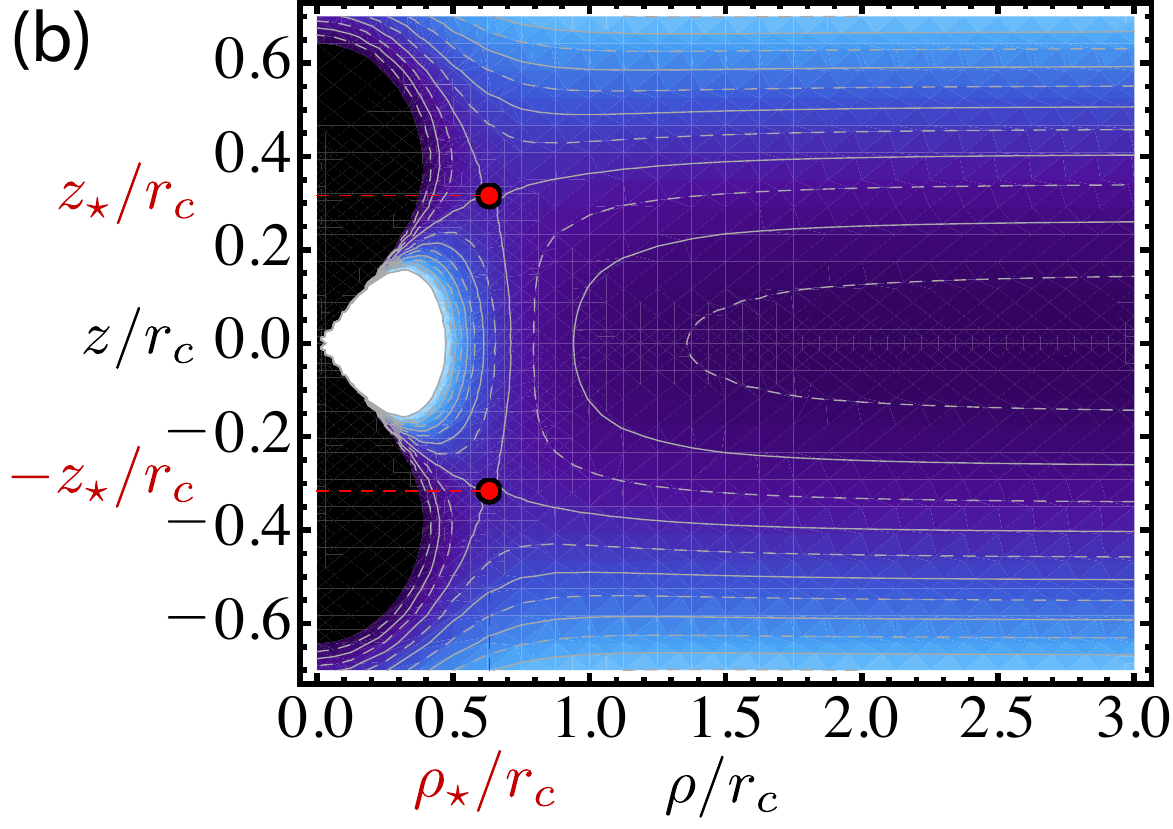}
\caption{(color online) Contour plots of the 3D-potentials $V_{3D}(\mathbf{r})$ and $V_{3D}^{\rm mol}(\mathbf{r})$  of Eq.~\eqref{eq:3DPotFull} and~\eqref{eq:3DPot} are shown in panels (a) and (b), respectively, in units of $V_0=2\Delta_r\left(\Omega_r/2\Delta_r\right)^4$. Here, $\mathbf{r}=(\rho\cos\varphi,\rho\sin\varphi,z)$ is the relative distance between two atoms, with $\rho$ the in-plane radial coordinate and $z$ the transversal coordinate. Brighter regions represent stronger repulsive interactions. Two saddle points (red circles) located at $(\rho_\star,\pm z_\star)$ separate the repulsive from the attractive short-range region.}
\label{fig:2dStability}
\end{figure}

Fig.~\ref{fig:2dStability}(a) shows a contour plot of the 3D potential $V_{3D}(\mathbf{r})$ of Eq.~\eqref{eq:3DPotFull} in the $(\rho$-$z)$-plane where $\rho=\sqrt{x^2+y^2}$. We consider $^{85}$Rb atoms with $\Delta_r=2\pi\times 250$ MHz, $\Omega_r=2\pi\times 100$ MHz, $r_c=520$ nm and $\omega_\perp=2\pi\times 200$ kHz. In the figure, darker color corresponds to deeper potentials. The potential exhibits two saddle-points, at $(\rho_\star,z_\star)=(0.78, \pm0.39)\, r_c$, with a height $V_{3D}(\rho_\star,z_\star)=220\;\mu\text{K}\cdot k_B$, which serve as an energy barrier separating the repulsive long-range dipole-dipole interaction regime from an attractive short-distance regime \cite{Micheli2007, Buchler2007}. For relative kinetic energies smaller than the height of the potential barrier, the interaction is purely repulsive and the system can be stabilized against collapse due to the attractive part of the interaction. For blue detuning of the dressing laser there is a resonant Condon point at $\rho=r_c$ for $z=0$, where we observe a rapid increase of the interaction potential, which is discussed in Sec.~\ref{sec:dcdressing}. Along the axial direction $z$ (for $\rho=0$)  there is no resonant point and the potential approaches smoothly zero where inelastic or reactive collisions will occur. 

For relative distances larger than the Condon radius, $r_c$ [Eq.~\eqref{eq:rc}], the interaction can be approximated by dipole-dipole interaction. In this case, the 3D potential can be rewritten as 
\begin{equation}
\begin{split}\label{eq:3DPot}
V_{3D}^{\rm mol}(\mathbf{r})=V_0\left[\kappa \tilde{z}^2+%
\frac{\tilde\rho^2-2 \tilde{z}^2}{(\tilde\rho^2+\tilde z^2)^{5/2}}\right],
\end{split}%
\end{equation}
which is reminiscent of polar molecules \cite{Buchler2007,Micheli2007}. Here, $\tilde z=z/r_c$ and $\tilde
\rho=(x^2+y^2)^{1/2}/r_c$.  The corresponding contour plot is shown in Fig.~\ref{fig:2dStability}(b). The dimensionless parameter $\kappa$ characterizes the strength of the confinement relative to the interaction
\begin{equation}
\begin{split}
\kappa=\left(\frac{2\Delta_r}{\Omega_r}\right)^4\frac{M\omega_\perp^2 r_c^2}{%
8\hbar \Delta_r}= 2\left(\frac{\Delta_r}{\Omega_r}\right)^3\frac{%
\omega_\perp }{\Omega_r}\left(\frac{r_c}{a_\mathrm{ho}}\right)^2,
\end{split}%
\end{equation}
where $a_\mathrm{ho}=\sqrt{\hbar/M\omega_\perp}$ is the harmonic oscillator
length in transversal direction. For $\kappa\gg 1$ the confinement in $z$-direction dominates over the dipole-dipole interaction. Analytic analysis of Eq.~\eqref{eq:3DPot} yields the saddle points
\begin{equation}
\begin{split}
\rho_\star=\pm 2 z_\star\qquad\text{and}\qquad
z_\star=\frac{3^{1/5}}{\sqrt{5}\kappa^{1/5}}.
\end{split}
\end{equation}
The height of the potential at the saddle point, corresponding to the height of the energy barrier separating the repulsive region from the attractive region, is 
\begin{equation}
\begin{split}
V_{3D}^{\rm mol}(z_\star,\rho_\star)=V_0\left(\frac{\kappa}{3}\right)^{3/5},
\end{split}
\end{equation}
which is independent of $\Delta_r$ for a fixed ratio of $\Omega_r/\Delta_r$. For the same parameters used above we obtain $\kappa=30$ which yields a potential barrier at the saddle point corresponding to $T_\star=150\,\mu$K. Comparing Fig.~\ref{fig:2dStability}(a) and (b) one can see the rapid change of the interaction strength at the Condon radius for Eq.~\eqref{eq:3DPotFull} (panel (a)) leads to a repulsive shield which leads to a slightly higher potential barrier corresponding to $T_\star=220\,\mu$K compared to the case of a  pure inverse-cubic interaction potential (panel (b)) with $T_\star=150\,\mu$K. Additionally we find that the saddle-point in panel (a) is shifted to larger $\rho$  values than in panel (b).

In a 3D scenario two particles undergo collapse when they have large enough collisional energy to overcome the energy barrier at the saddle points. In the numerical simulations of Sec.~\ref{sec:dissipation} we will treat the system as purely 2D with a $1/r^3$-potential (see Eq.~\eqref{eq:fij}), assuming particles are being lost if they collapse in 3D. For a specific set of parameters we use the latter analysis to calculate the height of the potential at the saddle point, $V_{3D}(\rho_\star,z_\star)$, in a 3D scenario. The height of the saddle point will then be translated into a critical 2D distance, $\rho_{\rm loss}$, at which $E_{\tilde g\tilde g}(\rho_{\rm loss})=V_{3D}(\rho_\star,z_\star)$, with  $E_{\tilde g\tilde g}$ defined in Eq.~\eqref{eq:Pert}(b).

\section{Laser cooling of dressed Rydberg atoms}\label{sec:lasercooling}
In this section we examine in detail the dissipative processes in Eq.~\eqref{eq:mastereq} and study laser cooling of interacting atoms in the presence of Rydberg dressing. As an example we consider Doppler cooling but the model can be extended to sub-Doppler cooling schemes. For simplicity of the \textit{analytic} treatment in Sec.~\ref{sec:fpeq}, we will first consider model atoms where we neglect decay from the Rydberg state in order to derive semiclassical Fokker-Planck equations which describe the cooling dynamics in the presence of Rydberg-dressing and interactions. In Sec.~\ref{sec:rateeq} we will include decay from the Rydberg state which couples the dressed ground state via rate equations to one of the intermediate states. Again for simplicity of the analytic treatment in this section, we will first consider model atoms with a single intermediate state $|m\rangle$. In Sec.~\ref{sec:dissipation} we will numerically investigate the interplay of laser-cooling and heating due to population of intermediate states of an ensemble of dressed Rydberg atoms taking into a large number of internal states. We show that laser cooling can alleviate the heating dynamics for interacting Rydberg atoms described above, extending the lifetime of strongly interacting phases in these systems, as for self-assembled crystals.

\subsection{Fokker-Planck equation}\label{sec:fpeq}
In the following we will derive equations of motion for the external dynamics (position and momentum) of model atoms consisting of a ground state $|g\rangle$ which is coupled with a far detuned laser to a Rydberg state $|r\rangle$ with Rabi frequency $\Omega_r$ and detuning $\Delta_r\gg\Omega_r$. Additionally, the ground state $|g\rangle$ is coupled to a lower-lying excited state $|e\rangle$ using a counter-propagating laser with Rabi frequency $\Omega_e$ and detuning $\Delta_e$. The state $|e\rangle$ decays directly to the ground state $|g\rangle$ with a fast decay rate $\Gamma_e$, realizing a closed cycle transition. Atoms which are both in the Rydberg state interact via dipole-dipole interaction described by the Hamiltonian Eq.~\eqref{eq:inthamdc}.

The derivation of equations of motion for the external degrees of freedom is done in four steps \cite{Dalibard1985b, Minogin1987}: (i) first we use the Wigner function formalism to map the density operator $\rho$ onto a quasi-probability distribution in phase space
\begin{widetext}
\begin{multline}\label{eq:wigner}
W^{(N)}(\mathbf{r}_1\dots\mathbf{r}_N;\mathbf{p}_1\dots\mathbf{p}_N;t)=\\
\int \frac{d
\mathbf{u}_1}{h^3}\ldots \frac{d\mathbf{u}_N}{h^3} \braket{{\textstyle\mathbf{r}_1+\frac{\mathbf{u}_1}{2}\dots
\mathbf{r}_N+\frac{\mathbf{u}_N}{2}}|\rho|{\textstyle\mathbf{r}_1-\frac{\mathbf{u}_1}{2}
\ldots\mathbf{r}_N-\frac{\mathbf{u}_N}{2}}}e^{-i \mathbf{u}_1 \cdot
\mathbf{p}_1/\hbar}\ldots e^{-i \mathbf{u}_N\cdot \mathbf{p}_N/\hbar}.
\end{multline}
\end{widetext}
(ii) We derive the equation of motion for $W$ using the master equation Eq.~\eqref{eq:mastereq} and expand the resulting equation of motion up to second order in terms of the photon momentum in order to obtain a positive probability function in the semiclassical limit. (iii) We adiabatically eliminate the Rydberg state $\ket r$ in the limit of a large detuning $\Delta_r$. The resulting two-level systems - consisting of a dressed ground state $\ket{\tilde g}=|g\rangle+(\Omega_r/2\Delta_r)|r\rangle$ with dipole moment $d_{\tilde g}$ and an excited state $\ket e$ - interact only when the atoms are in the dressed ground state.  (iv) Finally, we adiabatically eliminate the fast internal degrees of freedom of this effective two-level system in favor of the (much slower) external dynamics. Due to the Rabi oscillations between the dressed ground state $|\tilde g\rangle$ (with dipole moment $d_{\tilde g}$) and the lower-lying excited state $|e\rangle$ (with negligible dipole moment) the dipole of the atom is fluctuating in time. This leads to an additional diffusion term in the equation of motion in addition to the standard Doppler-cooling diffusion terms. After a lengthy calculation one  obtains for two atoms in their dressed ground state, $W_{ \tilde g \tilde g}^{(2)}=\langle \tilde g\tilde g|W|\tilde g\tilde g\rangle$, the following equation of motion
\begin{multline}\label{eq:fpeq}
\left( \frac{\partial }{\partial t}+\sum_{i=1}^{2}\frac{\mathbf{p}_{i}}{M}
 \frac{\partial }{\partial \mathbf{r}_{i}}\right) W_{ \tilde g \tilde g}^{(2)}=\sum_{i=1}^{2}
\left[ \frac{\partial }{\partial \mathbf{p}_{i}}\mathbf{f}_c+\mathcal{D}_1^{(i)}\right]  W_{ \tilde g \tilde g}^{(2)}\\
-\mathbf{f}_{12}\cdot \left( \frac{\partial }{\partial \mathbf{
p}_{1}}-\frac{\partial }{\partial \mathbf{p}_{2}}\right) W_{ \tilde g \tilde g}^{(2)}+\mathcal{D}_{12}  W_{\tilde g\tilde g}^{(2)},
\end{multline}
which is valid when $\Delta_r\gg\Omega_r,\Omega_e,\Gamma_e\gg\hbar k_{eg}^2/m$, where $k_{eg}$ is the wave number of the ($|e\rangle$-$|g\rangle$)-transition.
Additional to the free motion term on the left hand side, the first term on the right hand side describes single particle Doppler cooling with the cooling force $\mathbf{f}_c=\beta\,\mathbf{\hat k}_{Le}\;\mathbf{k}_{Le}\cdot \mathbf{p}$ and the standard diffusion operator $\mathcal{D}_1=\sum_k D_{k}\partial^2_{p_k}$ accounting for spontaneous emission and diffusion due to the cooling laser in various spatial directions $k\in\{x,y,z\}$ \cite{Minogin1987}. Here we assumed two counter-propagating laser beams and expanded the resulting radiation pressure forces up to second order in $\mathbf{k}_{Le}$.
The third term of Eq.~\eqref{eq:fpeq} proportional to
\begin{equation}
\mathbf{f}_{12}=\frac{\Omega_r^4}{4\Delta_r^2(V-2\Delta_r)^2}\frac{\partial V}{\partial \mathbf{r}_1}=-\frac{\partial E_{\tilde g\tilde g}}{\partial \mathbf{r}_1}
\end{equation}
accounts for the interaction between the two atoms. We note that this term can be rewritten,
using the potential of the dressed ground state, Eq.~\eqref{eq:Pert}, obtained in section \ref{sec:twoparticleham}. Besides those standard laser cooling terms, \cite{Dalibard1985b, Minogin1987}, the term
\begin{equation}\label{eq:crossnoise}
\mathcal{D}_{12} =\frac{4\Omega_e^2\Gamma_e}{(\Gamma_e^2+4\Delta_e^2)^2}\left[\mathbf{f}_{12}\cdot\left( \frac{\partial }{\partial \mathbf{
p}_{1}}-\frac{\partial }{\partial \mathbf{p}_{2}}\right)\right]^2
\end{equation}
is a two-body diffusion term which accounts for the fluctuations of the force (which is present only in the dressed ground state), due to
Rabi-oscillations between $\ket e$ and $\ket{\tilde g}$. For near-resonant laser light it is proportional to the population and lifetime of the excited state. In the limit $\mathbf{f}_{12}\ll\hbar \Gamma_e \mathbf{k}_{Le}$ this diffusion term is small compared to the single-particle diffusion terms and one can approximate the system with two interacting atom, which are independently laser cooled.

\subsection{Quantum jumps and rate equations}\label{sec:rateeq}
A quantum jump occurs when the atom in the Rydberg state $|r\rangle$ decays to an intermediate state $| m\rangle$ due to spontaneous emission or blackbody radiation. In the following, we allow for the possibility that the intermediate state $\ket m$ is long-lived, with a lifetime $\sim 1/\Gamma_{mg}$ comparable with the external dynamics and thus cannot be adiabatically eliminated. This will lead to two coupled equations of motion: (i) an equation for laser-cooled atoms in the dressed ground state, with dipole moment $d_{\tilde g}=(\Omega_r/2\Delta_r)^2d_r$, and (ii) an equation of motion for atoms in the $\ket m$-state (and possibly with a dipole moment $d_m$), which do not experience a cooling force.

To simplify the notation for the analytic treatment, we consider only a single atom in the presence of the cooling and dressing lasers. The generalization to two atoms is straightforward. After preforming steps (i)-(iv) of the latter section we find the following coupled FPEs for the dynamics of the dressed ground state $\ket{\tilde g}$ and the intermediate state $\ket m$
\begin{widetext}
\begin{subequations}\label{eq:fpe1a}
\begin{align}
\left(\frac{\partial}{\partial t}+\frac{\mathbf{p}}{M} \cdot\frac{\partial}{
\partial \mathbf{r}} \right) W^{(1)}_{\tilde g}=&-\Gamma_{\tilde g}W^{(1)}_{\tilde g}+(\Gamma_{m}+\mathcal{D}_{mg}) W^{(1)}_{m} +\left[ \frac{\partial}{\partial \mathbf{p}} \cdot\mathbf{f}_c+ \mathcal{D}_1\right]  W^{(1)}_{\tilde g}, \label{eq:FP1a}\\
\label{eq:FP1b}
\left(\frac{\partial}{\partial t}+\frac{\mathbf{p}}{M} \cdot\frac{\partial}{
\partial \mathbf{r}} \right)W^{(1)}_{m}=&-\Gamma_{m} W^{(1)}_{m}+\left[\Gamma_{\tilde g}+\left(\frac{ \Omega_r}{2 \Delta_r}\right)^2
\mathcal{D}_{rm}\right]W^{(1)}_{\tilde g},
\end{align}\end{subequations}
\end{widetext}
where $W^{(1)}_{\alpha}=\langle \alpha| W^{(1)}|\alpha\rangle$ is the single particle Wigner function for the atom in state $|\alpha\rangle$ ($\alpha\in\{\tilde g,m\}$). The equations are coupled by the effective ground state decay-rate 
\begin{equation}
\Gamma_{\tilde g }=\left(\frac{ \Omega_r}{2 \Delta_r}\right)^2\Gamma_{rm}
\end{equation}
due to optical pumping via the Rydberg state $\ket r$. The terms $\mathcal{D}_{\alpha\beta}$ are standard diffusion operators~\cite{Minogin1987} accounting for spontaneous emission from state $|\alpha\rangle$ to $|\beta\rangle$.

\section{Dissipative dynamics of dressed Rydberg atoms}\label{sec:dissipation}
We will now analyze numerically the dissipative processes of Eq.~\eqref{eq:mastereq}, which are associated with the finite lifetime of excited Rydberg states.  In Sec.~\ref{sec:decoherencesingle} below, we consider the case of a {\it single} atom, in which we investigate the population of intermediate states due to spontaneous emission or blackbody radiation. As a consequence, the atom acquires a time-dependent dipole moment. In addition, spontaneous emission and blackbody radiation act as small heating sources due to the photon recoil. 
In Sec.~\ref{sec:moldyn} we will summarize our previous discussion of the driven-dissipative dynamics of laser-cooled and interacting Rydberg-dressed atoms as a prescription for a molecular dynamics simulation.
The effect of population in intermediate states will be more substantial in the case of an {\it ensemble} of Rydberg-dressed atoms, as we discuss in Sec.~\ref{sec:decoherencetwo}. In particular, we find that the dominant heating effect originates from the time-dependent dipole moment which induces strong mechanical effects in the many-body system.
A crucial point is that, the time-dependence of the dipole moment exhibits characteristically different behavior in the DC-electric field case compared to the AC-microwave one. For example, in the case of a static electric field ($F>0$) the states $|\alpha_F\rangle$ can have a large {\it parallel} or {\it antiparallel} dipoles, $\mathbf{d}_\alpha=\langle \alpha_F|\mathbf{\hat d}|\alpha_F\rangle$, on the order of kilo-Debye, causing strong dipole-dipole interactions with other Rydberg-dressed atoms. Whereas, in the case of a AC-microwave field (and no DC-electric field) the intermediate states do not have any dipole moment.

\subsection{Decoherence of a single atom}\label{sec:decoherencesingle}

\begin{figure}[tb]
\centering
\includegraphics[width=8.5 cm]{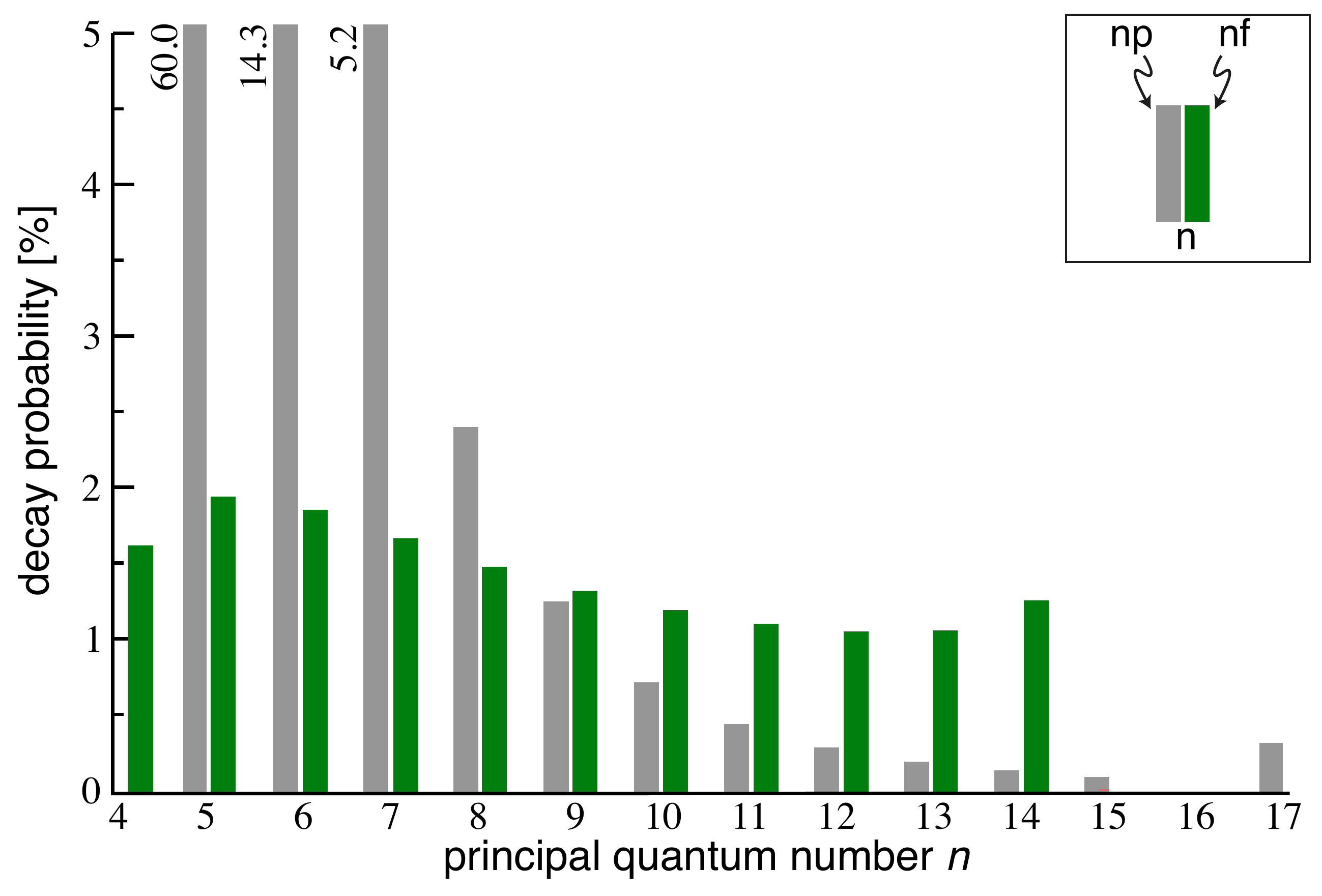}
\includegraphics[width=8.5 cm]{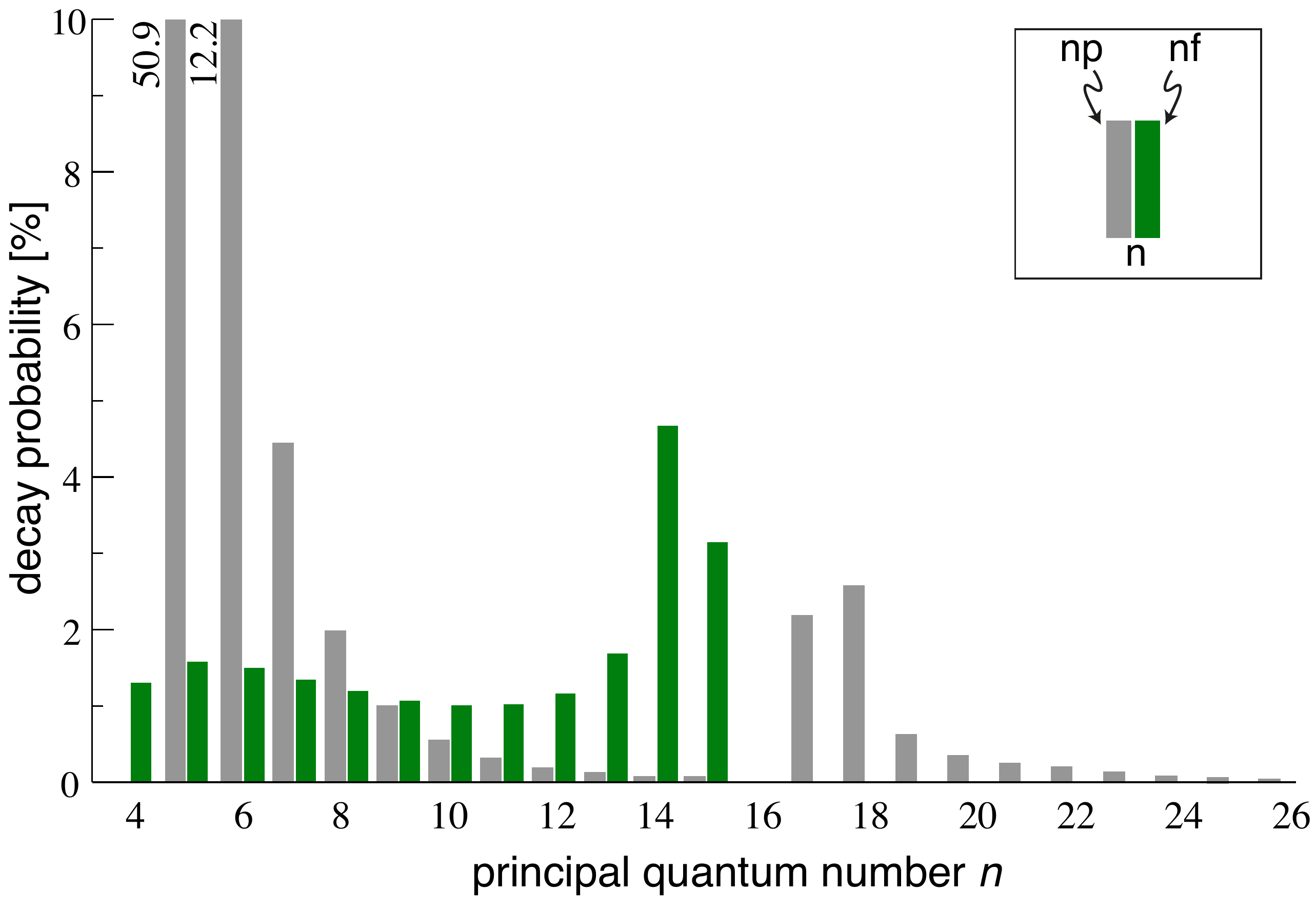}
\caption{(color online) Relative decay rates $\Gamma_{r,n\ell}/\Gamma_r$ of the $|r\rangle=|16\, d_{5/2},\,m_j=-5/2\rangle$ state of $^{85}$Rb. Green and gray bars indicate decay rates to $\ket{n p}$- and $\ket{n f}$-states, respectively. In both cases we summed over $j$ and $m_j$. The upper and lower panel show the relative decay rates for $T=0$~K and $T=300$~K, respectively. The total decay rate of $\Gamma_{16d}=2 \pi \times$ 51 kHz (43 kHz) for $T=300$~K ($T=0$) agrees well with calculations carried out in \cite{Beterov2009a}. Note that in the upper and lower panel the $y$-axes are cut at 5~\% and 10~\%, respectively. For states with a higher decay rate percentage numbers on the left side of the bars indicate their value.}
\label{fig:branchingratio}
\end{figure}

The dressed ground state $\ket {\tilde g}$ has a finite lifetime $\tau_{\tilde g} =
\Gamma_{\tilde g}^{-1}$, where $ \Gamma_{\tilde g} =
(\Omega_r/2\Delta_r)^2 \Gamma_{ r}$, and $1/\Gamma_{ r}$ is the
lifetime of the Rydberg state. The latter is in general given by
blackbody radiation and spontaneous emission, which redistribute population
from the $\ket {r}$ ($\ket {r_F}$)-state to all possible intermediate 
states $|m\rangle$. Such a decay event from the Rydberg state is followed by a cascade process where several intermediate states $|m\rangle$ can be populated. It is crucial to note  that, the cascade process  in general does not happen instantaneously, due to the finite lifetime of the intermediate states. As we discuss below this has far reaching consequences in the long-time dynamics of Rydberg-dressed crystals.

We calculated the decay constants $\Gamma_{\alpha\beta}$ given in Eqs. (\ref{decayc1}) and  (\ref{decayc2}) numerically using both quantum defect theory~\cite{Bates1949, Gounand1979, Branden2010} and a model potential method~\cite{Marinescu1994}. With the first method we calculated the decay matrix including all angular momentum states up to $n=18$, while with the second method we obtained the decay matrix up to $n=110$ including $s$, $p$, $d$, $f$ and $g$-angular momentum states.  With $\Gamma_\alpha=\sum_\beta\Gamma_{\alpha\beta}$ we denote the total decay rate of the state $\ket{\alpha}$ and $\Gamma_0\equiv\Gamma_{\tilde g}=(\Omega_r/2\Delta_r)^2\Gamma_{ r}=(\Omega_r/2\Delta_r)^2\sum_\beta\Gamma_{r\beta}$ is the effective decay rate of the dressed ground state. Fig.~\ref{fig:branchingratio} shows the branching ratio of decay in the absence of external fields from the $|16d\rangle$ state to $|np\rangle$ and $|nf\rangle$ states, where we summed over $j$ and $m_j$ levels. For $T=0$ K (upper panel) the only contribution to the decay rate comes from spontaneous emission, which favors decay to low-lying states. About 60~\% of the population in the $|16d\rangle$ state decays to the $|5p\rangle$ state. The lower panel of Fig.~\ref{fig:branchingratio} shows the branching ratio for $T=300$~K. In this case the decay rate is determined not only by spontaneous emission but also by blackbody radiation, which in addition leads to substantial decay to neighboring states. Hence, only 51~\% of the population in the $|16d\rangle$ states decays directly to the $|5p\rangle$ state. From this it is clear that in current cold atom experiments performed in room temperature environments, the role of high-lying intermediate states cannot be neglected in the long-time limit.

\begin{figure}[tb]
\centering
\includegraphics[width=8 cm]{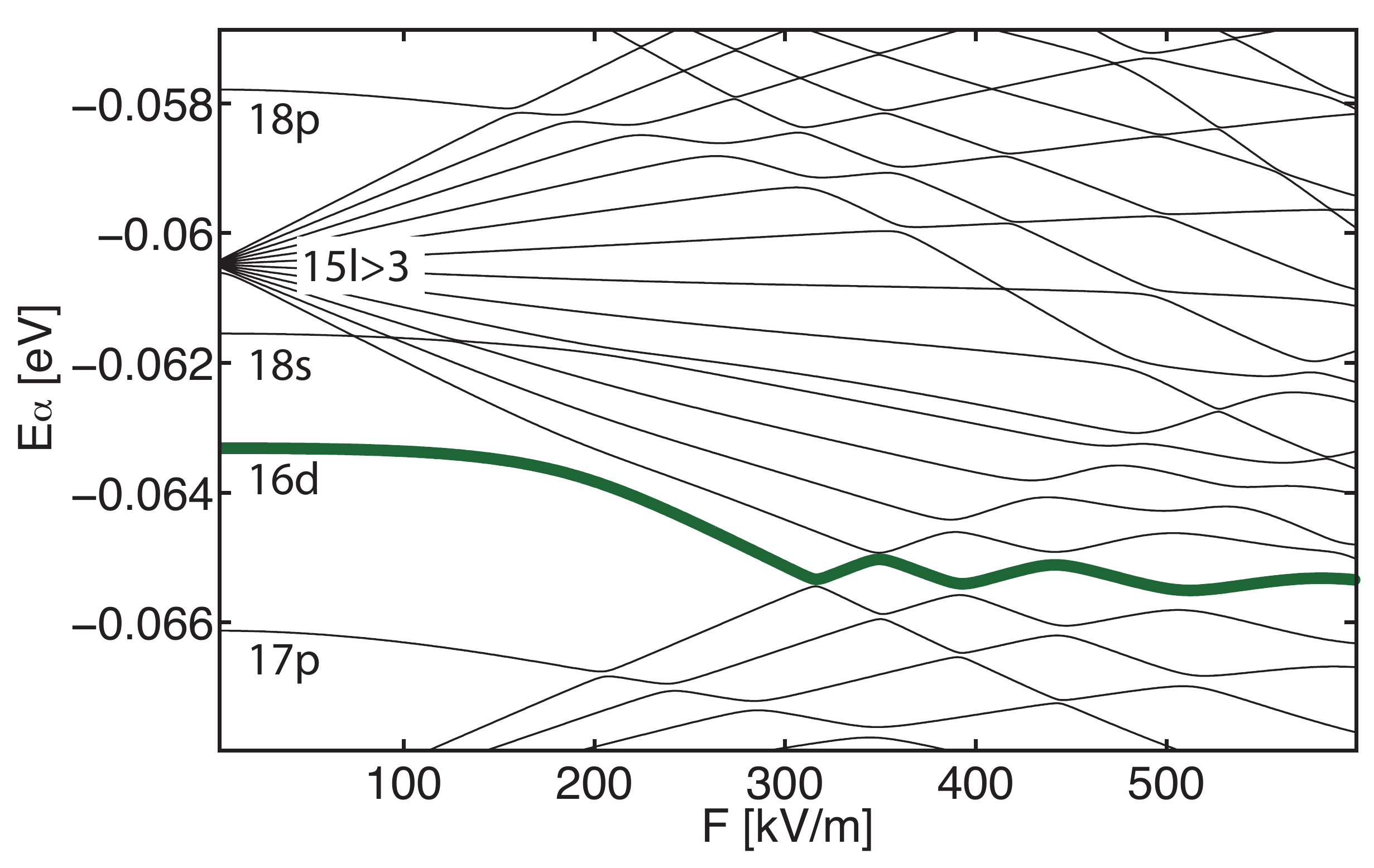}
\caption{(color online) Stark map: Atomic energy levels $E_\alpha(F)$ of $^{85}$Rb as a function of the field strength $F$ around the state $|16d, m=0\rangle$. Note that we plot only states with magnetic quantum number $m=0$. States with an angular quantum number $\ell>3$ have a negligible quantum defect and show a linear Stark effect, while $s$, $p$ and $d$ states show a quadratic.  The energy of the state which for $F\rightarrow 0$ connects to $|16d, m=0\rangle$ (thick green line) is well separated in energy from neighboring states up to a field strength of $F\approx 300$~kV/m.}
\label{fig:starkmap}
\end{figure}
\begin{figure}[tb]
\centering
\includegraphics[width=7 cm]{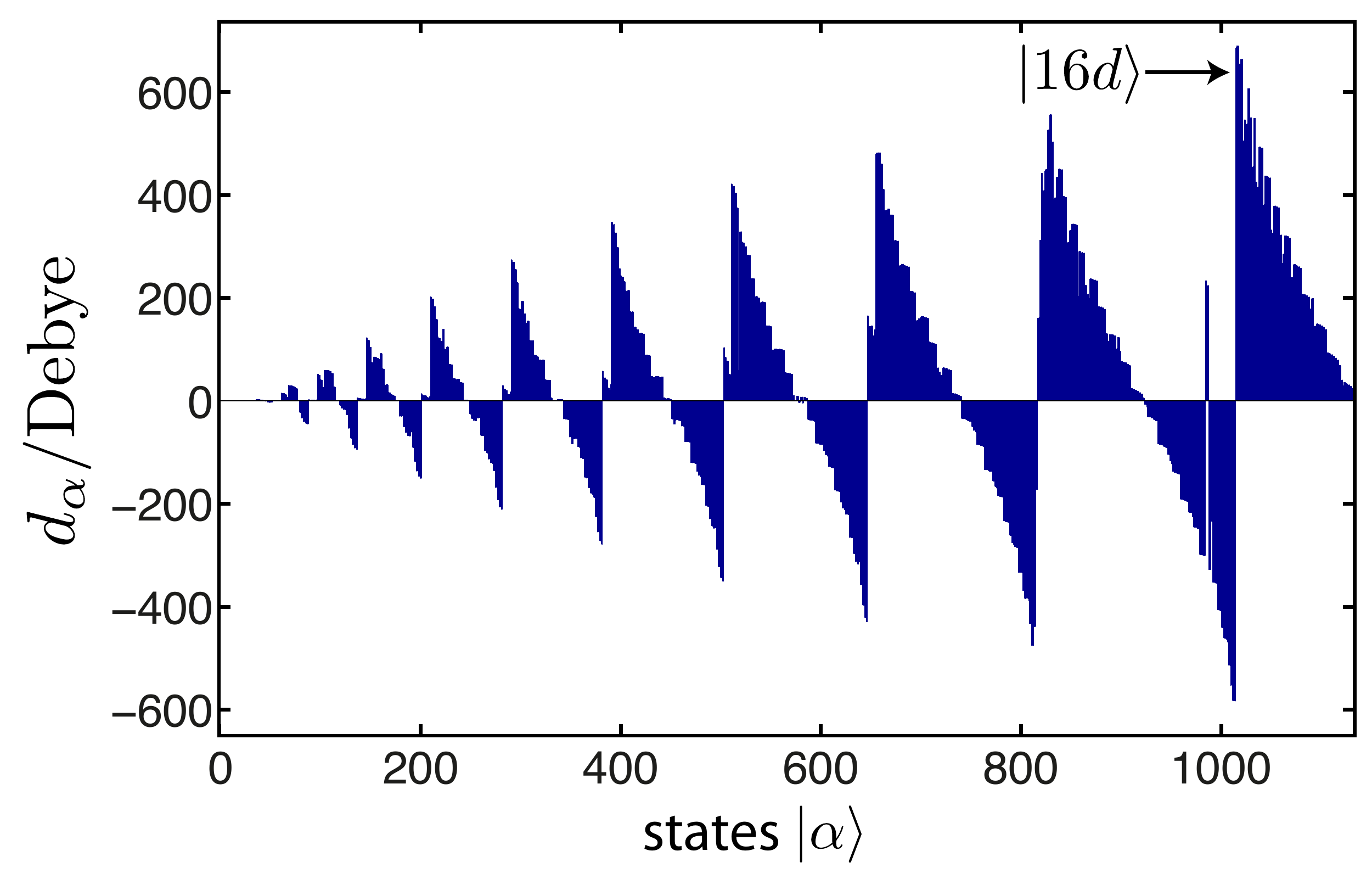}
\caption{(color online) Electric dipole moment $d_\alpha=\langle\alpha_F|\hat d_z|\alpha_F\rangle$ of the Stark split states $|\alpha_F\rangle$ of a $^{85}$Rb atom in an electric DC-field with strength $F=3$~kV/cm. States are numbered with increasing energy, e.g.~the state $|16d, m=0\rangle$ corresponds to $\alpha=1017$. The figure shows that depending on the state the dipole moment $d_\alpha$ can be positive (parallel) or negative (antiparallel) leading to a repulsive or attractive interaction.}
\label{fig:dipolestates}
\end{figure}

\textit{The internal dynamics} corresponding to a cascade process of the
electron towards the ground state following a decay event vim the Rydberg states is
given by 
\begin{equation}\label{eq:TimeEv}
\frac{d}{dt}\left(
\begin{array}{c}
p_{0} \\
p_{1} \\
p_{2} \\
\vdots  \\
\end{array}%
\right) =\left(
\begin{array}{ccccc}
-\Gamma _{0} & \Gamma _{10} & \Gamma _{20} &  \,    \\
\Gamma _{01} & -\Gamma _{1} & \Gamma _{21} &  \ldots\\
\Gamma _{02} & \Gamma _{12} & -\Gamma _{2} &  \,    \\
\,           & \vdots       & \,           &  \,    \\
\end{array}%
\right)  \left(
\begin{array}{c}
p_{0} \\
p_{1} \\
p_{2} \\
\vdots  \\
\end{array}%
\right) ,
\end{equation}%
where $p_{\alpha}$ is the probability of being in the state $\ket{\alpha_{(F)}}$ in the absence (presence) of a static electric field, $F=0$ ($F>0$). Note that in the case of a DC-electric field ($F>0$) the decay rates $\Gamma_{\alpha_F\beta_F}$ are a sum of several decay rates between bare states, $\Gamma_{\alpha\beta}$, according to the contributions of various different angular momentum states to the Stark split states $|\alpha_F\rangle$ and $|\beta_F\rangle$.

{\it DC-electric field:} The  Stark structure of atomic states (including only $m=0$ states) is shown in Fig.~\ref{fig:starkmap} for a electric field along $\mathbf{e}_z$. States with an angular momentum $\ell>3$ are almost degenerate and exhibit a linear Stark effect, while lower angular momentum states, which have a larger quantum defect which breaks the degeneracy, exhibit a quadratic Stark shift. Fig.~\ref{fig:starkmap} shows that the $|16d\rangle$ state (thick green line) is well separated from the neighboring states by an energy gap of $\sim 200$ GHz up to a field strength of $F\sim 3$ kV/cm and experiences a strong energy shift, corresponding to a large dipole moment. 
For an electric field $\mathbf{E}_{\rm dc}=F\,\mathbf{e}_z$ only $m=0$ states are coupled and the resulting dipole moments are polarized along the direction of the external field. Fig.~\ref{fig:dipolestates} shows the $z$-component of the dipole moment, $\mathbf{d}_{\alpha}$, for different states $|\alpha_F\rangle$ and a field strength of $F= 3$~kV/cm. The figure shows that states $|\alpha_F\rangle$ gain a large dipole moment of hundreds of Debye which can be {\it positive} and {\it negative}. The dipole moment of the state which for $F\rightarrow 0$ connects to $|16d, m=0\rangle$ is about 680~Debye.

\begin{figure}[tb]
\centering
\includegraphics[width=4 cm]{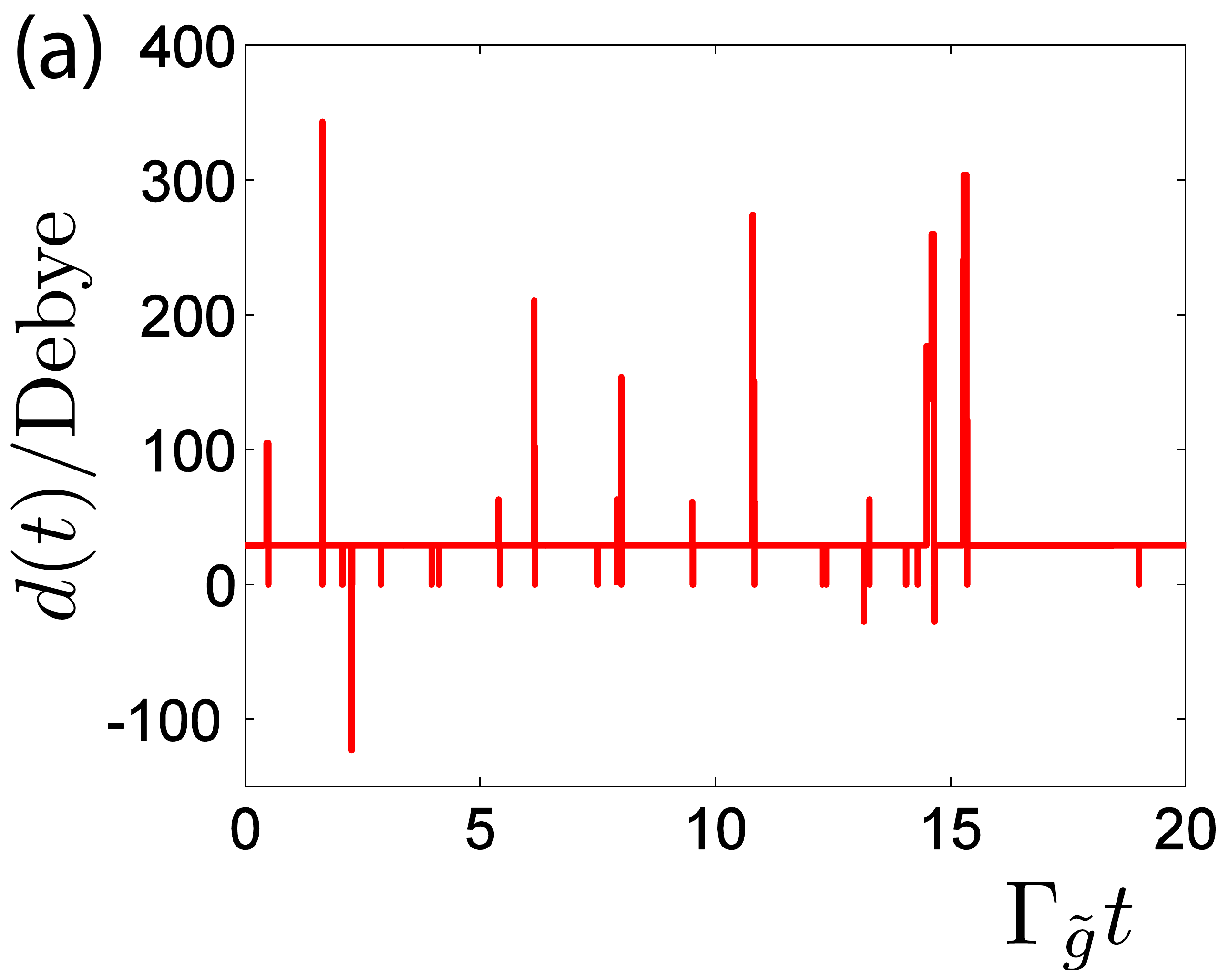}
\includegraphics[width=4 cm]{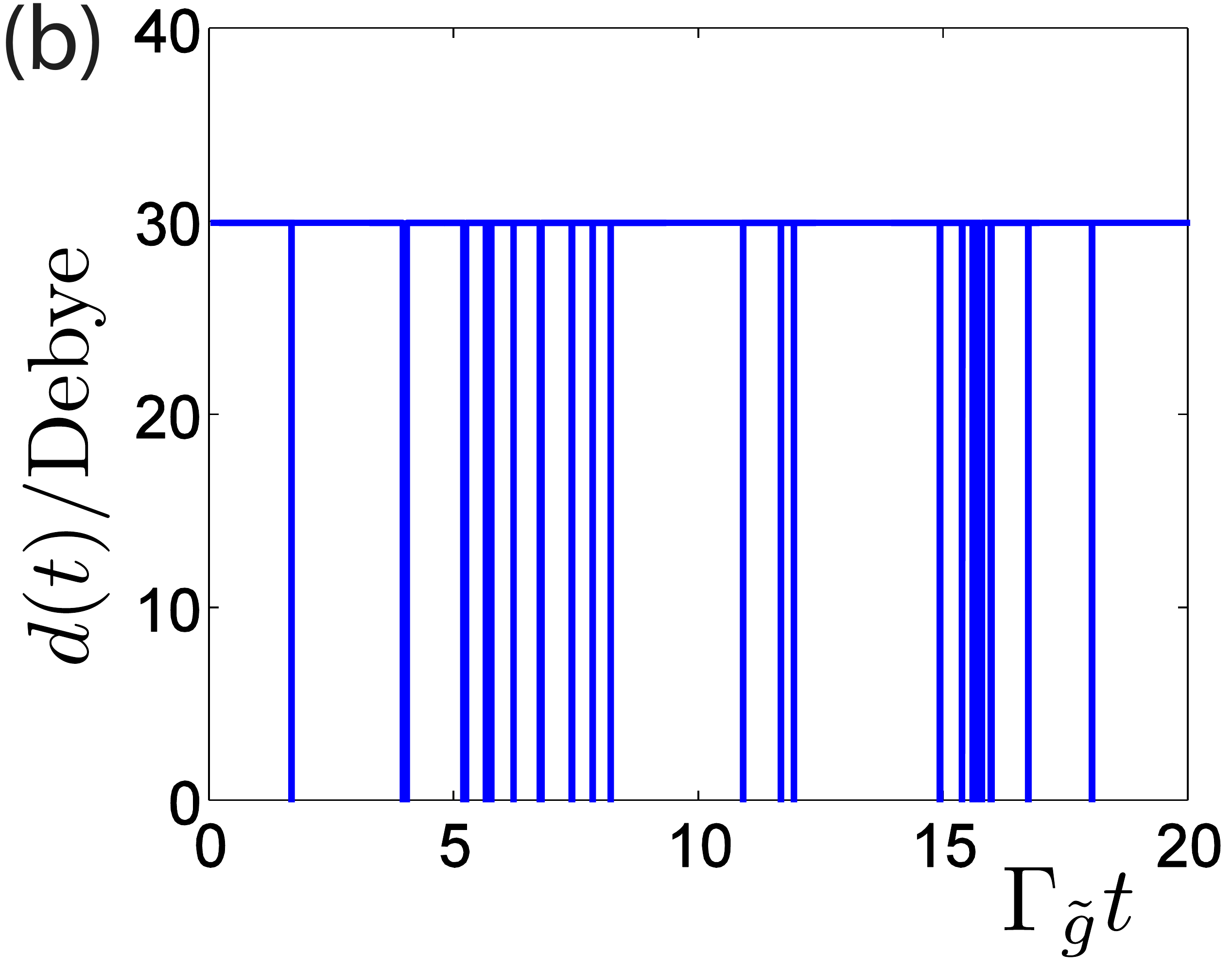}
\caption{(color online) Time evolution of the atomic dipole moment $d(t)$ according to population of various internal states governed by Eq.~\eqref{eq:TimeEv} for a $^{85}$Rb atom. In  panel (a) the atom is polarized using the DC-electric field dressing scheme while in panel (b) it is polarized with the AC-microwave dressing scheme using the same parameters as given in Sec.~\ref{sec:decoherencesingle}. In both cases the dipole moment of the dressed ground state is $d_{\tilde g}=30$~Debye. In the case of the DC-electric field dressing scheme (panel (a)) the dipole moment fluctuates between large positive and negative values due to population of intermediate states, while in the case of the AC-microwave dressing scheme (panel (b)) the dipole moment jumps between $d(t)=0$ and $d(t)=30$~Debye.}
\label{fig:dipoletime}
\end{figure}

The time evolution of Eq.~\eqref{eq:TimeEv} is readily simulated. We prepare the atom in the dressed ground state $|\tilde g\rangle$, that is $p_{\tilde g}=p_0=1$. After a time step $\Delta t$ much smaller than the timescales associated with the decay rates a random number determines if and to which internal state $|\alpha_{(F)}\rangle$ the atom makes a transition (quantum jump) according to the rates of Eq.~\eqref{eq:TimeEv}. During each time step the dipole moment $d(t)$ of the atom is equal to the dipole moment of the current state, $d_{\alpha_{(F)}}$. This new configuration then again propagated another time step $\Delta t$.

Figure~\ref{fig:dipoletime}(a) shows an example result for the time evolution of the atomic dipole moment $d(t)$ of a $^{85}$Rb atoms polarized by a DC-electric field with strength $F=3\text{ kV/cm}$. The ground state is coupled to the Rydberg state $\ket{r_{F}}$ which in the absence of a DC-electric field ($F=0$)  connects to the $\ket r=\ket{16 d, m=0}$-state using a Rydberg-laser with $(\Omega_{r}/2\Delta_{r})=0.21$. We initially prepare the atom in the dressed ground state $\ket {\tilde g_{F}}$, and  compute the time evolution using Eq.~\eqref{eq:TimeEv}. The red trajectory in Fig.~\ref{fig:dipoletime}(a) shows a typical result for the deviation of the atomic dipole from the dipole moment of the dressed ground state $d_{\tilde g}=30\;\text{Debye}$ as a function of time $t$, in units of the ground state lifetime $\tau _{\tilde{g}}=125\;\mu \text{s}$. The figure shows large positive and negative spikes for the values of $d$, followed
by long times where the system is in the dressed ground state, with dipole moment $d_{\tilde g}$. The large ``spikes'' correspond to the population of intermediate states $\ket {\alpha_F}$, during the cascade process towards the ground state, following a spontaneous emission event. These large positive and negative fluctuations of $d(t)$ will cause strong dipole-dipole interactions and hence large mechanical effects when a gas of interacting Rydberg atoms is considered.\\
\\
\textit{Microwave dressing and F=0:} In this case we couple the ground state $|5s\rangle$ of Rubidium to the state $\ket{17s}$ using a laser with $(\Omega_{r}/2\Delta_{r})^{2}=0.05$. An additional microwave field couples the states $\ket{17s}$ and $\ket{17p}$ such that the dressed ground state obtains a dipole moment of $d_{\tilde g}=30\;\text{Debye}$. The blue trajectory in Fig.~\ref{fig:dipoletime}(b) shows a typical trajectory for the time evolution of the dipole moment as a function of time in units of $%
\tau _{\tilde{g}}=90\;\mu \text{s}$. For this dressing scheme the
intermediate states $\ket m$ do not possess a dipole, and thus the dipole
fluctuates between long periods when it has the value $d_{\tilde{g}}$%
, corresponding to the atom in the dressed ground state, to periods where the
atom has no dipole, corresponding to the cascade processes following
spontaneous emission, like a ``blinking dipole''.

\begin{figure}[tb]
\centering
\includegraphics[width=6.3 cm]{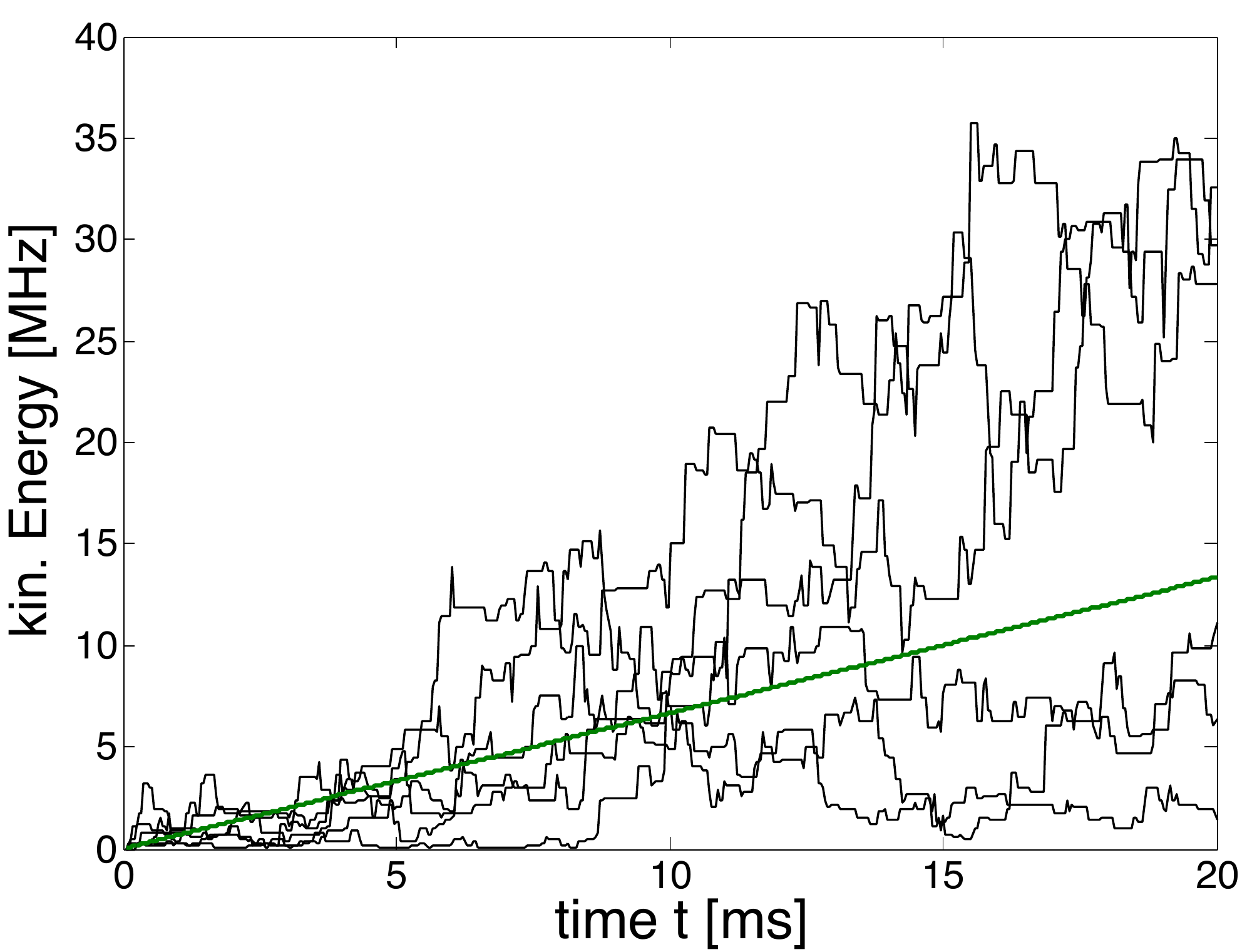}
\caption{(color online) Single kinetic energy trajectories, $E_{\rm kin}$, as a function of time $t$ for a {\it single} Rydberg-dressed $^{85}$Rb atom. We observe an increase in kinetic energy due to photon recoil due to decay events from the Rydberg state. The thick green line is a linear fit of 50 single trajectories (thin black lines) which gives a mean heating rate of $2\pi\times 106.3\,\text{kHz/ms}$.}
\label{fig:spontem}
\end{figure}

In comparison, in the DC-electric field case (blue trajectory in panel (a)) the dipole fluctuations are much larger and can take both positive and negative values. Whereas in the microwave case (red trajectory in panel (b)) the dipoles fluctuate between zero and $d_{\tilde g}$.

In the following we examine the  \textit{external dynamics} of a {\it single} atom. It is dominated by small momentum fluctuations associated with the photon recoil following a series of cascade decay events. As an example, we simulate heating due to decay from the $|r\rangle=|16d\rangle$ state using the same parameters as before. The result of a molecular dynamics simulation is shown in Fig.~\ref{fig:spontem}, where we plot single kinetic energy trajectories as a function of time (thin black lines). We propagate the internal dynamics according to Eq.~\eqref{eq:TimeEv}. Each decay event is associated with a momentum kick in a random direction corresponding to the transition, see next Section. It is shown in the figure that as time progresses the kinetic energy increases due to photon recoils. We perform a statistical average over 50 runs of simulations (thick green line) which yields an average heating rate of $2\pi\times 106.3$ kHz/ms. It will  be shown below that for high enough densities this single particle heating rate is much smaller than the heating rate due to interactions associated with fluctuations of the dipole moment in the many-body case, as discussed in the next section.

\subsection{Molecular dynamics simulation}\label{sec:moldyn}
In our semiclassical description the state of each atom at time $t$ is specified by $\{\mathbf{r}_{i}(t),\mathbf{p}_{i}(t), \alpha_{i}\}$, where $\mathbf{r}_{i}(t)$ and $\mathbf{p}_{i}(t)$ are the center-of-mass coordinate and momentum of the $i$-th atom, respectively, and $\alpha\in\{\tilde g, m, m',\ldots\}$ denotes the internal electronic state with $i=1,\ldots,N$. The internal state is either the ground state $|\tilde{g}\rangle$, dressed by the admixure of the Rydberg state, and the excited state $|e\rangle$ from the laser cooling (see Fig.~\ref{fig:levelscheme}), or one of the many intermediate states $|m\rangle$ populated during a decay cascade back to the ground state. The dynamics of the gas is described by a set of coupled equations, as written down for the case of one or two atoms in Eqs.\eqref{eq:fpe1a} and \eqref{eq:fpeq}, respectively. These results can be generalized immediately  to $N$ atoms. Mathematically, they are a set of coupled Fokker-Planck equations for the external motion, and rate equations for the internal electronic states, describing a combined diffusive and jump Markov process for the probability density $W^{(N)}(\mathbf{r}_{1},\mathbf{p}_{1},\alpha_{1};\ldots,\mathbf{r}_{N},\mathbf{p}_{N},\alpha_{N};t)$. This stochastic process is readily simulated.

Consider atom $i$, which we assume to be in the dressed ground state $|\tilde{g}\rangle$. Its center of mass motion obeys the Langevin equations
\begin{subequations}\label{eq:MolDyn}
\begin{align}
\mathbf{\dot r}_i&=\frac{\mathbf{p}_i}{M},\label{eq:MolDyn1}\\
\mathbf{\dot p}_i&=\sum_{j\neq i}\mathbf{f}^{(ji)}_{\rm int}-\beta\mathbf{p}_i  + \mathbf{f}^{(i)}_{\rm ex}+\mathbf{F}^{(i)},\label{eq:MolDyn2}
\end{align}
\end{subequations}
where on the right hand side of Eq.~\eqref{eq:MolDyn2} we sum over the forces from all the other ground state atoms, as well as the forces from atoms in one of the intermediate states $|m\rangle$, with
\begin{equation}
\begin{split}\label{eq:fij}
\mathbf{f}^{(ji)}_{\rm int}(t) =-\frac{d_{\alpha_i}d_{\alpha_j}}{4 \pi \epsilon_0}\frac{\partial}{\partial \mathbf{r}_i}\frac{1}{|\mathbf{r}_i-\mathbf{r}
_j|^3}.
\end{split}
\end{equation}
Here, $d_{\alpha_i}=d_{\tilde g}$ is the effective dipole moment of the dressed ground state (see Eqs. \eqref{eq:Pert} and~\eqref{eq:EggMWRcMW} for the DC- and AC-field cases, respectively) and $d_{\alpha_j}$ can either be equal to $d_{\tilde g}$ if the $j$-th atom is in the dressed ground state $|\tilde g_j\rangle$ or equal to $d_{m}$ if the $j$-th atom is in the state $|m\rangle$. In the DC-electric field case $d_m=\langle m_F|\mathbf{d}|m_F\rangle$, while in the AC-microwave field case $d_m=0$, see Sec.~\ref{sec:decoherencesingle}.
In addition, we have added in Eq.~\eqref{eq:MolDyn2} the familiar terms describing possible laser cooling~\cite{Minogin1987}, and an external trapping force $\mathbf{f}^{(i)}_{\rm ex}$. The last term in Eq.~\eqref{eq:MolDyn2} is a stochastic force~\cite{Gardiner2004} from quantum fluctuations due to the recoil of spontaneous emission events from both the Rydberg state repopulating the ground state, but also from laser cooling, obeying
\begin{equation}
\langle F^{(i)}_k(t)F^{(j)}_l(t')\rangle=D_{k}\delta_{kl}\delta_{ij} \delta(t-t'),
\end{equation}
with $k,l\in\{x,y,z\}$ and $i,j\in 1,\ldots,N$ and $D_{k}$ the diffusion coefficients of Eq.~\eqref{eq:fpeq}.
We make the simplifying assumption that the cross-noise term of Eq.~\eqref{eq:crossnoise} arising from the fluctuating ground state dipole due to laser cooling  discussed in Sec.~\ref{sec:lasercooling} is negligible. 

Similarly, atom $i$ in one of the intermediate states $|m\rangle$ obeys the equation of motion
\begin{equation}\label{eq:MolDynA}
\mathbf{\dot r}_i=\frac{\mathbf{p}_i}{M},\qquad
\mathbf{\dot p}_i=\sum_{j\neq i}\mathbf{f}^{(ji)}_{\rm int},
\end{equation}
where again $\mathbf{f}^{(ji)}_{\rm int}$ are the forces of Eq.~\eqref{eq:fij} and $d_{\alpha_i}=d_{m}$. Note that atoms in one of the (intermediate) Rydberg states $|m\rangle$ are assumed to be neither trapped, $\mathbf{f}_{\rm ex}^{(i)}=0$, nor laser cooled, $\beta=0$. 

Optical pumping from $| \tilde g\rangle$ to one of the intermediate states $|m\rangle$ and the following cascaded decay, $|\tilde g\rangle\rightarrow|m\rangle\rightarrow\ldots\rightarrow|m'\rangle\rightarrow|\tilde g\rangle$, back to the dressed ground state will redistribute the atomic populations according to the rate equations \eqref{eq:TimeEv}. We can simulate this many-body dynamics by starting with a given atomic configuration, and propagating Eqs.~\eqref{eq:MolDyn} and~\eqref{eq:MolDynA} for a time step $\Delta t$, much smaller than the timescales corresponding to the above mentioned rates. We then determine the probabilities for atom $i$ in state $\alpha_{i}$ to make a transition (jump) to another internal state according to the rate equations~\eqref{eq:TimeEv} for optical pumping and decay, and pick a new configuration according to these probabilities. Each cascade of quantum jump is associated with momentum kicks $\hbar \mathbf{k}_{rm}$, ..., $\hbar \mathbf{k}_{m' g}$ in random spatial directions according to the distribution $N_{\alpha\beta}$ of Eq.~\eqref{eq:Lind} and with $\mathbf{k}_{\alpha\beta}$ the wavevector of the corresponding transition. This new configuration is then again propagated another time step according to equations Eqs.~\eqref{eq:MolDyn} and~\eqref{eq:MolDynA}.

Due to population of intermediate states in the long-time limit and the resulting time-dependent dipole moments  we expect a rapid heating caused by fluctuations of the interaction force described above. In addition we also expect atomic losses. For example, when a ground state atom collides with an atom in an intermediate $\ket{\alpha}$-state with a large dipole moment the kinetic energy released in the collision can (i) be of the order of the height of the saddle point (see Sec.~\ref{sec:2dvalidity}). In this case the atoms will overcome the potential barrier and experience an attractive interaction. Or, (ii) the energy released can be of the order of the depth of external confinement in the transversal direction. A third loss mechanism is be Penning ionization at distances smaller than $4 n^2 a_0$ \cite{Olson1979,Penning2005,Reinhard2008}. Each of these processes leads to two-body or one-body losses. In the simulations, we utilize a model including two-particle losses, where two particles are lost whenever their relative distance $r$ becomes smaller than a certain critical radius $r_{\rm loss}$ (see Sec.~\ref{sec:2dvalidity}).

\subsection{Effects of decoherence on the dynamics of many interacting Rydberg atoms}\label{sec:decoherencetwo}

In this section we numerically investigate the effects of spontaneous emission and blackbody radiation on the long-time dynamics of an {\it ensemble} of interacting Rydberg atoms by performing semi-classical molecular dynamics simulations. As discussed above population of intermediate states $|m\rangle$ with different interaction properties, e.g. different dipole moments, will lead to strong fluctuations of the interparticle forces in the many-body case.

Using semi-classical molecular dynamics simulations we study the resulting non-equilibrium dynamics of an ensemble constitution of $N=67$ Rydberg-dressed $^{85}$Rb atoms. Initially they are prepared in a perfect triangular crystal in a box with hard walls at $T=0$.  For both dressing schemes we choose system parameters such that the ground state dipole moment is $d_{\tilde g}=30$~Debye, corresponding to a melting temperature of $T_{M}=0.6\;\mu$K ($2\pi\times$ 12 kHz) for $n_{2D}=1\;\mu$m$^{-2}$. Additionally, the effect of laser-cooling and a triangular in-plane optical lattice is analyzed in order to stabilize the atomic crystal. Note that, we assume only dressed ground state atoms experience both the cooling laser and the lattice with a depth of 50 $E_R$, with $E_R=2\pi\times 3.8$ kHz, while atoms in one of the intermediate states $|m\rangle$ do not, see Eqs.~\eqref{eq:MolDyn} and~\eqref{eq:MolDynA}.

\begin{figure}[tb]
\centering
\includegraphics[width=6.8 cm]{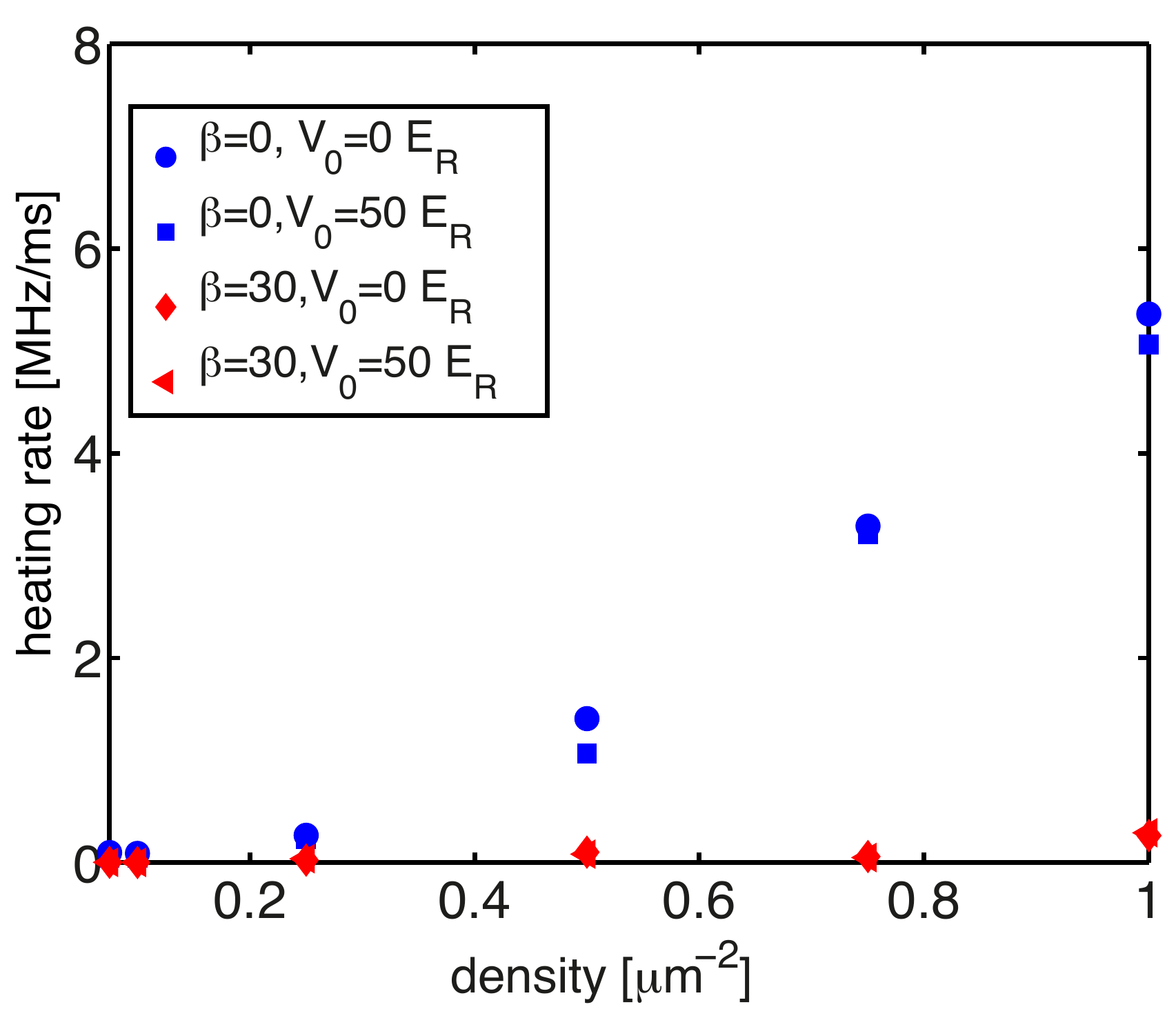}\\
\includegraphics[width=6.8 cm]{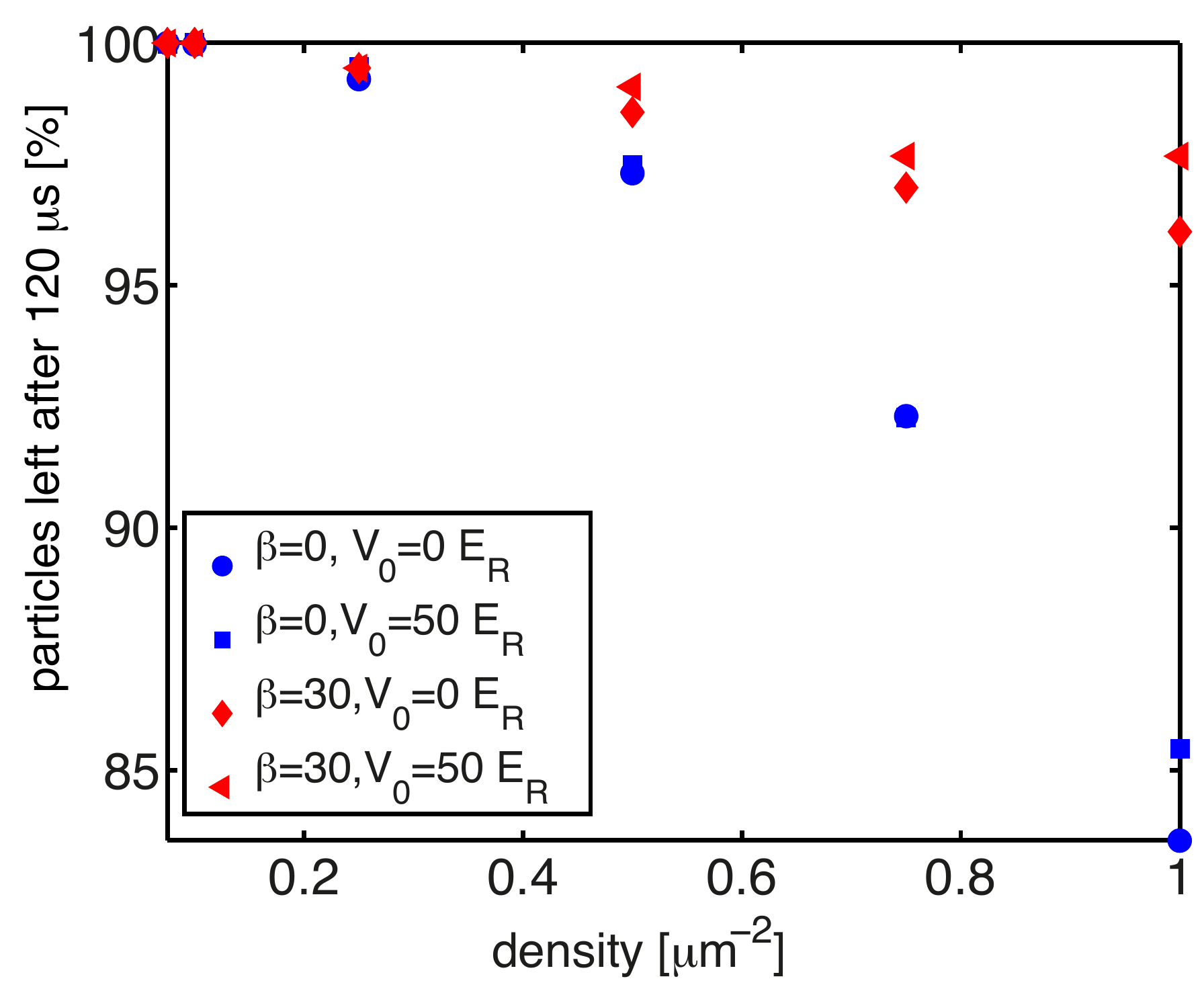}\\
\caption{(color online) DC-electric field case: We numerically investigate the heating rate (upper panel) and the remaining particle number after a time $\tau_{\tilde g}=125\;\mu$s (lower panel) as a function of the density $n_{2D}$ for different laser cooling parameters $\beta$ and lattice depths $V_0$. Parameters see text.}
\label{fig:heatingDC}
\end{figure}

\begin{figure}[tb]
\centering
\includegraphics[width=7 cm]{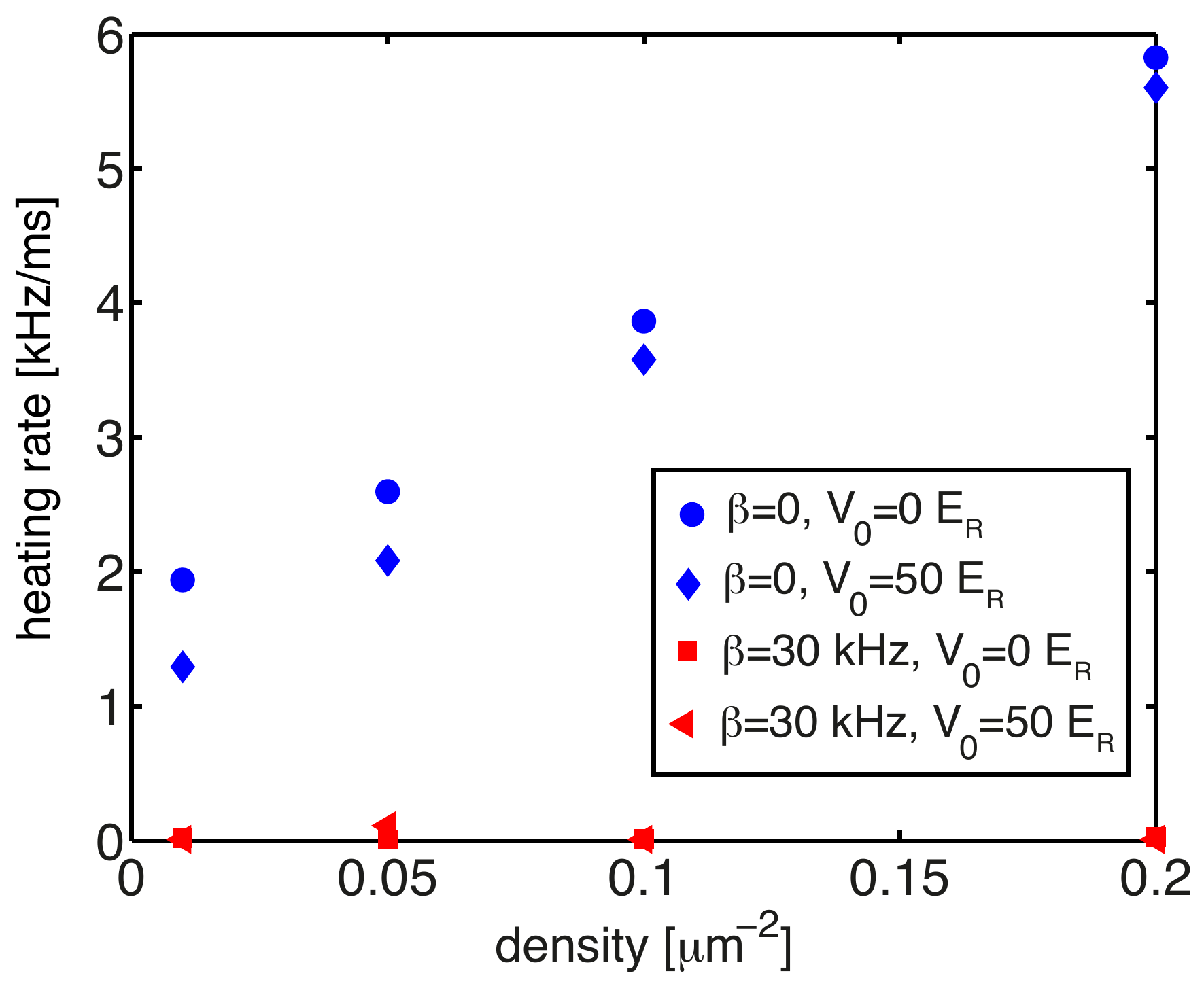}\\
\includegraphics[width=7 cm]{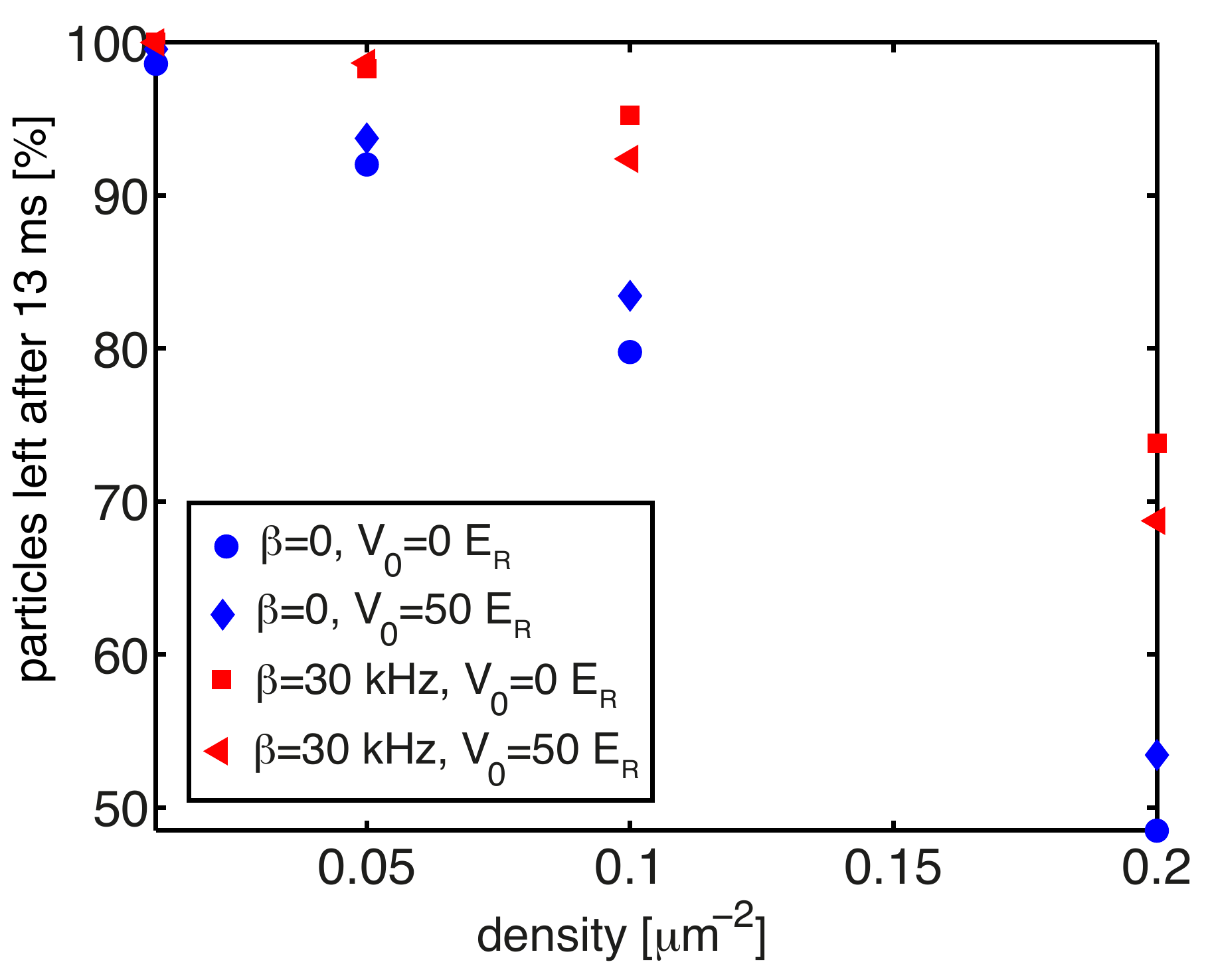}\\
\caption{(color online) AC-microwave field case: We numerically investigate the heating rate (upper panel) and the remaining particle number after a time $\tau_{\tilde g}=12.8$ ms (lower panel) as a function of the density $n_{2D}$ for different laser cooling parameters $\beta$ and lattice depths $V_0$. Parameters see text.}
\label{fig:heatingMW}
\end{figure}

\subsubsection{DC-electric field dressing}
Fig.~\ref{fig:heatingDC} analyzes heating in a gas of Rydberg dressed atoms using the DC-electric field scheme of Fig.~\ref{fig:levelscheme}(b) with the same parameters as in Sec.~\ref{sec:sumdc}. For $\Delta_r=2\pi\times 250$ MHz the Condon point is at $r_c\sim 520$ nm. For a transversal trapping frequency of $\omega_\perp=2\pi \times 200$ kHz the height of the potential barrier at the saddle point is $~\sim 150\;\mu$K. Each marker in Fig.~\ref{fig:heatingDC} is an average over 50 runs of the simulation, where we simulated the dynamics for a time $\tau_{\tilde g}$.

Fig.~\ref{fig:heatingDC}(a) shows the heating rate, $\gamma_{\tilde g}E_{\rm kin}(t=\tau_{\tilde g})$ as a function of the atomic density for different cooling and lattice parameter, with $E_{\rm kin}(t)$ the mean kinetic energy of all $N$ atoms. In the case of no laser cooling, $\beta=0$, there is a strong dependence of the heating rate on the density, while for a cooling rate of $\beta=2\pi\;\times\;30$~kHz the heating rate is almost zero and  a steady state is realized on a timescale $\beta^{-1}\ll\tau_{\tilde g}$. This steady state is due to the interplay of laser cooling and heating together with loss of high energy particles from the 2D confinement. This is shown in Fig.~\ref{fig:heatingDC}(b), where the number of remaining particles after a time $\tau_{\tilde g}$ is shown as a function of the atomic density. Again, there is a strong dependence on the density and also on the laser cooling rate $\beta$. For densities $n_{2D}\lesssim 0.04 \; \mu\text{m}^{-2}$ the effect of the fluctuating dipoles can be neglected on surrounding atoms, while for densities $n_{2D}\sim1\; \mu\text{m}^{-2}$ the particle number has been decreased by 3\% (with cooling) or 15 \% (no cooling) after a time $\tau_{\tilde g}=120\;\mu$s. Similar to Fig.~\ref{fig:singletrajectory} we observe an accelerated loss of particles, resulting in a dramatic decrease of the particle number for times $t>\tau_{\tilde g}$. The effect of an in-plane optical lattice is small and hardly changes the heating rate or particle number shown in Fig.~\ref{fig:heatingDC}.

\subsubsection{AC-microwave field dressing}
Fig.~\ref{fig:heatingMW} shows heating dynamics in a gas of Rydberg dressed atoms using the {\it microwave dressing scheme} of Fig.~\ref{fig:levelscheme}(c) with the same parameters as in Sec.~\ref{sec:sumac}. For $\Delta_r=2\pi\times 1.0$ GHz and $\alpha=0.5$ we find $r_c\sim 1.38\;\mu$m. With $\omega_\perp=2\pi\times 100$ kHz the hight of the barrier at the saddle points is $\sim 70\; \mu$K.

In our treatment, the basic heating process of a crystal comes from its spatial rearrangement, once one of the atoms decays to an intermediate state $|m\rangle$ with zero dipole moment. We explain this as follows: Due to the lack of dipole moment, particles in the intermediate state $|m\rangle$ can travel a distance $v_R/\Gamma_m$ without interacting with neighboring atoms, where $v_R$ is the recoil velocity and $\Gamma_m$ the mean lifetime of an intermediate state. (i) if $v_R/\Gamma_m$ is of the order of the mean particle distance in the crystal $n_{2D}^{-1/2}$   two particles will most probably come closer than the critical radius $r_{\rm loss}$ and get lost. We find that this latter process can result in an unusual {\it collisional} evaporative cooling effect. (ii) if $v_R/\Gamma_m$ is smaller than the mean particle distance in the crystal the atom may return to the ground state with a finite dipole moment, resulting in large time-dependent fluctuations of the dipole-dipole interaction between neighboring atoms. The net heating rate of the crystal is a competition between processes (i) and (ii).

We note that accidental vdW interactions may occur between atoms in one of the intermediate states $|m\rangle$, which may lead to additional heating. However, these processes are not considered here.

Fig.~\ref{fig:heatingMW}(a) shows the mean heating rate, $\gamma_{\tilde g}E_{\rm kin}(t=\tau_{\tilde g})$ as a function of the atomic density for different cooling and lattice parameter, with $E_{\rm kin}(t)$ the mean kinetic energy of all $N$ atoms. Again, each marker in Fig.~\ref{fig:heatingMW} is an average over 50 runs of the simulation, where we simulated the dynamics for a time $\tau_{\tilde g}=12.8$ ms. Similar as in the DC-electric field case of Fig.~\ref{fig:heatingDC}, there is a strong dependence of the heating rate on the density and on the laser cooling rate $\beta$, while the effect of an in-plane optical lattice is weak. Additional laser cooling of the atoms with a rate $\beta=30$~kHz again leads to a steady state of the mean kinetic energy (see panel (a)) and diminishes the particle loss rate, which is shown in Fig.~\ref{fig:heatingDC}(b). The number of remaining particles after a time $\tau_{\tilde g}$ is shown as a function of the atomic density. Remarkably, more than 70\% (50\%) of the particles are left with (without) laser cooling and densities of $n_{2D}=0.2/\mu$m$^2$ after $t=\tau_{\tilde g}$. 

\section{Conclusions and Outlook}
In this work we have investigated the long-time non-equilibrium dynamics of an ensemble of cold ground state atoms, weakly admixed with a Rydberg state using laser light. For times comparable to or larger than the effective lifetime of the ground state, the population of intermediate Rydberg states following spontaneous emission significantly affects the atomic motion in a strongly interacting gas, by providing a dominant heating and loss mechanism. Numerical simulations indicate that, due to the absence of dipole moments in the intermediate Rydberg states, ensembles of atoms polarized by AC-microwave fields exhibit a significantly different long-time dynamics, compared to atoms dressed by DC-electric fields.

Understanding this long-time dynamics is relevant for the experimental demonstration of laser-dressing techniques~\cite{Ott2011} Êas well as the creation of long-lived strongly correlated atomic phases such as self-assembled atomic crystals. Related to recent work on mixtures of polar molecules and Rydberg atoms~\cite{Huber2012, Zhao2012}, we speculate that long-lived cold crystals may be used as cold reservoirs for achieving, e.g., sympathetic cooling of polar molecules~\cite{Glaetzle}.

\section*{Acknowledgment}               
The authors thank  E. Arimondo, M. Baranov, I. Lesanovsky, M. D. Lukin and H. R. Sadeghpour for  stimulating discussions. PZ and AG thank ITAMP, the Harvard Physics Department and the Joint Quantum Institute, University of Maryland, for their hospitality where part of this work was done. Work was supported in parts by COHERENCE, EOARD and EPSRC through CHIST-ERA R-ION, and the Austrian Science Fund.


\begin{thebibliography}{83}
\expandafter\ifx\csname natexlab\endcsname\relax\def\natexlab#1{#1}\fi
\expandafter\ifx\csname bibnamefont\endcsname\relax
  \def\bibnamefont#1{#1}\fi
\expandafter\ifx\csname bibfnamefont\endcsname\relax
  \def\bibfnamefont#1{#1}\fi
\expandafter\ifx\csname citenamefont\endcsname\relax
  \def\citenamefont#1{#1}\fi
\expandafter\ifx\csname url\endcsname\relax
  \def\url#1{\texttt{#1}}\fi
\expandafter\ifx\csname urlprefix\endcsname\relax\def\urlprefix{URL }\fi
\providecommand{\bibinfo}[2]{#2}
\providecommand{\eprint}[2][]{\url{#2}}

\bibitem[{\citenamefont{Gallagher}(2005)}]{Gallagher2005}
\bibinfo{author}{\bibfnamefont{T.~F.} \bibnamefont{Gallagher}},
  \emph{\bibinfo{title}{{Rydberg Atoms}}} (\bibinfo{publisher}{Cambridge
  University Press}, \bibinfo{year}{2005}).

\bibitem[{\citenamefont{Saffman et~al.}(2010)\citenamefont{Saffman, Walker, and
  M{\o}lmer}}]{Saffman2010}
\bibinfo{author}{\bibfnamefont{M.}~\bibnamefont{Saffman}},
  \bibinfo{author}{\bibfnamefont{T.~G.} \bibnamefont{Walker}},
  \bibnamefont{and}
  \bibinfo{author}{\bibfnamefont{K.}~\bibnamefont{M{\o}lmer}},
  \bibinfo{journal}{Rev. Mod. Phys.} \textbf{\bibinfo{volume}{82}},
  \bibinfo{pages}{2313} (\bibinfo{year}{2010}).

\bibitem[{\citenamefont{Comparat and Pillet}(2010)}]{Comparat2010}
\bibinfo{author}{\bibfnamefont{D.}~\bibnamefont{Comparat}} \bibnamefont{and}
  \bibinfo{author}{\bibfnamefont{P.}~\bibnamefont{Pillet}},
  \bibinfo{journal}{J. Phys. B} \textbf{\bibinfo{volume}{27}},
  \bibinfo{pages}{208} (\bibinfo{year}{2010}).

\bibitem[{\citenamefont{Seaton}(1983)}]{Seaton1983}
\bibinfo{author}{\bibfnamefont{M.~J.} \bibnamefont{Seaton}},
  \bibinfo{journal}{Rep. Prog. Phys.} \textbf{\bibinfo{volume}{46}},
  \bibinfo{pages}{167} (\bibinfo{year}{1983}).

\bibitem[{\citenamefont{L\"{o}w et~al.}(2012)\citenamefont{L\"{o}w, Weimer,
  Nipper, Balewski, Butscher, B\"{u}chler, and Pfau}}]{Low2012}
\bibinfo{author}{\bibfnamefont{R.}~\bibnamefont{L\"{o}w}},
  \bibinfo{author}{\bibfnamefont{H.}~\bibnamefont{Weimer}},
  \bibinfo{author}{\bibfnamefont{J.}~\bibnamefont{Nipper}},
  \bibinfo{author}{\bibfnamefont{J.~B.} \bibnamefont{Balewski}},
  \bibinfo{author}{\bibfnamefont{B.}~\bibnamefont{Butscher}},
  \bibinfo{author}{\bibfnamefont{H.~P.} \bibnamefont{B\"{u}chler}},
  \bibnamefont{and} \bibinfo{author}{\bibfnamefont{T.}~\bibnamefont{Pfau}},
  \bibinfo{journal}{J. Phys. B} \textbf{\bibinfo{volume}{45}},
  \bibinfo{pages}{113001} (\bibinfo{year}{2012}).

\bibitem[{\citenamefont{Anderson et~al.}(1998)\citenamefont{Anderson, Veale,
  and Gallagher}}]{Anderson1998}
\bibinfo{author}{\bibfnamefont{W.~R.} \bibnamefont{Anderson}},
  \bibinfo{author}{\bibfnamefont{J.~R.} \bibnamefont{Veale}}, \bibnamefont{and}
  \bibinfo{author}{\bibfnamefont{T.~F.} \bibnamefont{Gallagher}},
  \bibinfo{journal}{Phys. Rev. Lett.} \textbf{\bibinfo{volume}{80}},
  \bibinfo{pages}{249} (\bibinfo{year}{1998}).

\bibitem[{\citenamefont{Singer et~al.}(2004)\citenamefont{Singer, Reetz-Lamour,
  Amthor, Marcassa, and Weidem\"{u}ller}}]{Singer2004}
\bibinfo{author}{\bibfnamefont{K.}~\bibnamefont{Singer}},
  \bibinfo{author}{\bibfnamefont{M.}~\bibnamefont{Reetz-Lamour}},
  \bibinfo{author}{\bibfnamefont{T.}~\bibnamefont{Amthor}},
  \bibinfo{author}{\bibfnamefont{L.~G.} \bibnamefont{Marcassa}},
  \bibnamefont{and}
  \bibinfo{author}{\bibfnamefont{M.}~\bibnamefont{Weidem\"{u}ller}},
  \bibinfo{journal}{Phys. Rev. Lett.} \textbf{\bibinfo{volume}{93}},
  \bibinfo{pages}{163001} (\bibinfo{year}{2004}).

\bibitem[{\citenamefont{Tong et~al.}(2004)\citenamefont{Tong, Farooqi,
  Stanojevic, Krishnan, Zhang, C\^{o}t\'{e}, Eyler, and Gould}}]{Tong2004}
\bibinfo{author}{\bibfnamefont{D.}~\bibnamefont{Tong}},
  \bibinfo{author}{\bibfnamefont{S.~M.} \bibnamefont{Farooqi}},
  \bibinfo{author}{\bibfnamefont{J.}~\bibnamefont{Stanojevic}},
  \bibinfo{author}{\bibfnamefont{S.}~\bibnamefont{Krishnan}},
  \bibinfo{author}{\bibfnamefont{Y.~P.} \bibnamefont{Zhang}},
  \bibinfo{author}{\bibfnamefont{R.}~\bibnamefont{C\^{o}t\'{e}}},
  \bibinfo{author}{\bibfnamefont{E.~E.} \bibnamefont{Eyler}}, \bibnamefont{and}
  \bibinfo{author}{\bibfnamefont{P.~L.} \bibnamefont{Gould}},
  \bibinfo{journal}{Phys. Rev. Lett.} \textbf{\bibinfo{volume}{93}},
  \bibinfo{pages}{063001} (\bibinfo{year}{2004}).

\bibitem[{\citenamefont{Liebisch et~al.}(2005)\citenamefont{Liebisch, Reinhard,
  Berman, and Raithel}}]{Liebisch2005}
\bibinfo{author}{\bibfnamefont{T.~C.} \bibnamefont{Liebisch}},
  \bibinfo{author}{\bibfnamefont{A.}~\bibnamefont{Reinhard}},
  \bibinfo{author}{\bibfnamefont{P.~R.} \bibnamefont{Berman}},
  \bibnamefont{and} \bibinfo{author}{\bibfnamefont{G.}~\bibnamefont{Raithel}},
  \bibinfo{journal}{Phys. Rev. Lett.} \textbf{\bibinfo{volume}{95}},
  \bibinfo{pages}{253002} (\bibinfo{year}{2005}).

\bibitem[{\citenamefont{Afrousheh et~al.}(2006)\citenamefont{Afrousheh,
  Bohlouli-Zanjani, Carter, Mugford, and Martin}}]{Afrousheh2006}
\bibinfo{author}{\bibfnamefont{K.}~\bibnamefont{Afrousheh}},
  \bibinfo{author}{\bibfnamefont{P.}~\bibnamefont{Bohlouli-Zanjani}},
  \bibinfo{author}{\bibfnamefont{J.}~\bibnamefont{Carter}},
  \bibinfo{author}{\bibfnamefont{A.}~\bibnamefont{Mugford}}, \bibnamefont{and}
  \bibinfo{author}{\bibfnamefont{J.}~\bibnamefont{Martin}},
  \bibinfo{journal}{Phys. Rev. A} \textbf{\bibinfo{volume}{73}},
  \bibinfo{pages}{063403} (\bibinfo{year}{2006}).

\bibitem[{\citenamefont{Carroll et~al.}(2006)\citenamefont{Carroll, Sunder, and
  Noel}}]{Carroll2006}
\bibinfo{author}{\bibfnamefont{T.}~\bibnamefont{Carroll}},
  \bibinfo{author}{\bibfnamefont{S.}~\bibnamefont{Sunder}}, \bibnamefont{and}
  \bibinfo{author}{\bibfnamefont{M.}~\bibnamefont{Noel}},
  \bibinfo{journal}{Phys. Rev. A} \textbf{\bibinfo{volume}{73}},
  \bibinfo{pages}{032725} (\bibinfo{year}{2006}).

\bibitem[{\citenamefont{Heidemann et~al.}(2007)\citenamefont{Heidemann,
  Raitzsch, Bendkowsky, Butscher, L\"{o}w, Santos, and Pfau}}]{Heidemann2007}
\bibinfo{author}{\bibfnamefont{R.}~\bibnamefont{Heidemann}},
  \bibinfo{author}{\bibfnamefont{U.}~\bibnamefont{Raitzsch}},
  \bibinfo{author}{\bibfnamefont{V.}~\bibnamefont{Bendkowsky}},
  \bibinfo{author}{\bibfnamefont{B.}~\bibnamefont{Butscher}},
  \bibinfo{author}{\bibfnamefont{R.}~\bibnamefont{L\"{o}w}},
  \bibinfo{author}{\bibfnamefont{L.}~\bibnamefont{Santos}}, \bibnamefont{and}
  \bibinfo{author}{\bibfnamefont{T.}~\bibnamefont{Pfau}},
  \bibinfo{journal}{Phys. Rev. Lett.} \textbf{\bibinfo{volume}{99}},
  \bibinfo{pages}{163601} (\bibinfo{year}{2007}).

\bibitem[{\citenamefont{Vogt et~al.}(2007)\citenamefont{Vogt, Viteau, Chotia,
  Zhao, Comparat, and Pillet}}]{Vogt2007}
\bibinfo{author}{\bibfnamefont{T.}~\bibnamefont{Vogt}},
  \bibinfo{author}{\bibfnamefont{M.}~\bibnamefont{Viteau}},
  \bibinfo{author}{\bibfnamefont{A.}~\bibnamefont{Chotia}},
  \bibinfo{author}{\bibfnamefont{J.}~\bibnamefont{Zhao}},
  \bibinfo{author}{\bibfnamefont{D.}~\bibnamefont{Comparat}}, \bibnamefont{and}
  \bibinfo{author}{\bibfnamefont{P.}~\bibnamefont{Pillet}},
  \bibinfo{journal}{Phys. Rev. Lett.} \textbf{\bibinfo{volume}{99}},
  \bibinfo{pages}{073002} (\bibinfo{year}{2007}).

\bibitem[{\citenamefont{Amthor et~al.}(2007)\citenamefont{Amthor, Reetz-Lamour,
  Westermann, Denskat, and Weidem\"{u}ller}}]{Amthor2007}
\bibinfo{author}{\bibfnamefont{T.}~\bibnamefont{Amthor}},
  \bibinfo{author}{\bibfnamefont{M.}~\bibnamefont{Reetz-Lamour}},
  \bibinfo{author}{\bibfnamefont{S.}~\bibnamefont{Westermann}},
  \bibinfo{author}{\bibfnamefont{J.}~\bibnamefont{Denskat}}, \bibnamefont{and}
  \bibinfo{author}{\bibfnamefont{M.}~\bibnamefont{Weidem\"{u}ller}},
  \bibinfo{journal}{Phys. Rev. Lett.} \textbf{\bibinfo{volume}{98}},
  \bibinfo{pages}{023004} (\bibinfo{year}{2007}).

\bibitem[{\citenamefont{van Ditzhuijzen et~al.}(2008)\citenamefont{van
  Ditzhuijzen, Koenderink, Hern\'{a}ndez, Robicheaux, Noordam, and {van den
  Linden van den Heuvell}}}]{VanDitzhuijzen2008}
\bibinfo{author}{\bibfnamefont{C.~S.~E.} \bibnamefont{van Ditzhuijzen}},
  \bibinfo{author}{\bibfnamefont{A.~F.} \bibnamefont{Koenderink}},
  \bibinfo{author}{\bibfnamefont{J.}~\bibnamefont{Hern\'{a}ndez}},
  \bibinfo{author}{\bibfnamefont{F.}~\bibnamefont{Robicheaux}},
  \bibinfo{author}{\bibfnamefont{L.}~\bibnamefont{Noordam}}, \bibnamefont{and}
  \bibinfo{author}{\bibfnamefont{H.~B.} \bibnamefont{{van den Linden van den
  Heuvell}}}, \bibinfo{journal}{Phys. Rev. Lett.}
  \textbf{\bibinfo{volume}{100}}, \bibinfo{pages}{243201}
  (\bibinfo{year}{2008}).

\bibitem[{\citenamefont{Johnson et~al.}(2008)\citenamefont{Johnson, Urban,
  Henage, Isenhower, Yavuz, Walker, and Saffman}}]{Johnson2008}
\bibinfo{author}{\bibfnamefont{T.}~\bibnamefont{Johnson}},
  \bibinfo{author}{\bibfnamefont{E.}~\bibnamefont{Urban}},
  \bibinfo{author}{\bibfnamefont{T.}~\bibnamefont{Henage}},
  \bibinfo{author}{\bibfnamefont{L.}~\bibnamefont{Isenhower}},
  \bibinfo{author}{\bibfnamefont{D.}~\bibnamefont{Yavuz}},
  \bibinfo{author}{\bibfnamefont{T.}~\bibnamefont{Walker}}, \bibnamefont{and}
  \bibinfo{author}{\bibfnamefont{M.}~\bibnamefont{Saffman}},
  \bibinfo{journal}{Phys. Rev. Lett.} \textbf{\bibinfo{volume}{100}},
  \bibinfo{pages}{113003} (\bibinfo{year}{2008}).

\bibitem[{\citenamefont{Pritchard et~al.}(2010)\citenamefont{Pritchard,
  Maxwell, Gauguet, Weatherill, Jones, and Adams}}]{Pritchard2010}
\bibinfo{author}{\bibfnamefont{J.}~\bibnamefont{Pritchard}},
  \bibinfo{author}{\bibfnamefont{D.}~\bibnamefont{Maxwell}},
  \bibinfo{author}{\bibfnamefont{A.}~\bibnamefont{Gauguet}},
  \bibinfo{author}{\bibfnamefont{K.}~\bibnamefont{Weatherill}},
  \bibinfo{author}{\bibfnamefont{M.}~\bibnamefont{Jones}}, \bibnamefont{and}
  \bibinfo{author}{\bibfnamefont{C.}~\bibnamefont{Adams}},
  \bibinfo{journal}{Phys. Rev. Lett.} \textbf{\bibinfo{volume}{105}},
  \bibinfo{pages}{193603} (\bibinfo{year}{2010}).

\bibitem[{\citenamefont{Sevin\c{c}li et~al.}(2011)\citenamefont{Sevin\c{c}li,
  Ates, Pohl, Schempp, Hofmann, G\"{u}nter, Amthor, Weidem\"{u}ller, Pritchard,
  Maxwell et~al.}}]{Sevincli2011a}
\bibinfo{author}{\bibfnamefont{S.}~\bibnamefont{Sevin\c{c}li}},
  \bibinfo{author}{\bibfnamefont{C.}~\bibnamefont{Ates}},
  \bibinfo{author}{\bibfnamefont{T.}~\bibnamefont{Pohl}},
  \bibinfo{author}{\bibfnamefont{H.}~\bibnamefont{Schempp}},
  \bibinfo{author}{\bibfnamefont{C.~S.} \bibnamefont{Hofmann}},
  \bibinfo{author}{\bibfnamefont{G.}~\bibnamefont{G\"{u}nter}},
  \bibinfo{author}{\bibfnamefont{T.}~\bibnamefont{Amthor}},
  \bibinfo{author}{\bibfnamefont{M.}~\bibnamefont{Weidem\"{u}ller}},
  \bibinfo{author}{\bibfnamefont{J.~D.} \bibnamefont{Pritchard}},
  \bibinfo{author}{\bibfnamefont{D.}~\bibnamefont{Maxwell}},
  \bibnamefont{et~al.}, \bibinfo{journal}{J. Phys. B}
  \textbf{\bibinfo{volume}{44}}, \bibinfo{pages}{184018}
  (\bibinfo{year}{2011}).

\bibitem[{\citenamefont{Dudin and Kuzmich}(2012)}]{Dudin2012}
\bibinfo{author}{\bibfnamefont{Y.~O.} \bibnamefont{Dudin}} \bibnamefont{and}
  \bibinfo{author}{\bibfnamefont{A.}~\bibnamefont{Kuzmich}},
  \bibinfo{journal}{Science} \textbf{\bibinfo{volume}{336}},
  \bibinfo{pages}{887} (\bibinfo{year}{2012}).

\bibitem[{\citenamefont{Nipper et~al.}(2012)\citenamefont{Nipper, Balewski,
  Krupp, Butscher, L\"{o}w, and Pfau}}]{Nipper2012}
\bibinfo{author}{\bibfnamefont{J.}~\bibnamefont{Nipper}},
  \bibinfo{author}{\bibfnamefont{J.~B.} \bibnamefont{Balewski}},
  \bibinfo{author}{\bibfnamefont{A.~T.} \bibnamefont{Krupp}},
  \bibinfo{author}{\bibfnamefont{B.}~\bibnamefont{Butscher}},
  \bibinfo{author}{\bibfnamefont{R.}~\bibnamefont{L\"{o}w}}, \bibnamefont{and}
  \bibinfo{author}{\bibfnamefont{T.}~\bibnamefont{Pfau}},
  \bibinfo{journal}{Phys. Rev. Lett.} \textbf{\bibinfo{volume}{108}},
  \bibinfo{pages}{113001} (\bibinfo{year}{2012}).

\bibitem[{\citenamefont{Weimer et~al.}(2008)\citenamefont{Weimer, L\"{o}w,
  Pfau, and B\"{u}chler}}]{Weimer2008}
\bibinfo{author}{\bibfnamefont{H.}~\bibnamefont{Weimer}},
  \bibinfo{author}{\bibfnamefont{R.}~\bibnamefont{L\"{o}w}},
  \bibinfo{author}{\bibfnamefont{T.}~\bibnamefont{Pfau}}, \bibnamefont{and}
  \bibinfo{author}{\bibfnamefont{H.~P.} \bibnamefont{B\"{u}chler}},
  \bibinfo{journal}{Phys. Rev. Lett.} \textbf{\bibinfo{volume}{101}},
  \bibinfo{pages}{250601} (\bibinfo{year}{2008}).

\bibitem[{\citenamefont{L\"{o}w et~al.}(2009)\citenamefont{L\"{o}w, Weimer,
  Krohn, Heidemann, Bendkowsky, Butscher, B\"{u}chler, and Pfau}}]{Low2009}
\bibinfo{author}{\bibfnamefont{R.}~\bibnamefont{L\"{o}w}},
  \bibinfo{author}{\bibfnamefont{H.}~\bibnamefont{Weimer}},
  \bibinfo{author}{\bibfnamefont{U.}~\bibnamefont{Krohn}},
  \bibinfo{author}{\bibfnamefont{R.}~\bibnamefont{Heidemann}},
  \bibinfo{author}{\bibfnamefont{V.}~\bibnamefont{Bendkowsky}},
  \bibinfo{author}{\bibfnamefont{B.}~\bibnamefont{Butscher}},
  \bibinfo{author}{\bibfnamefont{H.}~\bibnamefont{B\"{u}chler}},
  \bibnamefont{and} \bibinfo{author}{\bibfnamefont{T.}~\bibnamefont{Pfau}},
  \bibinfo{journal}{Phys. Rev. A} \textbf{\bibinfo{volume}{80}},
  \bibinfo{pages}{033422} (\bibinfo{year}{2009}).

\bibitem[{\citenamefont{Pohl et~al.}(2010)\citenamefont{Pohl, Demler, and
  Lukin}}]{Pohl2010}
\bibinfo{author}{\bibfnamefont{T.}~\bibnamefont{Pohl}},
  \bibinfo{author}{\bibfnamefont{E.}~\bibnamefont{Demler}}, \bibnamefont{and}
  \bibinfo{author}{\bibfnamefont{M.~D.} \bibnamefont{Lukin}},
  \bibinfo{journal}{Phys. Rev. Lett.} \textbf{\bibinfo{volume}{104}},
  \bibinfo{pages}{043002} (\bibinfo{year}{2010}).

\bibitem[{\citenamefont{Weimer and B\"{u}chler}(2010)}]{Weimer2010}
\bibinfo{author}{\bibfnamefont{H.}~\bibnamefont{Weimer}} \bibnamefont{and}
  \bibinfo{author}{\bibfnamefont{H.~P.} \bibnamefont{B\"{u}chler}},
  \bibinfo{journal}{Phys. Rev. Lett.} \textbf{\bibinfo{volume}{105}},
  \bibinfo{pages}{230403} (\bibinfo{year}{2010}).

\bibitem[{\citenamefont{Honer et~al.}(2010)\citenamefont{Honer, Weimer, Pfau,
  and B\"{u}chler}}]{Honer2010}
\bibinfo{author}{\bibfnamefont{J.}~\bibnamefont{Honer}},
  \bibinfo{author}{\bibfnamefont{H.}~\bibnamefont{Weimer}},
  \bibinfo{author}{\bibfnamefont{T.}~\bibnamefont{Pfau}}, \bibnamefont{and}
  \bibinfo{author}{\bibfnamefont{H.~P.} \bibnamefont{B\"{u}chler}},
  \bibinfo{journal}{Phys. Rev. Lett.} \textbf{\bibinfo{volume}{105}},
  \bibinfo{pages}{160404} (\bibinfo{year}{2010}).

\bibitem[{\citenamefont{Pupillo et~al.}(2010)\citenamefont{Pupillo, Micheli,
  Boninsegni, Lesanovsky, and Zoller}}]{Pupillo2010}
\bibinfo{author}{\bibfnamefont{G.}~\bibnamefont{Pupillo}},
  \bibinfo{author}{\bibfnamefont{A.}~\bibnamefont{Micheli}},
  \bibinfo{author}{\bibfnamefont{M.}~\bibnamefont{Boninsegni}},
  \bibinfo{author}{\bibfnamefont{I.}~\bibnamefont{Lesanovsky}},
  \bibnamefont{and} \bibinfo{author}{\bibfnamefont{P.}~\bibnamefont{Zoller}},
  \bibinfo{journal}{Phys. Rev. Lett.} \textbf{\bibinfo{volume}{104}},
  \bibinfo{pages}{223002} (\bibinfo{year}{2010}).

\bibitem[{\citenamefont{Henkel et~al.}(2010)\citenamefont{Henkel, Nath, and
  Pohl}}]{Henkel2010}
\bibinfo{author}{\bibfnamefont{N.}~\bibnamefont{Henkel}},
  \bibinfo{author}{\bibfnamefont{R.}~\bibnamefont{Nath}}, \bibnamefont{and}
  \bibinfo{author}{\bibfnamefont{T.}~\bibnamefont{Pohl}},
  \bibinfo{journal}{Phys. Rev. Lett.} \textbf{\bibinfo{volume}{104}},
  \bibinfo{pages}{195302} (\bibinfo{year}{2010}).

\bibitem[{\citenamefont{Maucher et~al.}(2011)\citenamefont{Maucher, Henkel,
  Saffman, Kr\'{o}likowski, Skupin, and Pohl}}]{Maucher2011}
\bibinfo{author}{\bibfnamefont{F.}~\bibnamefont{Maucher}},
  \bibinfo{author}{\bibfnamefont{N.}~\bibnamefont{Henkel}},
  \bibinfo{author}{\bibfnamefont{M.}~\bibnamefont{Saffman}},
  \bibinfo{author}{\bibfnamefont{W.}~\bibnamefont{Kr\'{o}likowski}},
  \bibinfo{author}{\bibfnamefont{S.}~\bibnamefont{Skupin}}, \bibnamefont{and}
  \bibinfo{author}{\bibfnamefont{T.}~\bibnamefont{Pohl}},
  \bibinfo{journal}{Phys. Rev. Lett.} \textbf{\bibinfo{volume}{106}},
  \bibinfo{pages}{170401} (\bibinfo{year}{2011}).

\bibitem[{\citenamefont{Sela et~al.}(2011)\citenamefont{Sela, Punk, and
  Garst}}]{Sela2011}
\bibinfo{author}{\bibfnamefont{E.}~\bibnamefont{Sela}},
  \bibinfo{author}{\bibfnamefont{M.}~\bibnamefont{Punk}}, \bibnamefont{and}
  \bibinfo{author}{\bibfnamefont{M.}~\bibnamefont{Garst}},
  \bibinfo{journal}{Phys. Rev. B} \textbf{\bibinfo{volume}{84}},
  \bibinfo{pages}{085434} (\bibinfo{year}{2011}).

\bibitem[{\citenamefont{Schachenmayer et~al.}(2010)\citenamefont{Schachenmayer,
  Lesanovsky, Micheli, and Daley}}]{Schachenmayer2011}
\bibinfo{author}{\bibfnamefont{J.}~\bibnamefont{Schachenmayer}},
  \bibinfo{author}{\bibfnamefont{I.}~\bibnamefont{Lesanovsky}},
  \bibinfo{author}{\bibfnamefont{A.}~\bibnamefont{Micheli}}, \bibnamefont{and}
  \bibinfo{author}{\bibfnamefont{A.~J.} \bibnamefont{Daley}},
  \bibinfo{journal}{New J. Phys.} \textbf{\bibinfo{volume}{12}},
  \bibinfo{pages}{103044} (\bibinfo{year}{2010}).

\bibitem[{\citenamefont{Weimer et~al.}(2012)\citenamefont{Weimer, Yao, Laumann,
  and Lukin}}]{Weimer2012}
\bibinfo{author}{\bibfnamefont{H.}~\bibnamefont{Weimer}},
  \bibinfo{author}{\bibfnamefont{N.~Y.} \bibnamefont{Yao}},
  \bibinfo{author}{\bibfnamefont{C.~R.} \bibnamefont{Laumann}},
  \bibnamefont{and} \bibinfo{author}{\bibfnamefont{M.~D.} \bibnamefont{Lukin}},
  \bibinfo{journal}{Phys. Rev. Lett.} \textbf{\bibinfo{volume}{108}},
  \bibinfo{pages}{100501} (\bibinfo{year}{2012}).

\bibitem[{\citenamefont{Jaksch et~al.}(2000)\citenamefont{Jaksch, Cirac,
  Zoller, Rolston, Cote, and Lukin}}]{Jaksch2000}
\bibinfo{author}{\bibfnamefont{D.}~\bibnamefont{Jaksch}},
  \bibinfo{author}{\bibfnamefont{J.}~\bibnamefont{Cirac}},
  \bibinfo{author}{\bibfnamefont{P.}~\bibnamefont{Zoller}},
  \bibinfo{author}{\bibfnamefont{S.}~\bibnamefont{Rolston}},
  \bibinfo{author}{\bibfnamefont{R.}~\bibnamefont{Cote}}, \bibnamefont{and}
  \bibinfo{author}{\bibfnamefont{M.}~\bibnamefont{Lukin}},
  \bibinfo{journal}{Phys. Rev. Lett.} \textbf{\bibinfo{volume}{85}},
  \bibinfo{pages}{2208} (\bibinfo{year}{2000}).

\bibitem[{\citenamefont{Lukin et~al.}(2001)\citenamefont{Lukin, Fleischhauer,
  Cote, Duan, Jaksch, Cirac, and Zoller}}]{Lukin2001}
\bibinfo{author}{\bibfnamefont{M.~D.} \bibnamefont{Lukin}},
  \bibinfo{author}{\bibfnamefont{M.}~\bibnamefont{Fleischhauer}},
  \bibinfo{author}{\bibfnamefont{R.}~\bibnamefont{Cote}},
  \bibinfo{author}{\bibfnamefont{L.}~\bibnamefont{Duan}},
  \bibinfo{author}{\bibfnamefont{D.}~\bibnamefont{Jaksch}},
  \bibinfo{author}{\bibfnamefont{J.}~\bibnamefont{Cirac}}, \bibnamefont{and}
  \bibinfo{author}{\bibfnamefont{P.}~\bibnamefont{Zoller}},
  \bibinfo{journal}{Phys. Rev. Lett.} \textbf{\bibinfo{volume}{87}},
  \bibinfo{pages}{037901} (\bibinfo{year}{2001}).

\bibitem[{\citenamefont{Urban et~al.}(2009)\citenamefont{Urban, Johnson,
  Henage, Isenhower, Yavuz, Walker, and Saffman}}]{Urban2009a}
\bibinfo{author}{\bibfnamefont{E.}~\bibnamefont{Urban}},
  \bibinfo{author}{\bibfnamefont{T.~A.} \bibnamefont{Johnson}},
  \bibinfo{author}{\bibfnamefont{T.}~\bibnamefont{Henage}},
  \bibinfo{author}{\bibfnamefont{L.}~\bibnamefont{Isenhower}},
  \bibinfo{author}{\bibfnamefont{D.~D.} \bibnamefont{Yavuz}},
  \bibinfo{author}{\bibfnamefont{T.~G.} \bibnamefont{Walker}},
  \bibnamefont{and} \bibinfo{author}{\bibfnamefont{M.}~\bibnamefont{Saffman}},
  \bibinfo{journal}{Nature Phys.} \textbf{\bibinfo{volume}{5}},
  \bibinfo{pages}{110} (\bibinfo{year}{2009}).

\bibitem[{\citenamefont{Ga\"{e}tan et~al.}(2009)\citenamefont{Ga\"{e}tan,
  Miroshnychenko, Wilk, Chotia, Viteau, Comparat, Pillet, Browaeys, and
  Grangier}}]{Gaetan2009}
\bibinfo{author}{\bibfnamefont{A.}~\bibnamefont{Ga\"{e}tan}},
  \bibinfo{author}{\bibfnamefont{Y.}~\bibnamefont{Miroshnychenko}},
  \bibinfo{author}{\bibfnamefont{T.}~\bibnamefont{Wilk}},
  \bibinfo{author}{\bibfnamefont{A.}~\bibnamefont{Chotia}},
  \bibinfo{author}{\bibfnamefont{M.}~\bibnamefont{Viteau}},
  \bibinfo{author}{\bibfnamefont{D.}~\bibnamefont{Comparat}},
  \bibinfo{author}{\bibfnamefont{P.}~\bibnamefont{Pillet}},
  \bibinfo{author}{\bibfnamefont{A.}~\bibnamefont{Browaeys}}, \bibnamefont{and}
  \bibinfo{author}{\bibfnamefont{P.}~\bibnamefont{Grangier}},
  \bibinfo{journal}{Nature Phys.} \textbf{\bibinfo{volume}{5}},
  \bibinfo{pages}{115} (\bibinfo{year}{2009}).

\bibitem[{\citenamefont{Isenhower et~al.}(2010)\citenamefont{Isenhower, Urban,
  Zhang, Gill, Henage, Johnson, Walker, and Saffman}}]{Isenhower2010}
\bibinfo{author}{\bibfnamefont{L.}~\bibnamefont{Isenhower}},
  \bibinfo{author}{\bibfnamefont{E.}~\bibnamefont{Urban}},
  \bibinfo{author}{\bibfnamefont{X.~L.} \bibnamefont{Zhang}},
  \bibinfo{author}{\bibfnamefont{A.~T.} \bibnamefont{Gill}},
  \bibinfo{author}{\bibfnamefont{T.}~\bibnamefont{Henage}},
  \bibinfo{author}{\bibfnamefont{T.~A.} \bibnamefont{Johnson}},
  \bibinfo{author}{\bibfnamefont{T.~G.} \bibnamefont{Walker}},
  \bibnamefont{and} \bibinfo{author}{\bibfnamefont{M.}~\bibnamefont{Saffman}},
  \bibinfo{journal}{Phys. Rev. Lett.} \textbf{\bibinfo{volume}{104}},
  \bibinfo{pages}{010503} (\bibinfo{year}{2010}).

\bibitem[{\citenamefont{Wilk et~al.}(2010)\citenamefont{Wilk, Ga\"{e}tan,
  Evellin, Wolters, Miroshnychenko, Grangier, and Browaeys}}]{Wilk2010}
\bibinfo{author}{\bibfnamefont{T.}~\bibnamefont{Wilk}},
  \bibinfo{author}{\bibfnamefont{A.}~\bibnamefont{Ga\"{e}tan}},
  \bibinfo{author}{\bibfnamefont{C.}~\bibnamefont{Evellin}},
  \bibinfo{author}{\bibfnamefont{J.}~\bibnamefont{Wolters}},
  \bibinfo{author}{\bibfnamefont{Y.}~\bibnamefont{Miroshnychenko}},
  \bibinfo{author}{\bibfnamefont{P.}~\bibnamefont{Grangier}}, \bibnamefont{and}
  \bibinfo{author}{\bibfnamefont{A.}~\bibnamefont{Browaeys}},
  \bibinfo{journal}{Phys. Rev. Lett.} \textbf{\bibinfo{volume}{104}},
  \bibinfo{pages}{010502} (\bibinfo{year}{2010}).

\bibitem[{\citenamefont{Brion et~al.}(2007)\citenamefont{Brion, M{\o}lmer, and
  Saffman}}]{Brion2007}
\bibinfo{author}{\bibfnamefont{E.}~\bibnamefont{Brion}},
  \bibinfo{author}{\bibfnamefont{K.}~\bibnamefont{M{\o}lmer}},
  \bibnamefont{and} \bibinfo{author}{\bibfnamefont{M.}~\bibnamefont{Saffman}},
  \bibinfo{journal}{Phys. Rev. Lett.} \textbf{\bibinfo{volume}{99}},
  \bibinfo{pages}{260501} (\bibinfo{year}{2007}).

\bibitem[{\citenamefont{Saffman and M{\o}lmer}(2008)}]{Saffman2008}
\bibinfo{author}{\bibfnamefont{M.}~\bibnamefont{Saffman}} \bibnamefont{and}
  \bibinfo{author}{\bibfnamefont{K.}~\bibnamefont{M{\o}lmer}},
  \bibinfo{journal}{Phys. Rev. A} \textbf{\bibinfo{volume}{78}},
  \bibinfo{pages}{012336} (\bibinfo{year}{2008}).

\bibitem[{\citenamefont{M{\o }ller et~al.}(2008)\citenamefont{M{\o }ller,
  Madsen, and M{\o}lmer}}]{Moller2008}
\bibinfo{author}{\bibfnamefont{D.}~\bibnamefont{M{\o }ller}},
  \bibinfo{author}{\bibfnamefont{L.}~\bibnamefont{Madsen}}, \bibnamefont{and}
  \bibinfo{author}{\bibfnamefont{K.}~\bibnamefont{M{\o}lmer}},
  \bibinfo{journal}{Phys. Rev. Lett.} \textbf{\bibinfo{volume}{100}},
  \bibinfo{pages}{170504} (\bibinfo{year}{2008}).

\bibitem[{\citenamefont{M\"{u}ller et~al.}(2009)\citenamefont{M\"{u}ller,
  Lesanovsky, Weimer, B\"{u}chler, and Zoller}}]{Muller2009}
\bibinfo{author}{\bibfnamefont{M.}~\bibnamefont{M\"{u}ller}},
  \bibinfo{author}{\bibfnamefont{I.}~\bibnamefont{Lesanovsky}},
  \bibinfo{author}{\bibfnamefont{H.}~\bibnamefont{Weimer}},
  \bibinfo{author}{\bibfnamefont{H.~P.} \bibnamefont{B\"{u}chler}},
  \bibnamefont{and} \bibinfo{author}{\bibfnamefont{P.}~\bibnamefont{Zoller}},
  \bibinfo{journal}{Phys. Rev. Lett.} \textbf{\bibinfo{volume}{102}},
  \bibinfo{pages}{170502} (\bibinfo{year}{2009}).

\bibitem[{\citenamefont{Olmos et~al.}(2009)\citenamefont{Olmos,
  Gonz\'{a}lez-F\'{e}rez, and Lesanovsky}}]{Olmos2009a}
\bibinfo{author}{\bibfnamefont{B.}~\bibnamefont{Olmos}},
  \bibinfo{author}{\bibfnamefont{R.}~\bibnamefont{Gonz\'{a}lez-F\'{e}rez}},
  \bibnamefont{and}
  \bibinfo{author}{\bibfnamefont{I.}~\bibnamefont{Lesanovsky}},
  \bibinfo{journal}{Phys. Rev. Lett.} \textbf{\bibinfo{volume}{103}},
  \bibinfo{pages}{185302} (\bibinfo{year}{2009}).

\bibitem[{\citenamefont{Weimer et~al.}(2010)\citenamefont{Weimer, M\"{u}ller,
  Lesanovsky, Zoller, and B\"{u}chler}}]{Weimer2010a}
\bibinfo{author}{\bibfnamefont{H.}~\bibnamefont{Weimer}},
  \bibinfo{author}{\bibfnamefont{M.}~\bibnamefont{M\"{u}ller}},
  \bibinfo{author}{\bibfnamefont{I.}~\bibnamefont{Lesanovsky}},
  \bibinfo{author}{\bibfnamefont{P.}~\bibnamefont{Zoller}}, \bibnamefont{and}
  \bibinfo{author}{\bibfnamefont{H.}~\bibnamefont{B\"{u}chler}},
  \bibinfo{journal}{Nature Phys.} \textbf{\bibinfo{volume}{6}},
  \bibinfo{pages}{382} (\bibinfo{year}{2010}).

\bibitem[{\citenamefont{Han et~al.}(2010)\citenamefont{Han, He, Heshami, Li,
  and Simon}}]{Han2010}
\bibinfo{author}{\bibfnamefont{Y.}~\bibnamefont{Han}},
  \bibinfo{author}{\bibfnamefont{B.}~\bibnamefont{He}},
  \bibinfo{author}{\bibfnamefont{K.}~\bibnamefont{Heshami}},
  \bibinfo{author}{\bibfnamefont{C.-Z.} \bibnamefont{Li}}, \bibnamefont{and}
  \bibinfo{author}{\bibfnamefont{C.}~\bibnamefont{Simon}},
  \bibinfo{journal}{Phys. Rev. A} \textbf{\bibinfo{volume}{81}},
  \bibinfo{pages}{052311} (\bibinfo{year}{2010}).

\bibitem[{\citenamefont{Zhao et~al.}(2010)\citenamefont{Zhao, M\"uller,
  Hammerer, and Zoller}}]{Zhao2010}
\bibinfo{author}{\bibfnamefont{B.}~\bibnamefont{Zhao}},
  \bibinfo{author}{\bibfnamefont{M.}~\bibnamefont{M\"uller}},
  \bibinfo{author}{\bibfnamefont{K.}~\bibnamefont{Hammerer}}, \bibnamefont{and}
  \bibinfo{author}{\bibfnamefont{P.}~\bibnamefont{Zoller}},
  \bibinfo{journal}{Phys. Rev. A} \textbf{\bibinfo{volume}{81}},
  \bibinfo{pages}{052329} (\bibinfo{year}{2010}).

\bibitem[{\citenamefont{Kuznetsova et~al.}(2011)\citenamefont{Kuznetsova,
  Rittenhouse, Sadeghpour, and Yelin}}]{Kuznetsova2011}
\bibinfo{author}{\bibfnamefont{E.}~\bibnamefont{Kuznetsova}},
  \bibinfo{author}{\bibfnamefont{S.~T.} \bibnamefont{Rittenhouse}},
  \bibinfo{author}{\bibfnamefont{H.~R.} \bibnamefont{Sadeghpour}},
  \bibnamefont{and} \bibinfo{author}{\bibfnamefont{S.~F.} \bibnamefont{Yelin}},
  \bibinfo{journal}{Phys. Chem. Chem. Phys.} \textbf{\bibinfo{volume}{13}},
  \bibinfo{pages}{17115} (\bibinfo{year}{2011}).

\bibitem[{\citenamefont{Gorshkov et~al.}(2011)\citenamefont{Gorshkov,
  Otterbach, Fleischhauer, Pohl, and Lukin}}]{Gorshkov2011}
\bibinfo{author}{\bibfnamefont{A.~V.} \bibnamefont{Gorshkov}},
  \bibinfo{author}{\bibfnamefont{J.}~\bibnamefont{Otterbach}},
  \bibinfo{author}{\bibfnamefont{M.}~\bibnamefont{Fleischhauer}},
  \bibinfo{author}{\bibfnamefont{T.}~\bibnamefont{Pohl}}, \bibnamefont{and}
  \bibinfo{author}{\bibfnamefont{M.~D.} \bibnamefont{Lukin}},
  \bibinfo{journal}{Phys. Rev. Lett.} \textbf{\bibinfo{volume}{107}},
  \bibinfo{pages}{133602} (\bibinfo{year}{2011}).

\bibitem[{\citenamefont{Heidemann et~al.}(2008)\citenamefont{Heidemann,
  Raitzsch, Bendkowsky, Butscher, L\"{o}w, and Pfau}}]{Heidemann2008}
\bibinfo{author}{\bibfnamefont{R.}~\bibnamefont{Heidemann}},
  \bibinfo{author}{\bibfnamefont{U.}~\bibnamefont{Raitzsch}},
  \bibinfo{author}{\bibfnamefont{V.}~\bibnamefont{Bendkowsky}},
  \bibinfo{author}{\bibfnamefont{B.}~\bibnamefont{Butscher}},
  \bibinfo{author}{\bibfnamefont{R.}~\bibnamefont{L\"{o}w}}, \bibnamefont{and}
  \bibinfo{author}{\bibfnamefont{T.}~\bibnamefont{Pfau}},
  \bibinfo{journal}{Phys. Rev. Lett.} \textbf{\bibinfo{volume}{100}},
  \bibinfo{pages}{033601} (\bibinfo{year}{2008}).

\bibitem[{\citenamefont{Viteau et~al.}(2011)\citenamefont{Viteau, Bason,
  Radogostowicz, Malossi, Ciampini, Morsch, and Arimondo}}]{Viteau2011}
\bibinfo{author}{\bibfnamefont{M.}~\bibnamefont{Viteau}},
  \bibinfo{author}{\bibfnamefont{M.}~\bibnamefont{Bason}},
  \bibinfo{author}{\bibfnamefont{J.}~\bibnamefont{Radogostowicz}},
  \bibinfo{author}{\bibfnamefont{N.}~\bibnamefont{Malossi}},
  \bibinfo{author}{\bibfnamefont{D.}~\bibnamefont{Ciampini}},
  \bibinfo{author}{\bibfnamefont{O.}~\bibnamefont{Morsch}}, \bibnamefont{and}
  \bibinfo{author}{\bibfnamefont{E.}~\bibnamefont{Arimondo}},
  \bibinfo{journal}{Phys. Rev. Lett.} \textbf{\bibinfo{volume}{107}},
  \bibinfo{pages}{060402} (\bibinfo{year}{2011}).

\bibitem[{\citenamefont{Amthor et~al.}(2010)\citenamefont{Amthor, Giese,
  Hofmann, and Weidem\"uller}}]{Amthor2010}
\bibinfo{author}{\bibfnamefont{T.}~\bibnamefont{Amthor}},
  \bibinfo{author}{\bibfnamefont{C.}~\bibnamefont{Giese}},
  \bibinfo{author}{\bibfnamefont{C.~S.} \bibnamefont{Hofmann}},
  \bibnamefont{and}
  \bibinfo{author}{\bibfnamefont{M.}~\bibnamefont{Weidem\"uller}},
  \bibinfo{journal}{Phys. Rev. Lett.} \textbf{\bibinfo{volume}{104}},
  \bibinfo{pages}{013001} (\bibinfo{year}{2010}).

\bibitem[{\citenamefont{K\"{u}bler et~al.}(2010)\citenamefont{K\"{u}bler,
  Shaffer, Baluktsian, L\"{o}w, and Pfau}}]{Kubler2010}
\bibinfo{author}{\bibfnamefont{H.}~\bibnamefont{K\"{u}bler}},
  \bibinfo{author}{\bibfnamefont{J.~P.} \bibnamefont{Shaffer}},
  \bibinfo{author}{\bibfnamefont{T.}~\bibnamefont{Baluktsian}},
  \bibinfo{author}{\bibfnamefont{R.}~\bibnamefont{L\"{o}w}}, \bibnamefont{and}
  \bibinfo{author}{\bibfnamefont{T.}~\bibnamefont{Pfau}},
  \bibinfo{journal}{Nature Photon.} \textbf{\bibinfo{volume}{4}},
  \bibinfo{pages}{112} (\bibinfo{year}{2010}).

\bibitem[{\citenamefont{Mourachko et~al.}(1998)\citenamefont{Mourachko,
  Comparat, de~Tomasi, Fioretti, Nosbaum, Akulin, and Pillet}}]{Mourachko1998}
\bibinfo{author}{\bibfnamefont{I.}~\bibnamefont{Mourachko}},
  \bibinfo{author}{\bibfnamefont{D.}~\bibnamefont{Comparat}},
  \bibinfo{author}{\bibfnamefont{F.}~\bibnamefont{de~Tomasi}},
  \bibinfo{author}{\bibfnamefont{A.}~\bibnamefont{Fioretti}},
  \bibinfo{author}{\bibfnamefont{P.}~\bibnamefont{Nosbaum}},
  \bibinfo{author}{\bibfnamefont{V.~M.} \bibnamefont{Akulin}},
  \bibnamefont{and} \bibinfo{author}{\bibfnamefont{P.}~\bibnamefont{Pillet}},
  \bibinfo{journal}{Phys. Rev. Lett.} \textbf{\bibinfo{volume}{80}},
  \bibinfo{pages}{253} (\bibinfo{year}{1998}).

\bibitem[{\citenamefont{Lesanovsky}(2011)}]{Lesanovsky2011}
\bibinfo{author}{\bibfnamefont{I.}~\bibnamefont{Lesanovsky}},
  \bibinfo{journal}{Phys. Rev. Lett.} \textbf{\bibinfo{volume}{106}},
  \bibinfo{pages}{025301} (\bibinfo{year}{2011}).

\bibitem[{\citenamefont{Li et~al.}(2012)\citenamefont{Li, Hamadeh, and
  Lesanovsky}}]{Li2012}
\bibinfo{author}{\bibfnamefont{W.}~\bibnamefont{Li}},
  \bibinfo{author}{\bibfnamefont{L.}~\bibnamefont{Hamadeh}}, \bibnamefont{and}
  \bibinfo{author}{\bibfnamefont{I.}~\bibnamefont{Lesanovsky}},
  \bibinfo{journal}{Phys. Rev. A} \textbf{\bibinfo{volume}{85}},
  \bibinfo{pages}{053615} (\bibinfo{year}{2012}).

\bibitem[{\citenamefont{Santos et~al.}(2000)\citenamefont{Santos, Shlyapnikov,
  Zoller, and Lewenstein}}]{Santos2000}
\bibinfo{author}{\bibfnamefont{L.}~\bibnamefont{Santos}},
  \bibinfo{author}{\bibfnamefont{G.}~\bibnamefont{Shlyapnikov}},
  \bibinfo{author}{\bibfnamefont{P.}~\bibnamefont{Zoller}}, \bibnamefont{and}
  \bibinfo{author}{\bibfnamefont{M.}~\bibnamefont{Lewenstein}},
  \bibinfo{journal}{Phys. Rev. Lett.} \textbf{\bibinfo{volume}{85}},
  \bibinfo{pages}{1791} (\bibinfo{year}{2000}).

\bibitem[{\citenamefont{Baranov}(2008)}]{Baranov2008}
\bibinfo{author}{\bibfnamefont{M.}~\bibnamefont{Baranov}},
  \bibinfo{journal}{Phys. Rep.} \textbf{\bibinfo{volume}{464}},
  \bibinfo{pages}{71 } (\bibinfo{year}{2008}).

\bibitem[{\citenamefont{Carr et~al.}(2009)\citenamefont{Carr, DeMille, Krems,
  and Ye}}]{Carr2009}
\bibinfo{author}{\bibfnamefont{L.}~\bibnamefont{Carr}},
  \bibinfo{author}{\bibfnamefont{D.}~\bibnamefont{DeMille}},
  \bibinfo{author}{\bibfnamefont{R.}~\bibnamefont{Krems}}, \bibnamefont{and}
  \bibinfo{author}{\bibfnamefont{J.}~\bibnamefont{Ye}}, \bibinfo{journal}{New
  J. Phys.} \textbf{\bibinfo{volume}{11}}, \bibinfo{pages}{055049}
  (\bibinfo{year}{2009}).

\bibitem[{\citenamefont{Lahaye et~al.}(2009)\citenamefont{Lahaye, Menotti,
  Santos, Lewenstein, and Pfau}}]{Lahaye2009}
\bibinfo{author}{\bibfnamefont{T.}~\bibnamefont{Lahaye}},
  \bibinfo{author}{\bibfnamefont{C.}~\bibnamefont{Menotti}},
  \bibinfo{author}{\bibfnamefont{L.}~\bibnamefont{Santos}},
  \bibinfo{author}{\bibfnamefont{M.}~\bibnamefont{Lewenstein}},
  \bibnamefont{and} \bibinfo{author}{\bibfnamefont{T.}~\bibnamefont{Pfau}},
  \bibinfo{journal}{Rep. Prog. Phys.} \textbf{\bibinfo{volume}{72}},
  \bibinfo{pages}{126401} (\bibinfo{year}{2009}).

\bibitem[{\citenamefont{Boninsegni and Prokof’ev}(2012)}]{Boninsegni2012}
\bibinfo{author}{\bibfnamefont{M.}~\bibnamefont{Boninsegni}} \bibnamefont{and}
  \bibinfo{author}{\bibfnamefont{N.}~\bibnamefont{Prokof’ev}},
  \bibinfo{journal}{Rev. Mod. Phys.} \textbf{\bibinfo{volume}{84}},
  \bibinfo{pages}{759} (\bibinfo{year}{2012}).

\bibitem[{\citenamefont{{Baranov} et~al.}(2012)\citenamefont{{Baranov},
  {Dalmonte}, {Pupillo}, and {Zoller}}}]{Baranov2012}
\bibinfo{author}{\bibfnamefont{M.~A.} \bibnamefont{{Baranov}}},
  \bibinfo{author}{\bibfnamefont{M.}~\bibnamefont{{Dalmonte}}},
  \bibinfo{author}{\bibfnamefont{G.}~\bibnamefont{{Pupillo}}},
  \bibnamefont{and} \bibinfo{author}{\bibfnamefont{P.}~\bibnamefont{{Zoller}}},
  \bibinfo{journal}{arXiv:1207.1914}  (\bibinfo{year}{2012}).

\bibitem[{\citenamefont{B\"{u}chler et~al.}(2007)\citenamefont{B\"{u}chler,
  Demler, Lukin, Micheli, Prokof’ev, Pupillo, and Zoller}}]{Buchler2007}
\bibinfo{author}{\bibfnamefont{H.~P.} \bibnamefont{B\"{u}chler}},
  \bibinfo{author}{\bibfnamefont{E.}~\bibnamefont{Demler}},
  \bibinfo{author}{\bibfnamefont{M.}~\bibnamefont{Lukin}},
  \bibinfo{author}{\bibfnamefont{A.}~\bibnamefont{Micheli}},
  \bibinfo{author}{\bibfnamefont{N.}~\bibnamefont{Prokof’ev}},
  \bibinfo{author}{\bibfnamefont{G.}~\bibnamefont{Pupillo}}, \bibnamefont{and}
  \bibinfo{author}{\bibfnamefont{P.}~\bibnamefont{Zoller}},
  \bibinfo{journal}{Phys. Rev. Lett.} \textbf{\bibinfo{volume}{98}},
  \bibinfo{pages}{060404} (\bibinfo{year}{2007}).

\bibitem[{\citenamefont{Huber and B\"{u}chler}(2012)}]{Huber2012}
\bibinfo{author}{\bibfnamefont{S.~D.} \bibnamefont{Huber}} \bibnamefont{and}
  \bibinfo{author}{\bibfnamefont{H.~P.} \bibnamefont{B\"{u}chler}},
  \bibinfo{journal}{Phys. Rev. Lett.} \textbf{\bibinfo{volume}{108}},
  \bibinfo{pages}{193006} (\bibinfo{year}{2012}).

\bibitem[{\citenamefont{Zhao et~al.}(2012)\citenamefont{Zhao, Glaetzle,
  Pupillo, and Zoller}}]{Zhao2012}
\bibinfo{author}{\bibfnamefont{B.}~\bibnamefont{Zhao}},
  \bibinfo{author}{\bibfnamefont{A.~W.} \bibnamefont{Glaetzle}},
  \bibinfo{author}{\bibfnamefont{G.}~\bibnamefont{Pupillo}}, \bibnamefont{and}
  \bibinfo{author}{\bibfnamefont{P.}~\bibnamefont{Zoller}},
  \bibinfo{journal}{Phys. Rev. Lett.} \textbf{\bibinfo{volume}{108}},
  \bibinfo{pages}{193007} (\bibinfo{year}{2012}).

\bibitem[{\citenamefont{Micheli et~al.}(2007)\citenamefont{Micheli, Pupillo,
  B\"{u}chler, and Zoller}}]{Micheli2007}
\bibinfo{author}{\bibfnamefont{A.}~\bibnamefont{Micheli}},
  \bibinfo{author}{\bibfnamefont{G.}~\bibnamefont{Pupillo}},
  \bibinfo{author}{\bibfnamefont{H.~P.} \bibnamefont{B\"{u}chler}},
  \bibnamefont{and} \bibinfo{author}{\bibfnamefont{P.}~\bibnamefont{Zoller}},
  \bibinfo{journal}{Phys. Rev. A} \textbf{\bibinfo{volume}{76}},
  \bibinfo{pages}{043604} (\bibinfo{year}{2007}).

\bibitem[{\citenamefont{Gorshkov et~al.}(2008)\citenamefont{Gorshkov, Rabl,
  Pupillo, Micheli, Zoller, Lukin, and B\"{u}chler}}]{Gorshkov2008}
\bibinfo{author}{\bibfnamefont{A.~V.} \bibnamefont{Gorshkov}},
  \bibinfo{author}{\bibfnamefont{P.}~\bibnamefont{Rabl}},
  \bibinfo{author}{\bibfnamefont{G.}~\bibnamefont{Pupillo}},
  \bibinfo{author}{\bibfnamefont{A.}~\bibnamefont{Micheli}},
  \bibinfo{author}{\bibfnamefont{P.}~\bibnamefont{Zoller}},
  \bibinfo{author}{\bibfnamefont{M.~D.} \bibnamefont{Lukin}}, \bibnamefont{and}
  \bibinfo{author}{\bibfnamefont{H.~P.} \bibnamefont{B\"{u}chler}},
  \bibinfo{journal}{Phys. Rev. Lett.} \textbf{\bibinfo{volume}{101}},
  \bibinfo{pages}{073201} (\bibinfo{year}{2008}).

\bibitem[{\citenamefont{Kalia and Vashishta}(1981)}]{Kalia1981}
\bibinfo{author}{\bibfnamefont{R.}~\bibnamefont{Kalia}} \bibnamefont{and}
  \bibinfo{author}{\bibfnamefont{P.}~\bibnamefont{Vashishta}},
  \bibinfo{journal}{J. Phys. C} \textbf{\bibinfo{volume}{14}},
  \bibinfo{pages}{643} (\bibinfo{year}{1981}).

\bibitem[{\citenamefont{Glaetzle~et. al.}()}]{Glaetzle}
\bibinfo{author}{\bibfnamefont{A.~W.} \bibnamefont{Glaetzle~et. al.}},
  \bibinfo{note}{in preperation}.

\bibitem[{\citenamefont{Cirac et~al.}(1992)\citenamefont{Cirac, Blatt, Zoller,
  and Phillips}}]{Cirac1992}
\bibinfo{author}{\bibfnamefont{J.}~\bibnamefont{Cirac}},
  \bibinfo{author}{\bibfnamefont{R.}~\bibnamefont{Blatt}},
  \bibinfo{author}{\bibfnamefont{P.}~\bibnamefont{Zoller}}, \bibnamefont{and}
  \bibinfo{author}{\bibfnamefont{W.}~\bibnamefont{Phillips}},
  \bibinfo{journal}{Phys. Rev. A} \textbf{\bibinfo{volume}{46}},
  \bibinfo{pages}{2668} (\bibinfo{year}{1992}).

\bibitem[{\citenamefont{Dalibard and Cohen-Tannoudji}(1985)}]{Dalibard1985b}
\bibinfo{author}{\bibfnamefont{J.}~\bibnamefont{Dalibard}} \bibnamefont{and}
  \bibinfo{author}{\bibfnamefont{C.}~\bibnamefont{Cohen-Tannoudji}},
  \bibinfo{journal}{J. Phys. B} \textbf{\bibinfo{volume}{18}},
  \bibinfo{pages}{1661} (\bibinfo{year}{1985}).

\bibitem[{\citenamefont{Beterov et~al.}(2009)\citenamefont{Beterov, Ryabtsev,
  Tretyakov, and Entin}}]{Beterov2009a}
\bibinfo{author}{\bibfnamefont{I.}~\bibnamefont{Beterov}},
  \bibinfo{author}{\bibfnamefont{I.}~\bibnamefont{Ryabtsev}},
  \bibinfo{author}{\bibfnamefont{D.}~\bibnamefont{Tretyakov}},
  \bibnamefont{and} \bibinfo{author}{\bibfnamefont{V.}~\bibnamefont{Entin}},
  \bibinfo{journal}{Phys. Rev. A} \textbf{\bibinfo{volume}{79}},
  \bibinfo{pages}{052504} (\bibinfo{year}{2009}).

\bibitem[{\citenamefont{Zimmerman et~al.}(1979)\citenamefont{Zimmerman,
  Littman, Kash, and Kleppner}}]{Zimmerman1979}
\bibinfo{author}{\bibfnamefont{M.}~\bibnamefont{Zimmerman}},
  \bibinfo{author}{\bibfnamefont{M.}~\bibnamefont{Littman}},
  \bibinfo{author}{\bibfnamefont{M.}~\bibnamefont{Kash}}, \bibnamefont{and}
  \bibinfo{author}{\bibfnamefont{D.}~\bibnamefont{Kleppner}},
  \bibinfo{journal}{Phys. Rev. A} \textbf{\bibinfo{volume}{20}},
  \bibinfo{pages}{2251} (\bibinfo{year}{1979}).

\bibitem[{\citenamefont{Weiner et~al.}(1999)\citenamefont{Weiner, Bagnato,
  Zilio, and Julienne}}]{Weiner1999}
\bibinfo{author}{\bibfnamefont{J.}~\bibnamefont{Weiner}},
  \bibinfo{author}{\bibfnamefont{V.~S.} \bibnamefont{Bagnato}},
  \bibinfo{author}{\bibfnamefont{S.}~\bibnamefont{Zilio}}, \bibnamefont{and}
  \bibinfo{author}{\bibfnamefont{P.~S.} \bibnamefont{Julienne}},
  \bibinfo{journal}{Rev. Mod. Phys.} \textbf{\bibinfo{volume}{71}},
  \bibinfo{pages}{1} (\bibinfo{year}{1999}).

\bibitem[{\citenamefont{Klimov et~al.}(2002)\citenamefont{Klimov, Sanchez-Soto,
  Navarro, and Yustas}}]{Klimov2002}
\bibinfo{author}{\bibfnamefont{A.~B.} \bibnamefont{Klimov}},
  \bibinfo{author}{\bibfnamefont{L.~L.} \bibnamefont{Sanchez-Soto}},
  \bibinfo{author}{\bibfnamefont{A.}~\bibnamefont{Navarro}}, \bibnamefont{and}
  \bibinfo{author}{\bibfnamefont{E.~C.} \bibnamefont{Yustas}},
  \bibinfo{journal}{J. Mod. Opt.} \textbf{\bibinfo{volume}{49}},
  \bibinfo{pages}{2211} (\bibinfo{year}{2002}).

\bibitem[{\citenamefont{Minogin and Letokhov}(1987)}]{Minogin1987}
\bibinfo{author}{\bibfnamefont{V.~G.} \bibnamefont{Minogin}} \bibnamefont{and}
  \bibinfo{author}{\bibfnamefont{V.~S.} \bibnamefont{Letokhov}},
  \emph{\bibinfo{title}{{Laser Light Pressure on Atoms}}}
  (\bibinfo{publisher}{Harwood Academic}, \bibinfo{year}{1987}).

\bibitem[{\citenamefont{Bates and Damgaard}(1949)}]{Bates1949}
\bibinfo{author}{\bibfnamefont{D.}~\bibnamefont{Bates}} \bibnamefont{and}
  \bibinfo{author}{\bibfnamefont{A.}~\bibnamefont{Damgaard}},
  \bibinfo{journal}{Phil. Trans. R. Soc. Lond. A}
  \textbf{\bibinfo{volume}{242}}, \bibinfo{pages}{101} (\bibinfo{year}{1949}).

\bibitem[{\citenamefont{Gounand}(1979)}]{Gounand1979}
\bibinfo{author}{\bibfnamefont{F.}~\bibnamefont{Gounand}}, \bibinfo{journal}{J.
  Phys. France} \textbf{\bibinfo{volume}{40}}, \bibinfo{pages}{457}
  (\bibinfo{year}{1979}).

\bibitem[{\citenamefont{Branden et~al.}(2010)\citenamefont{Branden, Juhasz,
  Mahlokozera, Vesa, Wilson, Zheng, Kortyna, and Tate}}]{Branden2010}
\bibinfo{author}{\bibfnamefont{D.}~\bibnamefont{Branden}},
  \bibinfo{author}{\bibfnamefont{T.}~\bibnamefont{Juhasz}},
  \bibinfo{author}{\bibfnamefont{T.}~\bibnamefont{Mahlokozera}},
  \bibinfo{author}{\bibfnamefont{C.}~\bibnamefont{Vesa}},
  \bibinfo{author}{\bibfnamefont{R.}~\bibnamefont{Wilson}},
  \bibinfo{author}{\bibfnamefont{M.}~\bibnamefont{Zheng}},
  \bibinfo{author}{\bibfnamefont{A.}~\bibnamefont{Kortyna}}, \bibnamefont{and}
  \bibinfo{author}{\bibfnamefont{D.}~\bibnamefont{Tate}}, \bibinfo{journal}{J.
  Phys. B} \textbf{\bibinfo{volume}{43}}, \bibinfo{pages}{015002}
  (\bibinfo{year}{2010}).

\bibitem[{\citenamefont{Marinescu et~al.}(1994)\citenamefont{Marinescu,
  Sadeghpour, and Dalgarno}}]{Marinescu1994}
\bibinfo{author}{\bibfnamefont{M.}~\bibnamefont{Marinescu}},
  \bibinfo{author}{\bibfnamefont{H.~R.} \bibnamefont{Sadeghpour}},
  \bibnamefont{and} \bibinfo{author}{\bibfnamefont{A.}~\bibnamefont{Dalgarno}},
  \bibinfo{journal}{Phys. Rev. A} \textbf{\bibinfo{volume}{49}},
  \bibinfo{pages}{982} (\bibinfo{year}{1994}).

\bibitem[{\citenamefont{Gardiner}(2004)}]{Gardiner2004}
\bibinfo{author}{\bibfnamefont{C.}~\bibnamefont{Gardiner}},
  \emph{\bibinfo{title}{{Handbook of Stochastic Methods: for Physics, Chemistry
  and the Natural Sciences}}} (\bibinfo{publisher}{Springer},
  \bibinfo{year}{2004}), \bibinfo{edition}{3rd} ed.

\bibitem[{\citenamefont{Olson}(1979)}]{Olson1979}
\bibinfo{author}{\bibfnamefont{R.~E.} \bibnamefont{Olson}},
  \bibinfo{journal}{Phys. Rev. Lett.} \textbf{\bibinfo{volume}{43}},
  \bibinfo{pages}{126} (\bibinfo{year}{1979}).

\bibitem[{\citenamefont{Robicheaux}(2005)}]{Penning2005}
\bibinfo{author}{\bibfnamefont{F.}~\bibnamefont{Robicheaux}},
  \bibinfo{journal}{J. Phys. B} \textbf{\bibinfo{volume}{38}},
  \bibinfo{pages}{333} (\bibinfo{year}{2005}).

\bibitem[{\citenamefont{Reinhard et~al.}(2008)\citenamefont{Reinhard, {Cubel
  Liebisch}, Younge, Berman, and Raithel}}]{Reinhard2008}
\bibinfo{author}{\bibfnamefont{A.}~\bibnamefont{Reinhard}},
  \bibinfo{author}{\bibfnamefont{T.}~\bibnamefont{{Cubel Liebisch}}},
  \bibinfo{author}{\bibfnamefont{K.~C.} \bibnamefont{Younge}},
  \bibinfo{author}{\bibfnamefont{P.~R.} \bibnamefont{Berman}},
  \bibnamefont{and} \bibinfo{author}{\bibfnamefont{G.}~\bibnamefont{Raithel}},
  \bibinfo{journal}{Phys. Rev. Lett.} \textbf{\bibinfo{volume}{100}},
  \bibinfo{pages}{123007} (\bibinfo{year}{2008}).

\bibitem[{\citenamefont{Weber et~al.}(2011)\citenamefont{Weber, Niederpr\"um,
  Manthey, Langer, Guarrera, Barontini, and Ott}}]{Ott2011}
\bibinfo{author}{\bibfnamefont{T.~M.} \bibnamefont{Weber}},
  \bibinfo{author}{\bibfnamefont{T.}~\bibnamefont{Niederpr\"um}},
  \bibinfo{author}{\bibfnamefont{T.}~\bibnamefont{Manthey}},
  \bibinfo{author}{\bibfnamefont{P.}~\bibnamefont{Langer}},
  \bibinfo{author}{\bibfnamefont{V.}~\bibnamefont{Guarrera}},
  \bibinfo{author}{\bibfnamefont{G.}~\bibnamefont{Barontini}},
  \bibnamefont{and} \bibinfo{author}{\bibfnamefont{H.}~\bibnamefont{Ott}},
  \bibinfo{journal}{arXiv:1112.5118v1}  (\bibinfo{year}{2011}).

\end{thebibliography}
\end{document}